\documentclass[useAMS,usenatbib]{mnras}
\usepackage[dvips]{graphicx}
\usepackage{amssymb,amsmath}
\usepackage{color}
\usepackage{epstopdf}
\usepackage{natbib}
\usepackage{multirow}

\newcommand{\Mpcc}{{\rm Mpc}\,{\rm h}^{-1}}

\title[Gas removal]{ Satellite galaxies in groups in the CIELO Project I. Gas removal from galaxies and its re-distribution in the intragroup medium} 
\author[Rodr\'iguez  et al.]{S. Rodr\'iguez$^1$ \thanks{E-mail:silvio.rodriguez@unc.edu.ar}, D. Garcia Lambas$^{1,2}$, N. D. Padilla$^{1}$, P. Tissera$^{3,4}$,  L. Bignone$^{5}$, \newauthor{R. Dominguez-Tenreiro$^{6}$, R. Gonzalez$^{4}$ and S. Pedrosa$^5$}\\
$^1$ Instituto de Astronom\'ia Te\'orica y Experimental, UNC-CONICET, C\'ordoba, X5000BGR, Argentina\\
$^2$ Observatorio Astron\'omico de C\'ordoba, Universidad Nacional de C\'ordoba, X5000BGR, Argentina\\
$^3$ Instituto de Astrof\'isica, Pontificia Universidad Cat\'olica de Chile, 8970117, Santiago, Chile\\
$^4$ Centro de Astro-Ingenier\'ia, Pontificia Universidad Cat\'olica de Chile, 8970117, Santiago, Chile\\
$^5$ Instituto de Astronom\'ia y F\'isica del Espacio, CONICET, C1428ZAA Buenos Aires, Argentina\\
$^6$ Departamento de F\'isica Te\'orica, Universidad Aut\'onoma de Madrid, ES-28049, Madrid, Espa\~na}

\begin{document}

\date{Accepted --. Received --; in original form \today}

\pagerange{\pageref{firstpage}--\pageref{lastpage}} \pubyear{2020}

\maketitle

\label{firstpage}

\begin{abstract}
We study the impact of the environment on galaxies as they fall in and orbit in the potential well of a Local Group (LG) analogue, following them with high cadence.
The analysis is performed on eight disc satellite galaxies from 
the CIELO suite of hydrodynamical simulations. All galaxies have stellar masses within the range $[10^{8.1} - 10^{9.56}] M_{\sun} $h$^{-1}$.
We measure tidal torques, ram pressure and specific star formation rates (sSFR) as a function of time, and correlate them with the amount of gas lost by satellites along their orbits. Stronger removal episodes occur when the disc plane is oriented perpendicular to the direction of motion. More than one peripassage is required to significantly modify the orientations of the discs with respect to the orbital plane. The  gas removed during the interaction with the central galaxies may be also found opposite to the direction of motion, depending on the orbital configuration. Satellites are not totally quenched when the galaxies reach their first peripassage, and continue forming about  $10\%$ of the final stellar mass after this event. 
The fraction of removed gas is found to be the product of the joint action of  tidal torque and ram pressure, which can also trigger new star formation activity and subsequent supernova feedback. 

\end{abstract}

\begin{keywords}
software: simulations -- galaxies: evolution -- galaxies: interactions -- galaxies: star formation -- intergalactic medium
\end{keywords}

\section{Introduction}
\label{intro}

As galaxies form and evolve, they can fall into larger structures, such as groups and clusters \citep{GunnAndGott72}. This brings a series of changes in their properties such as the removal of different types of material (stars, gas, dark matter, dust) due to external, environmental processes \citep{Boselli2006}. This, in turn, can lead to variations in the stellar formation and metallicity of both the infalling galaxies \citep{Kennicutt1983, Skillman1996} and their immediate intragroup gas \citep{DeGrandi2001}.

The external processes responsible for the removal of material are either hydrodynamical
or gravitational. Among the hydrodynamical effects ram pressure stripping \citep{GunnAndGott72} is one of the most efficient mechanisms at quenching galaxies \citep{Steinhauser2016}. Due to this process part of the gas component is removed from galaxies by the pressure exerted by the hot gas in clusters and groups. Observational evidence of the action of ram pressure includes traces of ionized gas in the removed gas, which could be an indication of star formation activity \citep{Kenney1999, Yoshida2008, Kenney2014, Jaffe2014},  detection of removed neutral gas \citep{Haynes1984, Rasmussen2006, Yoon2017}, or indirect measures of the removed dust \citep{rgpt20}.

Among the gravitational mechanisms, the most important are tidal effects due to close encounters with other galaxies, especially with the central galaxy of the group or cluster \citep[see for instance,][]{Merrit1983}. The effects of  tidal interactions are expected to be stronger during, and after, the first pericentre with respect to the central galaxy when the star formation activity might be boosted as suggested by numerical simulations \citep{rupke2010,perez2011,sillero17}. \citet{Upadhyay2021} studied 11 galaxies in the COMA cluster and reported that galaxies are quenched efficiently  within $ \leq 1$ Gyr, after the first pericentre. They suggest that ram pressure or tidal stripping are required to produce this effect. This result is in contradiction with other works that claim quenching to occur closer to the apocentre \citep[][]{Rhee2020}. Hence, there is still much to be learnt on the  evolution and quenching of galaxies in relation to  environmental effects \citep{Dutta2021}. 

There are also mechanisms that could lead to removal of material from the galaxies due to internal causes rather than interactions with the environment.  One example is the feedback by galactic winds triggered by supernova (SN) explosions as indicated by several observational works \citep[see, for instance][]{deYoung1978,Heckman1990, Weiner2009, Martin2013, Rubin2014}. 
SN feedback is higher when star formation increases. However, SF has been reported to be modulated by
ram pressure which at first produces an increase of the star formation activity, followed by a subsequent decrease  \citep{Kapferer2009, Steinhauser2012, Safarzadeh2019, Troncoso2020}. Additionally, as tidal interactions could also enhance the star formation activity \citep{Condon1982, Keel1985, Kennicutt1987, Hummel1990, lambas2003, cintio21}, the impact of SN will be more significant during these events.

Several works have been advocated to study the coupled effects of these mechanisms in high density environments using numerical simulations \citep{Mayer2006, Mayer2007, Tonnesen2007, Bahe2015, Martin2019, Jackson2021}.
Some of these works  studied the impact of the environment over a galaxy focused on the changes in the composition of the galaxy, with only a few observational works focused on the detection of the removed material and its evolution \citep{BravoAlfaro2009,Bellhouse2017,Jaffe2018}. 

In this work, we  study the remotion by the intra-group medium of gaseous material from disc-like satellite galaxies from a 
Local Group analogue (hereafter, LG), following them with high cadence. This allows us to study the history 
of gas removal during and after the infall stage by ram pressure, tydal stripping, galaxy-galaxy interactions and SN feedback. We study the spatial distribution of the gas stripped from the satellite galaxies and its relation to  removal time and the physical properties of the satellites. Additionally, we analyse the impact on the star formation activity along their orbits.

To achieve this goal, we use data from a hydrodynamical re-simulation of a Local Group analogue of the Chemo-dynamIcal propertiEs of gaLaxies and the cOsmic web ({\sc CIELO}) Project. This simulation was performed using a version of {\small GADGET-3}, which includes SN feedback \citep{Scannapieco2006} and chemical evolution  \citep[][]{Jimenez2015,Pedrosa2015}. From this simulation, we select a sample of eight satellite galaxies with a significant disc component since we are also interested on the role played by the orientation of the galactic disc along the orbit and its relation with the SFR \citep{Kennicutt1998}.  These satellites orbit the two most massive virialized haloes of the simulated LG analogue, and are taken here as
representative of satellite systems of galaxies of similar virial masses, in general.


This work is organised as follows. In Section~\ref{sample} we describe the main characteristics of the simulation and  the selected sample of galaxies. In Section~\ref{remgas} we describe the processes responsible of  the gas removal and its dependence on orientation. 
Finally, in Section~\ref{conc} we present a brief summary and guidelines for future analysis of observational data.

\section{Simulations}
\label{sample}
In this paper we use a simulation of a LG analogue  of the 
the {\sc CIELO} project, which  is a long-term project which aims to study the formation of galaxies in different environments hosted by haloes with virial masses in the range $M_{200} =10^{10}-10^{12} M_{\sun} $h$^{-1}$ (Tissera et al., in preparation). 

Briefly, the initial conditions  of LG zoom-in simulations are based on a dark matter only run of a cosmological periodic cubic box of side length  $L = 100$ Mpc h$^{-1}$
consistent with a Lambda ($\Lambda$) Cold Dark Matter universe model with $\Omega_0=0.317$, $\Omega_{\Lambda}=  0.6825$, $\Omega_B=0.049,$ h$=0.6711$.

The MUSIC code \citep{HahnandAbel2011}, which computes multi-scale cosmological initial conditions under different approximations and transfer functions, was applied to extract the objects of interest and increase the numerical resolution.  A first set of 20 LG analogues  were extracted. Two of them were selected by imposing constraints on  the relative velocity, physical separation and the mass of the dark matter haloes. In this work we perform a detailed analysis of one of these two LG analogues. The other LG analogue hosts slightly smaller haloes, whose central galaxies have  stellar masses lower than the Milky Way at $z=0$.

The two main haloes of the analysed LG analogue have a relative velocity of   $165 \rm ~km\,s^{-1}$
and a physical separation of $0.461~\Mpcc$ at $z=0$\footnote{This is consistent with the estimates for the Local Group in \cite{vanderMarel2012}, where the MW and M31 are separated by a distance of $\sim 0.51~\Mpcc$ and have  relative velocity of $\rm 110 \rm ~km\,s^{-1}$.}.
The LG was re-run with a dark matter particle resolution of  $1.2 \times 10^{6} {\rm M_{\sun} h^{-1}}$.
Baryons were added with an initial gas mass of $\rm 10^{5.3} ~ M_{\sun} $h$^{-1}$.

The LG was run from $z =100$ to $z=0$. There are 128 snapshots available to study the evolution of the properties of the structure and galaxies with time. This simulation provides us with a suitable cadence to follow the impact of different physical processes as satellite galaxies fall into their main haloes.

\subsection{Subgrid physics}
We used a version of {\sc GADGET-3} based on  {\rm GADGET-2 } \citep{springel2003,springel2005} to run these simulations, which includes
a multiphase model for the gas component, metal-dependent cooling, star
formation, and SN feedback, as described in \citet{scan05} and
\citet{scan06}. These multiphase and SN-feedback models have been
used to successfully reproduce the star-formation activity of galaxies
during quiescent and starburst phases, and are able to drive violent
mass-loaded galactic winds with a strength reflecting the depth of the
potential well \citep{scan05,scan06}. This physically-motivated thermal
SN-feedback scheme is
particularly well-suited for the study of galaxy formation in a
cosmological context.

We assumed an Initial Mass Function of \citet{chabrier2003}, with lower and upper mass cut-offs of 0.1
${\rm M_{\odot}}$ and 40 ${\rm M_{\odot}}$, respectively. The  chemical evolution model included in LG follows the enrichment by Type~II and
Type~Ia Supernovae (SNII and SNIa, respectively), keeping track of 12 different chemical elements \citep{mosc2001}. 

SNII are assumed to originate from stars more massive than 8 ${\rm
M_{\odot}}$. Their nucleosynthesis products are derived from the
metal-dependent yields of \citet{WW95}. The lifetimes of SNII are
estimated according to the metal-and-mass-dependent lifetime-fitting
formulae of \citet{rait1996}. For SNIa, the model adopts the W7 model of
\citet{iwamoto1999}. More detailed explanations are given in \citet{scan06} and \citet{Jimenez2015}. This version of {\sc GADGET-3} has been previously used by \citet{Pedrosa2015}  to study the mass-size relation and the specific angular momentum content of galaxies, and by \citet{tissera2016a,tissera2016b} to investigate the origin of the metallicity gradients of the gas-phase components and stellar populations of galaxies in  the the so-called FENIX simulation.

\subsection{The simulated satellite galaxies and their host galaxies}

We extract the galaxies and their merger trees from the {\sc CIELO} database.
The virial haloes   were identified using a Friends-of-Friends algorithm \citep[FoF,][]{davis1985}. The {\sc SUBFIND} algorithm  \citep{springel2001a, dolag2009} was then applied to identify substructures within each of the virial haloes at all available redshift. The merger trees were built by using the {\sc AMIGA} algorithm \citep{amiga2009}.

At $z=0$ the central galaxies of the LG analogue, g4337 and g4469, belong to the main virialised haloes\footnote{The virial radius, $r_{200}$, is defined as the radius of a sphere located at the centre of a given  halo at which the  mass density is 200 times the critical density of the universe. Hereafter, every mention to $r_{200}$  refers to the virial radius of the halo to which a satellite galaxy belongs to at $z=0$.}, h4671 and h4672. 
The central galaxies have stellar masses of $10^{10.52} M_{\sun} $h$^{-1}$ and $10^{9.71} M_{\sun} $h$^{-1}$, respectively. Both of them are surrounded by surviving satellites at $z=0$. 
Throughout this paper, we adopt the most bound particle (star or dark matter) in each subhalo as the centre of each galaxy (central or satellite).

Each central galaxy and their satellites are followed back in time along their merger trees, identifying all of their main progenitors. At each analysed redshift, the following parameters were calculated for all of the sample. The size of a galaxy,  $r_{\rm opt}$, which is defined as the radius that encloses 83\% of stellar and star-forming gas mass \citep{tissera2012}. All quantities, such as the stellar mass  ($M_{\star}$), gas mass ($M_{\rm gas}$), galaxy mass ($M$, the  sum of stellar, dark matter and gas masses), and  the star formation rate (SFR), are calculated within $r_{\rm opt}$. 

After a careful analysis of the stability of the angular momentum calculations, we found that 500 particles is the minimum number that allows a robust estimation of the angular momentum direction. Hence, only satellites with more than 500 stellar particles are considered, this condition is met for the selected galaxies at all analysed redshifts. The kinetic factor defined as the ratio of rotational kinetic energy to total kinetic energy, $K_{\rm rot}/K$, is used to select satellites dominated by rotation \citep{navarro93}.

Our final sample comprises eight satellite galaxies within the virial radius of the haloes at $z=0$. These galaxies are also required to have a well-defined disc component so that $K_{\rm rot}/K > 0.4$. Table \ref{table_basic} summarises the main characteristics of our satellite sample, including the identification names of both the virialized haloes and their central galaxy.

\begin{table*}
	\begin{center}
		\caption{\label{table_basic} The selected satellite galaxies at $z = 0$. Columns from left to right correspond to halo and  central galaxy names, the halo virial mass, the stellar mass of the central galaxy, satellite name, stellar mass, rotational kinetic fraction and sSFR at $z=0$ of the galaxies. 
		}
		\begin{tabular}{cccccccc}
			 Halo ID & Central ID &$\rm M_ {200}$ & M$_{\star \rm Central}$ & Satellite ID & M$_{\star}$ &   $K_{\rm rot}/K$ & sSFR$_{z=0}$\\
			 & & $\log(M_{\sun} $h$^{-1})$ & $\log(M_{\sun} $h$^{-1})$ & & $\log(M_{\sun} $h$^{-1})$ & & yr$^{-1}$ \\\hline
			\multirow{4}*{\large h4671} & \multirow{4}*{\large g4337} & \multirow{4}*{\large 11.98} & \multirow{4}*{\large 10.52}  & g4338 & 8.74 & 0.55 & $1.5 \times 10^{-11}$\\
			      &       &       & &g4339 & 8.31 & 0.59 & $1.47 \times 10^{-11}$\\
			      &       &       & & g4341 & 8.16 & 0.52 & $0.0$\\
				  &       &       & & g4343 & 8.02 & 0.42 & $0.0$\\\hline
            \multirow{4}*{\large h4672} & \multirow{4}*{\large g4469} & \multirow{4}*{\large 11.78} & \multirow{4}*{\large 9.71} & g4470 & 9.48 & 0.62 & $6.2 \times 10^{-12}$\\
			      &       &       & & g4471 & 9.48 & 0.71 & $1.46 \times 10^{-10}$\\
				  &       &       & & g4473 & 8.50 & 0.72 & $6.65 \times 10^{-11}$\\
	              &       &       & & g4474 & 8.63 & 0.72 & $2.87 \times 10^{-11}$\\\hline
		\end{tabular}
	\end{center}
\end{table*}

\begin{figure*}
	\begin{center}
		\includegraphics[width=0.334\textwidth]{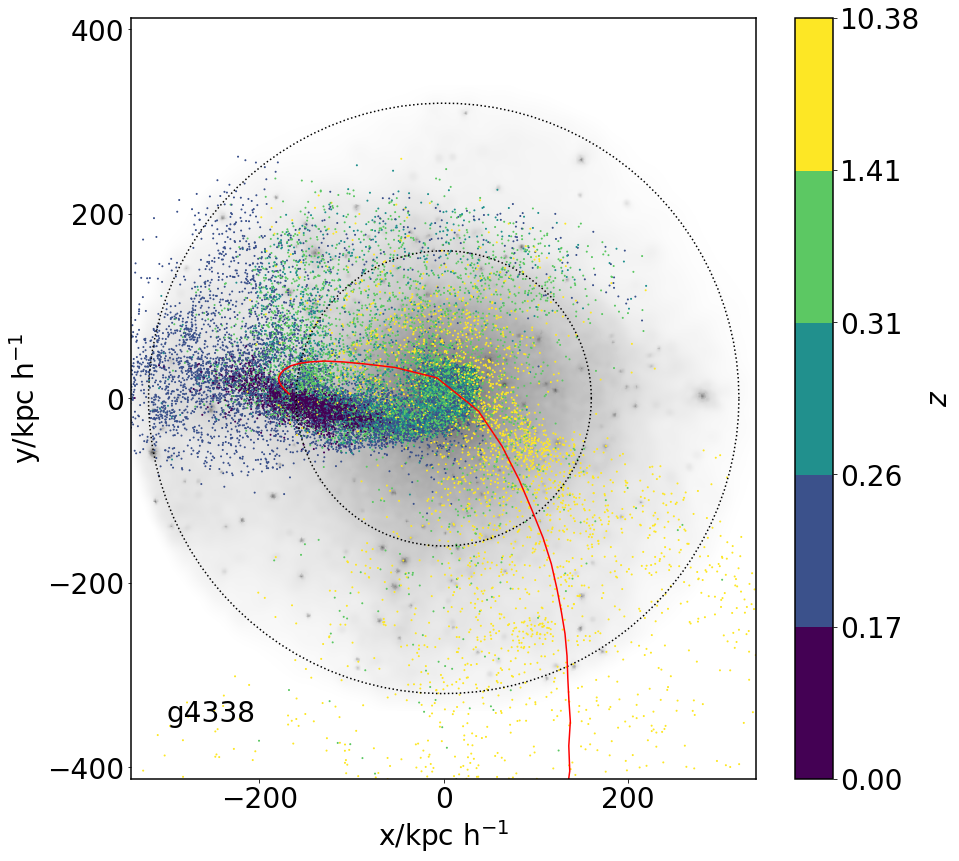}
		\includegraphics[width=0.334\textwidth]{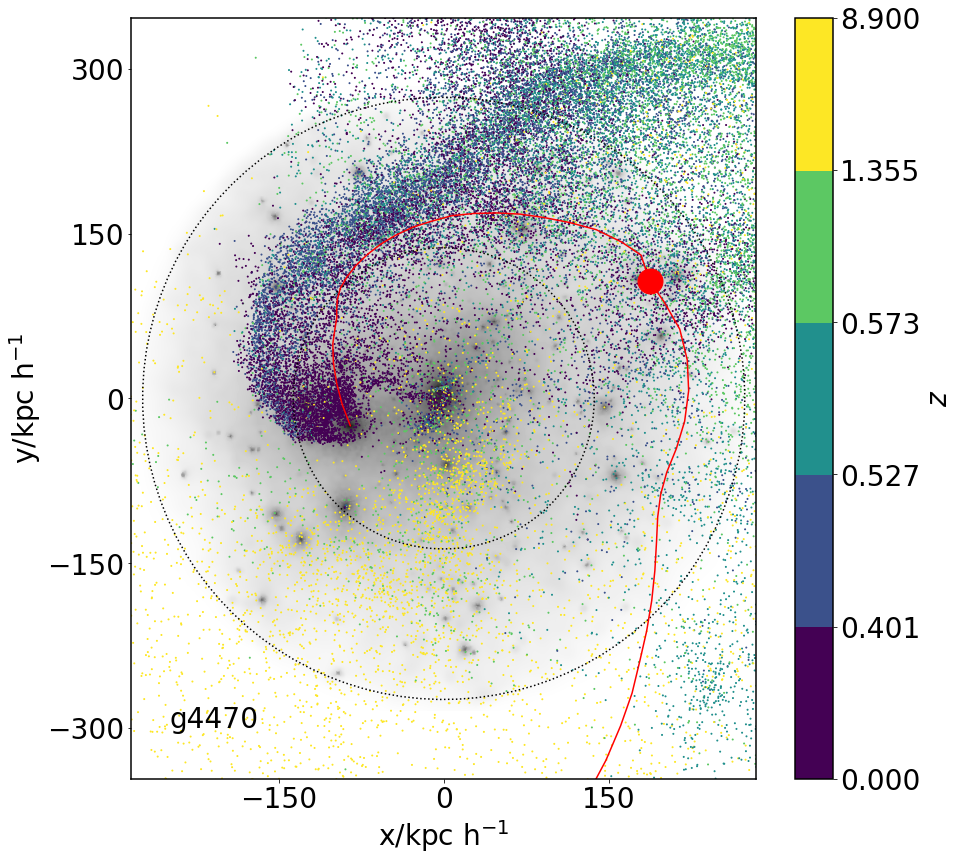}
		\includegraphics[width=0.334\textwidth]{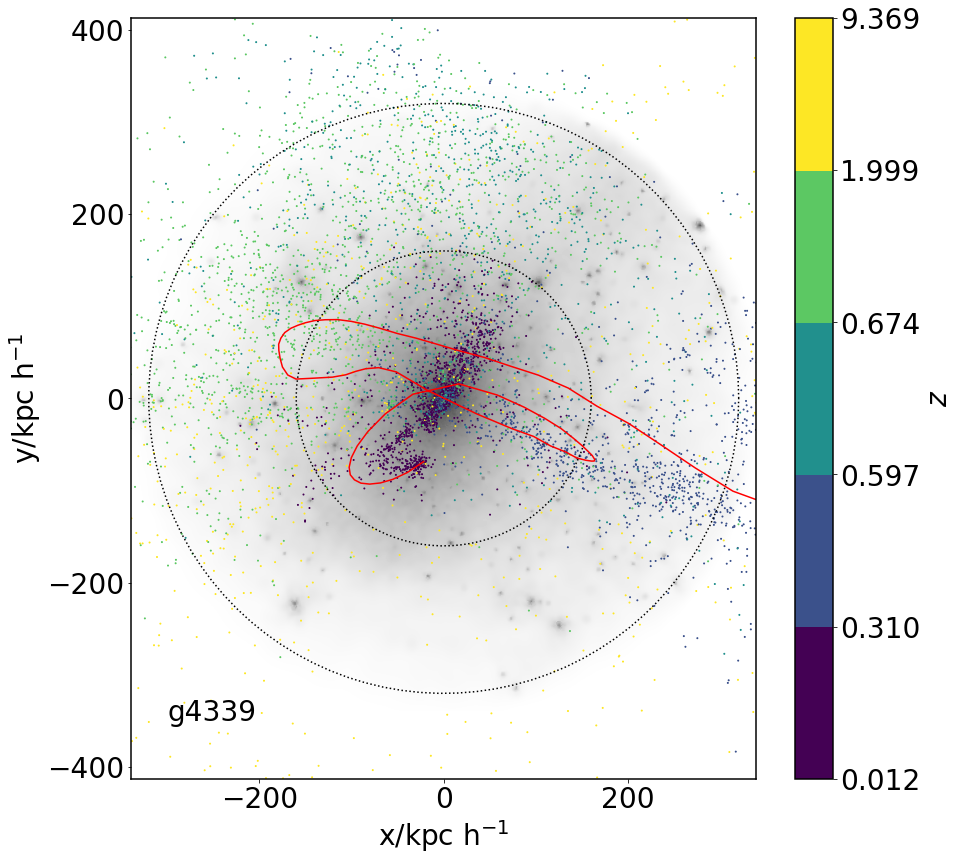}
		\includegraphics[width=0.334\textwidth]{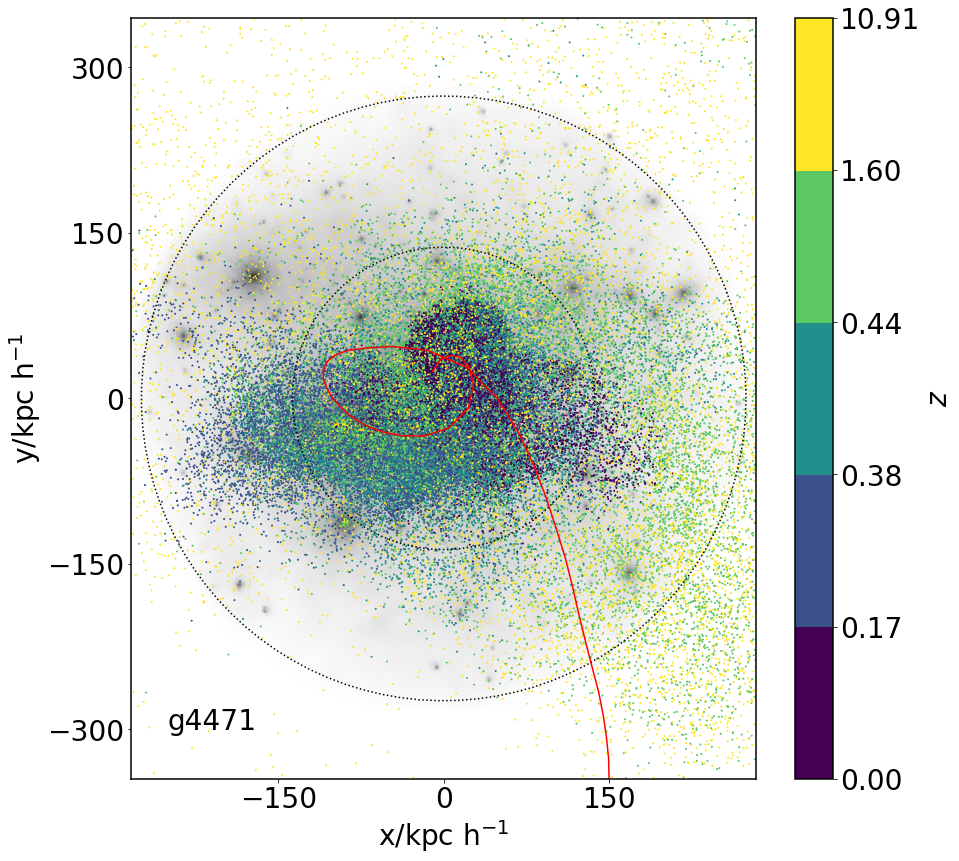}
		\includegraphics[width=0.334\textwidth]{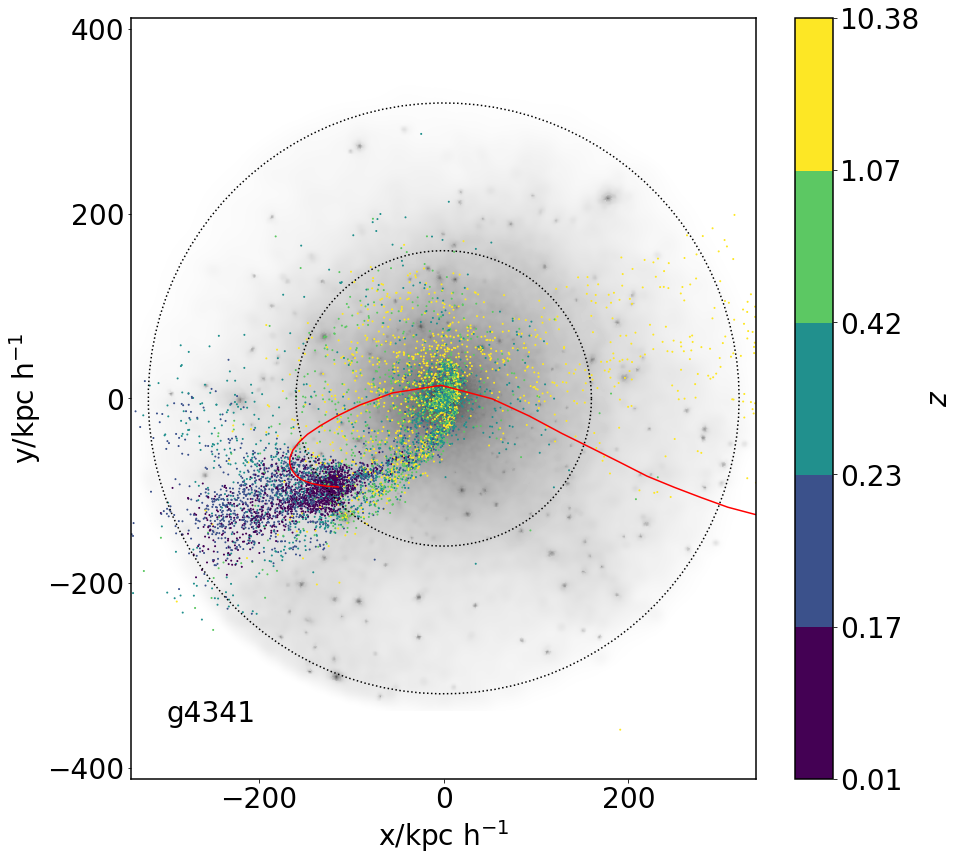}
		\includegraphics[width=0.334\textwidth]{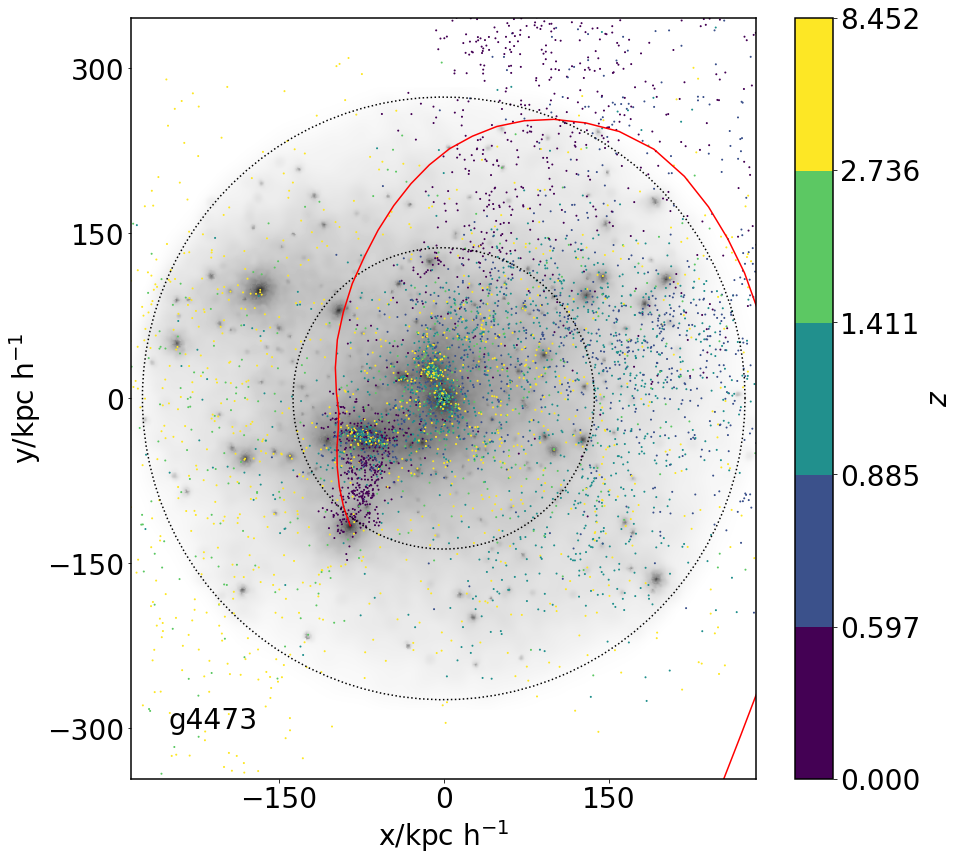}
		\includegraphics[width=0.334\textwidth]{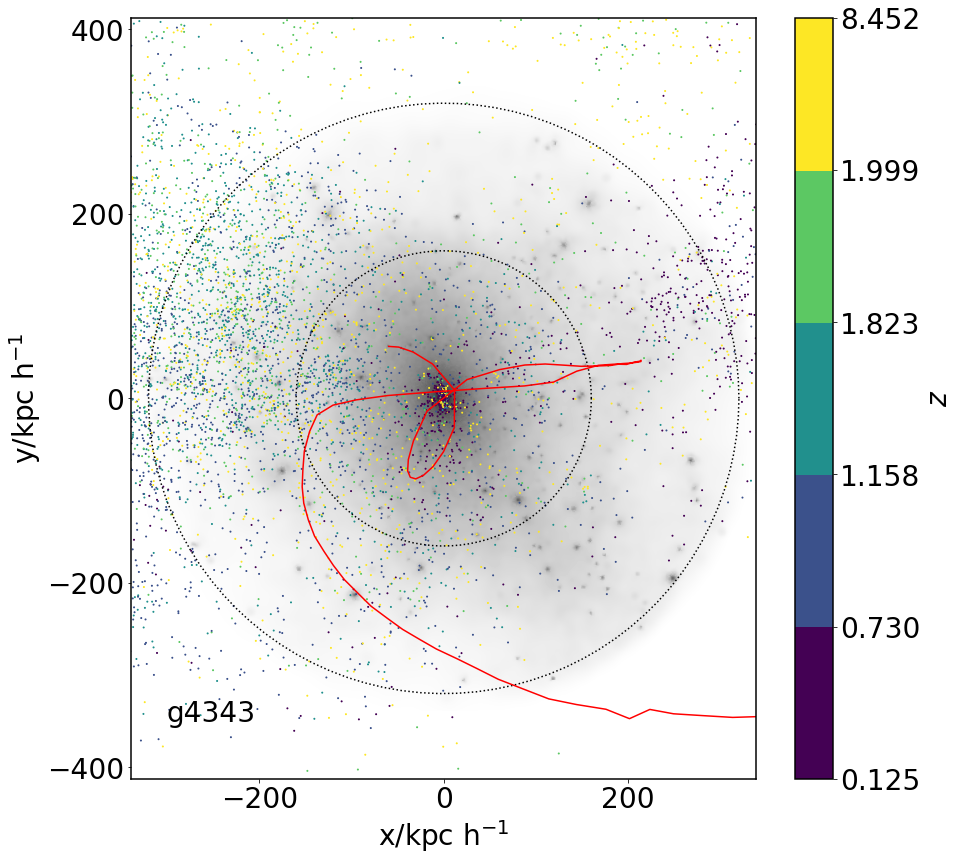}
		\includegraphics[width=0.334\textwidth]{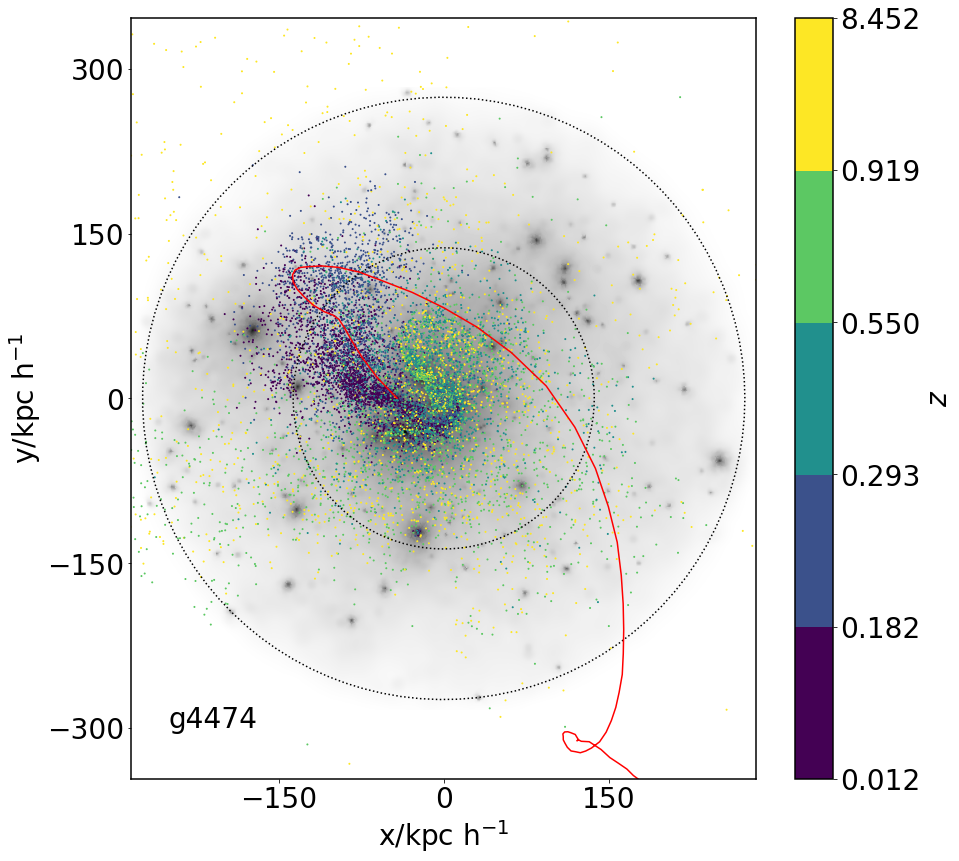}

	\end{center}
	\caption{\label{fig_orbits} Orbits of the selected satellite galaxies projected onto the orbital plane together with the position at $z=0$ of the gas that once belonged to each satellite galaxy, coloured by the redshift of removal from the galaxy (the colour ranges are separated in quintiles of the $z$ distribution of the particles). The grey shades show the $z=0$ dark matter density within a sphere of $2\times r_{200}$ centred at the central galaxy. The dashed circles correspond to one (inner) and two (outer) times $r_{200}$  from the central galaxy. The red dot in the orbit of g4770 indicates the position of the progenitor when it experienced the last significant merger, this is the only galaxy that had a recent important merger before entering $r_{200}$}.
\end{figure*}

\section{Analysis of the removed gas and cause for removal}
\label{remgas}
In order to analyse the properties of the gas removed from the simulated galaxies, we followed the merger trees for the central galaxies and their selected satellites as discussed in Section \ref{sample}. Then, we tracked the gas particles that are part of the main progenitors of our satellite galaxies at a given $z$, but that no longer belong to them at $z=0$. We also kept record of the last snapshot where those gas particles were still associated with a progenitor. In Fig. \ref{fig_orbits} we show the orbits described by the satellites in our sample, together with the distribution at $z=0$ of the removed gas, coloured by the latest redshift at which the gas was part of a satellite. The grey shades show the density of dark matter within $2 \times r_{200}$ from the central galaxy of the halo \citep[we use py-SPHviewer to estimate and plot the dark matter density, see][]{sphviewer}. We use comoving units for distances unless otherwise stated.

As can be seen from Fig. \ref{fig_orbits}, a significant fraction of the removed gas roughly follows the orbital path of the satellite galaxies. 
However, part of the removed gas from a given satellite can be found around other halo members, and especially around the central galaxy. Additionally, an important fraction of the removed gas is neither in the trail of the galaxy orbits nor residing in other galaxies, but preferentially found near the apocentre of the orbit,  at a further distance from the central galaxy \footnote{The evolution of these systems can be visualised in  the videos that are part of the additional material, available at \url{https://www.patriciatissera.com/post/cielo-simulations-satellites-orbiting-central-galaxies}}.

It is important to note that some satellites, in particular those in h4672, lost large amounts of gas at high redshift.  For example, for g4470, this occurred before the most important recent merger at $z=0.38$. As analysed in the following sections, many satellites fall onto groups with a large number of companions. Gas remotion  prior to entering the virial radius of a group could be a sign of pre-proccesing due to galaxy-galaxy interactions.

In the next section, we will  explore the evolution of different processes that can be responsible for the removal of gas from satellite galaxies, namely, ram pressure ($P_{\rm ram}$), tidal torques ($\tau$) and SN feedback.

\begin{table*}
	\begin{center}
		\caption{\label{table_derived} Observed characteristics of the satellite galaxies: merger ratio ($\mu$), stellar ratio ($f^{\rm e}_{\star}$), gas ratio ($f^{\rm e}_{\rm gas}$), number of companions, if the galaxy was the BHG, mass halo ratio ($f^{\rm e}_{\rm Halo}$), average ram pressure 
	($\langle P_{\rm ram}\rangle$)
	and average tidal torque over total mass 
	($ \langle \tau/M \rangle$ ). The latter two parameters are estimated at twice $r200$ as explained in the text.}
		\begin{tabular}{lcccccccc}
			Galaxy ID & $\mu$ & $f^{\rm e}_{\star}$ & $f^{\rm e}_{\rm gas}$ & $\langle P_{\rm ram} \rangle$ & $ \langle \tau/M \rangle$ & companions & BHG & $f^{\rm e}_{\rm Halo}$ \\
			       & &     & & $10^{-17}$ Pa & km$^2$ s$^{-2}$ & & & \\\hline
			g4338 & 0.0                  & 1.83 & 0.30 & 2.66  & 204.59 & 7  & Yes   & 0.58 \\
			g4339 & 0.0                  & 1.40 & 0.37 & 14.44 & 124.01 & 1  & Yes   & 0.81 \\
			g4341 & 0.0                  & 1.60 & 0.26 & 2.74  & 108.85 & 7  & g4338 & 0.09 \\
			g4343 & 0.0                  & 0.47 & 0.00 & 8.07  & 64.78  & 0  & Yes   & 1 \\\hline
			g4470 & 0.55                 & 3.12 & 0.65 & 1.97  & 106.62 & 16 & Yes   & 0.83 \\
			g4471 & 1.6 $\times 10^{-4}$ & 1.90 & 4.99 & 3.72  & 115.97 & 11 & Yes   & 0.77 \\
			g4473 & 1.5 $\times 10^{-3}$ & 1.10 & 0.98 & 4.60  & 32.41  & 16 & g4470 & 0.11 \\
			g4474 & 0.0                  & 1.64 & 2.07 & 3.94  & 36.24  & 11 & g4471 & 0.10 \\\hline
		\end{tabular}
	\end{center}
\end{table*}

Firstly, we estimate the effect of ram pressure stripping  on each satellite galaxy along its orbit following \citet[][eq. 61]{GunnAndGott72}, so that $P_{\rm ram}=\rho_{\rm IGM} v^2$, where $\rho_{\rm IGM}$ is the intra-group/cluster gas density, and $v$ is the velocity of the galaxy with respect to the gas. In order to estimate the average density of the gaseous medium we adopt a comoving sphere of 100 kpc h$^{-1}$ centred on the satellite galaxy, and select the gas particles within that sphere that a) do not belong to any galaxy and b) that have temperatures $T >10^6 K$. Those gas particles are then used to compute $\rho_{\rm IGM}$. The velocity term in this equation corresponds to the satellite galaxy velocity with respect to the average velocity weighted by the host gas mass.

We also derive the tidal torques exerted on satellite galaxies during their evolution ($\tau$). This is accomplished by estimating  the gravitational force that the central galaxy exerts on the satellites as well as the  gravitational force from other satellites located within a 100~$\rm h^{-1}$~kpc. To calculate the tidal torque we use,

\begin{equation}
	\vec{\tau} = \sum_{j=1}^n\sum_{i=1}^{l} G\frac{m_i M_j\vec{r_i}\times \vec{r_j}}{|\vec{r_j}-\vec{r_i}|^3},
	\label{eq_torque}
\end{equation}
where the sub--index $i$ indicates  gas, star or dark matter particle of a satellite galaxy (with a $l$ number of particles), $m_i$ is the mass of the $i$-th particle, $\vec{r_i}$ is the distance vector that goes from a given $i$-th particle to the centre of the satellite galaxy. The sub--index $j$ indicates a perturbing galaxy (this is either a central galaxy or any other galaxy within 100 kpc h$^{-1}$, with $n$ perturbing objects), where $M_j$ is the total mass of the perturbing galaxy (i.e. gas, stars and dark matter) and $\vec{r_j}$ is the vector that goes from the perturbing galaxy to the satellite galaxy. To simplify this calculation, we treat  other galaxies as point masses. In this analysis we only use particles that comply with $|\vec{r_i}|<|\vec{r_j}-\vec{r_i}|$. 

Finally, it is expected that gas removal from a satellite galaxy is not only due to external mechanisms, but that could also be affected by internal processes such as SN feedback. This process could expel gas  efficiently during starburst episodes. Since the number of SNII is related to the number of newborn massive stars, which have life-times of $\sim 10^6 - 10^7$ yrs \citep[][shorter than the time between snapshots]{rait1996}, we use the sSFR  as a proxy of SN feedback strength. 

Galaxies may have been involved in a recent merger event, or may have interacted with close companions before entering into the current halo. As a consequence, gas removal or/and gas consumption could have been induced previously. This process is generally referred to as pre-processing \citep{Zabludoff1998, Berrier2009, Hou2014, Pallero2019}. Some of the satellite galaxies hosted by  our two main haloes at $z=0$ were in fact part  of separate haloes before infall. Even more, these  systems  also hosted their own satellite systems as discussed below. 

To evaluate the impact of pre and post processing, we have estimated the following parameters for each satellite galaxy which are shown in Table \ref{table_derived}:
\begin{itemize}
\item $\mu$: the merger ratio  between the stellar mass of a satellite galaxy and of the satellite with which it merged after entering the region demarked by  $2\times r_{200}$ from the central galaxy for the first time (i.e. $\mu =0$ for no mergers).
\item $f^{\rm e}_{\star}$: the stellar ratio between the stellar mass of a satellite galaxy at $z=0$ and its stellar mass when it was at a distance of $2 \times r_{200}$ from the central galaxy.
\item $f^{\rm e}_{\rm gas}$: the same definition of $f^{\rm e}_{\star}$ but for the gas mass.
\item  $\langle P_{\rm ram} \rangle_{2r200}$: the average value of the ram pressure during the time interval between the moment when the galaxy crosses $2\times r_{200}$ for the first time and $z=0$.
\item $\langle \tau/M \rangle_{2r200}$: the mean value of the total tidal torque exerted on each satellite galaxy in the time interval between the time a galaxy crosses $2\times r_{200}$ for the first time and $z=0$.
\end{itemize}

We have also calculated different parameters related to the haloes hosting satellite galaxies before $t_{\rm e}$, which is the time when the satellite is associated with its current halo by the FOF algorithm for the first time.
\begin{itemize}
\item The number of companions resolved with at least 100 dark matter particles of a satellite galaxy.
\item Brightest Halo Galaxy (BHG), which indicates that a satellite galaxy was the central galaxy of its own halo before $t_{\rm e}$. Otherwise, it refers to the object that was the central galaxy at this time (given by the galaxy ID). 
\item The mass halo ratio, $f^{\rm e}_{\rm Halo}$, corresponding to the ratio between $M$ of a satellite galaxy and $M_{200}$ of its host halo before $t_e$ (recall that $M$ is the total mass of a galaxy).
\end{itemize}

In Table \ref{table_basic} we can see that our  galaxies belong to haloes with different characteristics. On the one hand,  the satellites of the most massive halo, h4671, are less massive than those of h4672.  This means that, overall, h4671 has a very dominant central galaxy and its satellites do not contribute significantly to the total mass. On the contrary, the satellites in halo h4672 contain a more significant fraction of the total halo matter. In fact,  according to the SUBFIND algorithm, 96\% of the dark matter in halo h4671 belongs to the central galaxy, whereas in halo h4672 only  61\% of the dark matter belongs to the central galaxy.

From Table \ref{table_derived} we can see that satellite g4470 is the only one that had a recent important merger before entering $r_{200}$ (in Fig. \ref{fig_orbits} the merger is marked with a red dot on the orbit), g4343 is the only satellite that lost stellar mass  ($f^{\rm e}_{\star} < 1$) since  entering the virial halo, that has been depleted of its gas reservoir, and that does not come with  companions.  Finally, g4338, g4339, g4470 and g4471 were  the BHGs of their own halo before $t_{\rm e}$, and their total masses represent more than the 50\% of their former halo mass (given by $f^{\rm e}_{\rm Halo}$).
Interestingly, in some cases the BHG and one of its satellite galaxies are part of the surviving satellite system  as can be seen from  Table \ref{table_derived} for example  g4338  and g4341 in h4671. Nevertheless, it is beyond the scope of this paper to further investigate the evolution of these systems.

\subsection{Orientations}
\label{orientations}

The orientation of the galactic disc with respect to the direction of motion or the position of nearby galaxies can play a critical role in modulating the impact of   ram pressure or tidal torque on the galaxy.

In the case of ram pressure, if the galaxy motion proceeds with the disc oriented face-on to the direction of the bulk velocity, a larger amount of gas is likely to be affected by this mechanism in comparison to a configuration where  the disc lies in the plane of motion since ram pressure 
can act over a larger area.
We expect that the effect of tidal torque exerted by another galaxy reaches a maximum when the orientation of the disc and the direction to the perturbing galaxy are near to 45 degrees, reaching a minimum when the disc is oriented parallel or perpendicular to the direction of the perturbing galaxy.

We adopt the direction of the angular momentum of stellar component of satellite galaxies as a measure of their disc orientation considering that they are all disc-dominated. Since all galaxies in our sample have at least 500 star particles in every snapshot analysed, we can reliably measure the total angular momentum.

\begin{figure*}
	\includegraphics[width=0.49\textwidth]{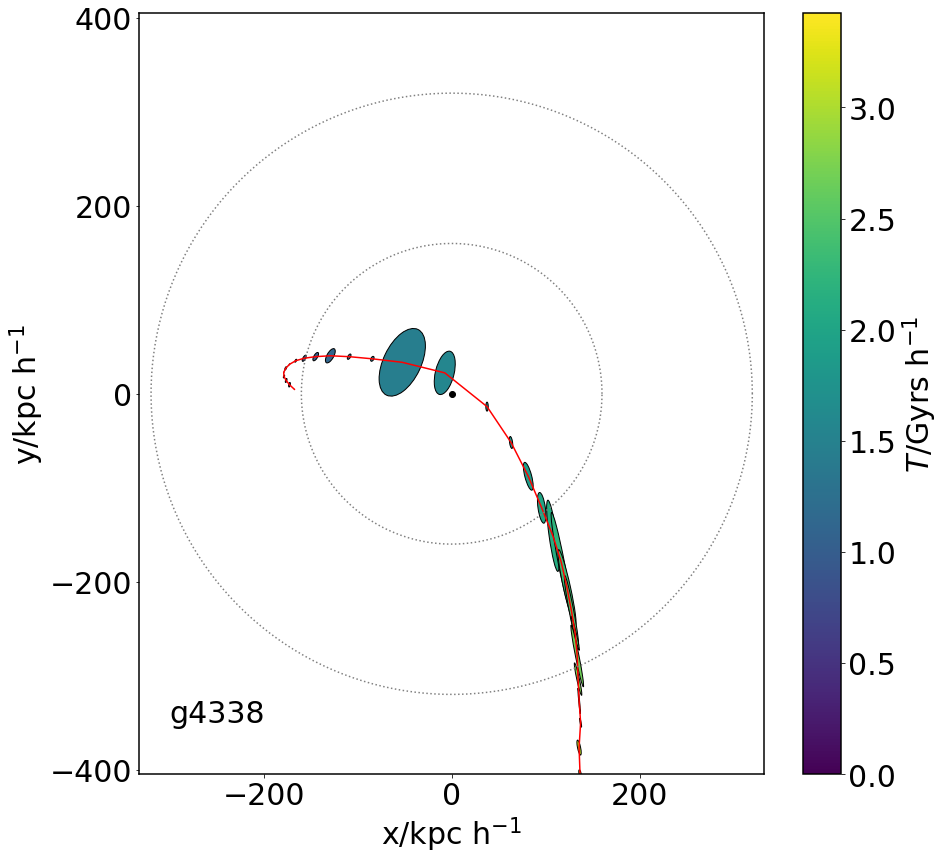}
	\includegraphics[width=0.49\textwidth]{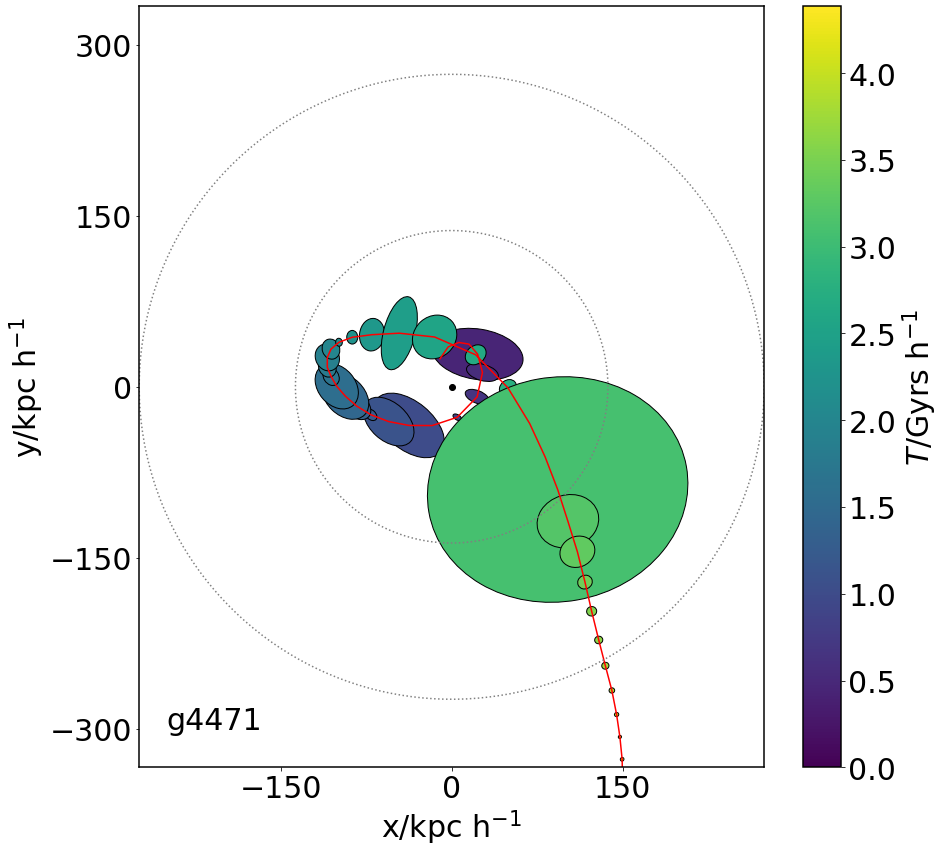}
	\caption{\label{fig_Orientation} Variation of the orientation of galaxies g4338 in h4671 (left panel) and g4471 in h4672 (right panel) along their orbits. At each time, the ratio between the semi-minor and semi-major axis of the ellipsoid that represents them in the figure corresponds to $1 - \cos(\theta_p)$, where $\theta_p$ is the angle defined by the angular momentum and the instantaneous orbital plane, the angle of the semi-major axis indicates the angle perpendicular to the angular momentum projected on this plane. The colour indicates look back time ($T=0$ indicates $z=0$)  and the size of the semi-major axis is proportional to the gas mass removed from the satellite galaxy in each snapshot. In Fig.~\ref{fig_Orientation_append}, the corresponding plots for the rest of the satellites galaxies are displayed.}
\end{figure*}

Figures \ref{fig_Orientation} and \ref{fig_Orientation_append}\footnote{ For illustration purposes, we select galaxies g4338 and g4471 as examples throughout the paper. These galaxies are chosen because  they are the more massive of their respective haloes that have not experienced an important, recent merger. In Section \ref{secapen}, the rest of the galaxies are displayed. We note that all satellite galaxies are taken into account in the analysis and interpretation of the results.} show the disc orientation with respect to the orbital plane (i.e. the plane perpendicular to the orbital angular momentum) for our sample of galaxies along their orbital motion. Satellite galaxies are represented by ellipses corresponding to projected circular discs with zero height oriented in the same direction as the galaxy rotational planes. The major semi-axes are proportional to the removed gas mass ($M_{\rm Rgas}$) while the ellipticities  illustrate the orientation of the discs with respect to the projected orbital plane.
In general, a large fraction of the gas is removed when the disc is oriented perpendicular to the direction of the galaxy motion. For most of the galaxies, the gas is removed when the angular momentum of the galaxy is aligned with the velocity vector (when the cosine of the angle between the two vectors is $>0.5$). For these galaxies, the average removed gas mass when the angular momentum is aligned is between $\sim 1.0$ to $\sim 6.5$ times that of when it is anti-aligned. Exceptions are g4341 and g4471, where this factor is $0.86$ and $0.68$, respectively.

\begin{figure*}
	\includegraphics[width=0.49\textwidth]{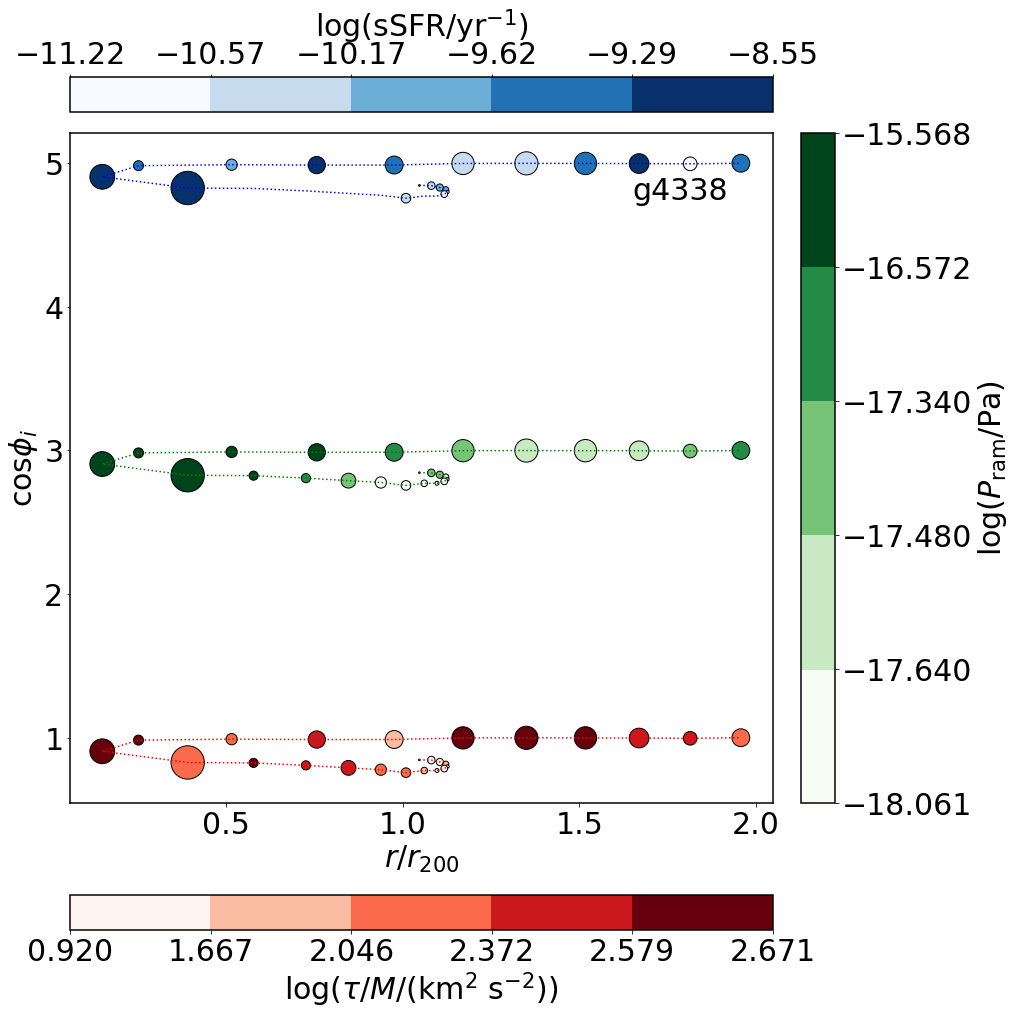}
	\includegraphics[width=0.49\textwidth]{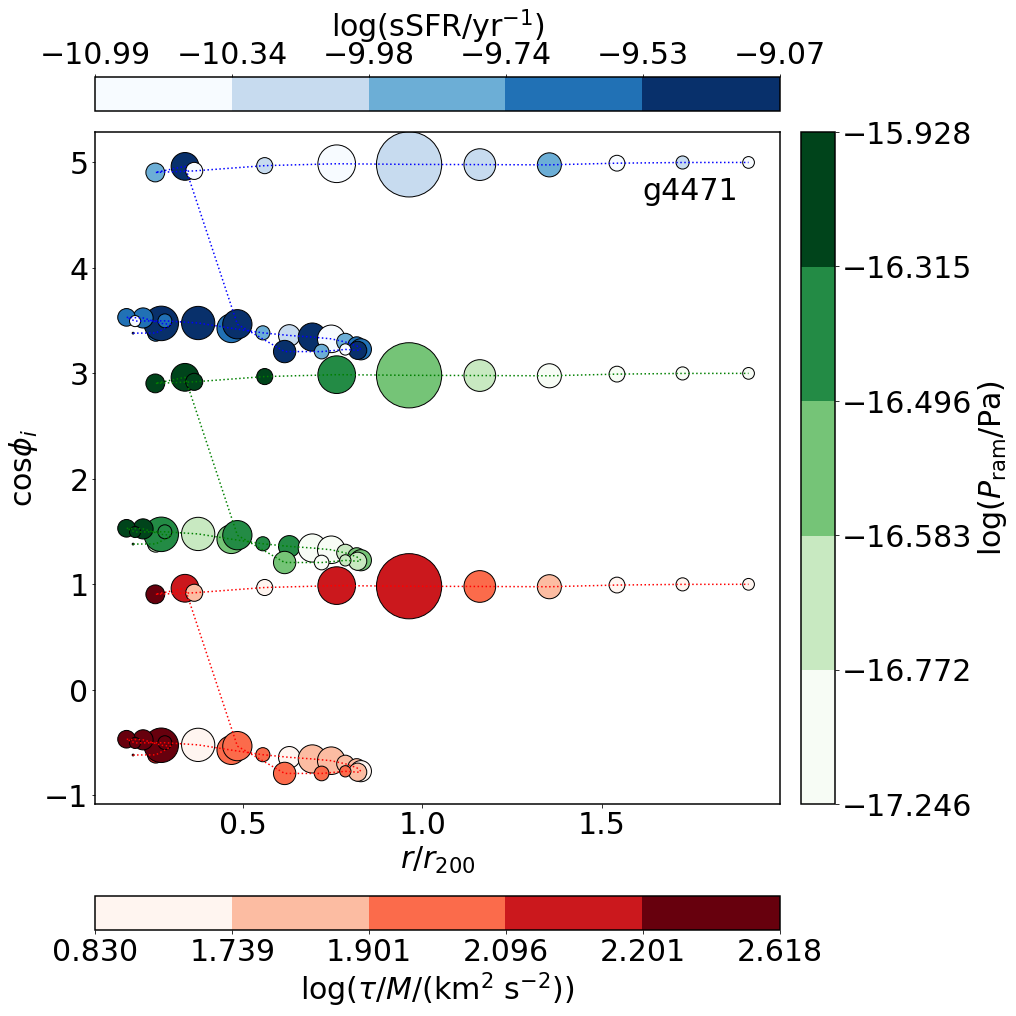}
	\caption{\label{fig_cosVSeffect} Variation of the angle between the angular momentum of a satellite galaxy at a given time and the angular momentum when the satellite galaxy was at a distance of $2\times r_{200}$ from the central galaxy for the first time ($\phi_i$) as a function of the distance to the central galaxy ($r$, normalised by $r_{200}$) for g4338 (right panel) and g4771 (left panel). The  blue shades  (top colour bar, top set of dots) indicate the variation of sSFR along the orbit. Green shades (right colour bar, middle set of dots) indicate the variation of $P_{\rm ram}$. Red shades (bottom colour bar and bottom set of dots) indicate the variation of $\tau/M$ along the orbit of the satellite galaxy. The size of the circles indicates $M_{\rm Rgas}$ in each snapshot, the dotted lines indicate the movement of the galaxies in this diagram. For ram pressure and sSFR a shift of +2 and +4, respectively, in the y-axis was added.  The plots for the rest of the galaxies are shown in Fig.~\ref{fig_cosVSeffect_append}.}
\end{figure*}

In order to further investigate the effect of galaxy orientations, we consider the relation between disc orientation and the different processes that induce gas removal.  Figures \ref{fig_cosVSeffect} and \ref{fig_cosVSeffect_append} show the relative angle ($\phi_i$) between the disc angular momentum and the direction of the angular momentum when the satellite was at a distance of $2 \times r_{200}$ for the first time, as a function of the distance to the central galaxy ($r$). This is a suitable measure of the changes in the disc orientation with respect to its original angular momentum. These figures also show the values of sSFR, $\tau$ and $P_{\rm ram}$ as well as $M_{\rm Rgas}$ along the orbits.

As can be seen in  Fig.\ref{fig_cosVSeffect} and Fig.\ref{fig_cosVSeffect_append}, only three of the satellite galaxies change significantly their disc orientations ($>45^{\circ}$). One of them ( g4343) shows a rapid variation while the other two satellites (g4339 and g4471) show a series of cumulative changes. In the analysed cases, we find that  all significant changes of satellite  orientations take place near their pericentre to the central galaxy. This is likely caused by the tidal torque exerted by the central galaxy. In Fig.~\ref{fig_orbits} we can see that those three galaxies have various passages near the central galaxy, while the other galaxies only have 1 or 0 close passages. Furthermore, g4471 is the only satellite in h4472 that reaches the closest position to the central galaxy. This trend suggests that the gravitational effect of the central galaxy over the orientation of  satellites may be cumulative, requiring several close encounters to make a significant impact, even for g4343, where the variation occurs in one of the later passages.

\subsection{Gas distribution}
\label{distribution}

\begin{figure}
    \centering
    \includegraphics[width=0.4\textwidth]{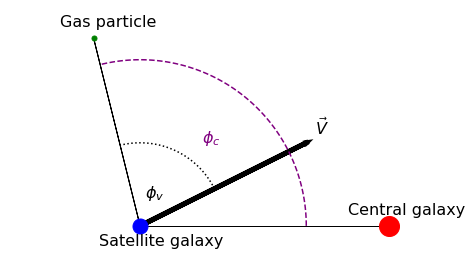}
    \caption{Schematic diagram that shows the angles $\phi_c$ (the purple dashed arc) and $\phi_v$ (the black dotted arc) determined by }the location of a gas particle, a satellite galaxy, a central galaxy and the velocity vector of the satellite galaxy ($\vec{V}$) in a simplified 2D representation. \label{fig_scheme}
\end{figure}

One possible way to study the removed gas from satellite galaxies in observations is to search for it in their vicinities.  In \cite{rgpt20} they use background sources around galaxy group satellites to make maps of foreground dust that they interpret as a signature of the gas removed from satellite galaxies.  To do this they select directions that are related to the geometry of the groups to make stacked dust maps and find tentative evidence of this removed gas.  Following their work we define two angles of reference adapted to the three-dimensional geometry of our simulation. The first one is determined between the vectors connecting the satellite and the removed gas particle position, and the satellite and the central galaxy ($\theta_c$). The second angle is defined between the vectors connecting the removed gas position and the satellite, and the velocity vector and the central galaxy ($\theta_v$). Figure \ref{fig_scheme} presents a simple diagram illustrating both angles.

\begin{figure*}
    \begin{center}
		\includegraphics[width=.7\textwidth]{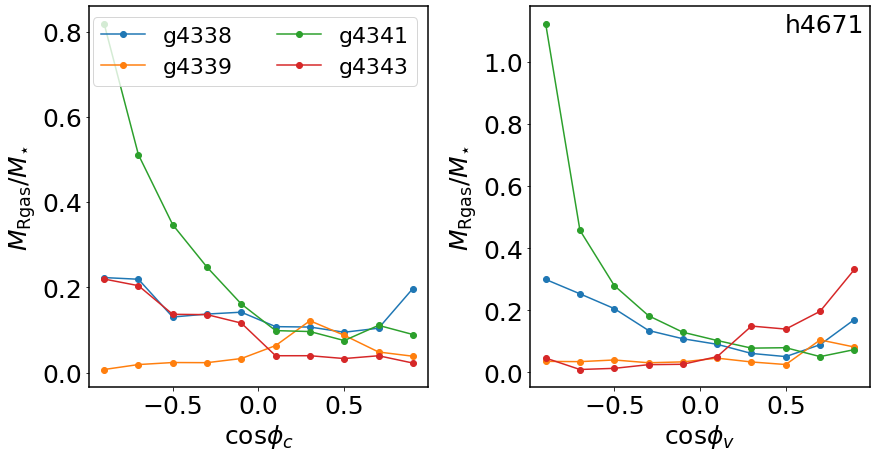}
		\includegraphics[width=.7\textwidth]{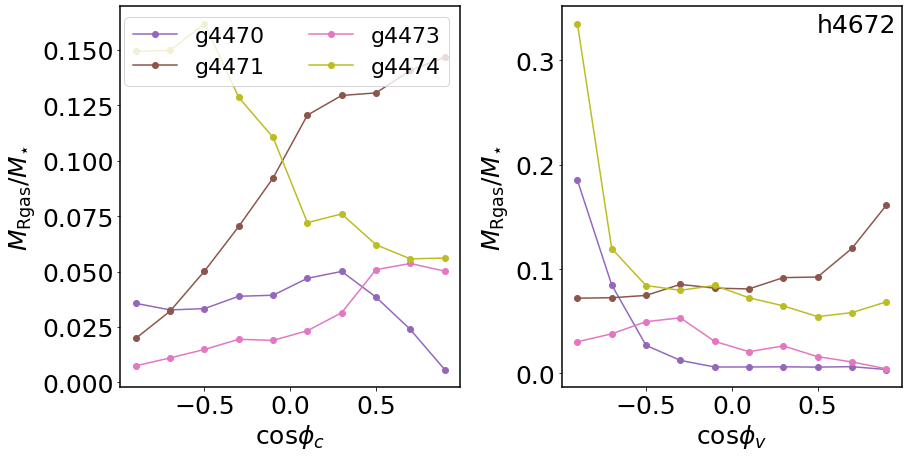}
    \end{center}
    \caption{\label{fig_MassDist} Removed gas mass fraction, $M_{\rm Rgas}/M_{\star}$, estimated within a comoving sphere of 200 kpc h$^{-1}$ from each satellite galaxy at $z=0$ as a function of the angles determined by the direction to the central galaxy ($\phi_c$; left panels) and velocity vectors ($\phi_v$; right panels) from the point of view of each satellite galaxy. In all cases the masses are normalised by $M_{\star}$ of the satellite galaxy at $z=0$. Top panels: Satellite galaxies of halo h4671. Bottom panels: Satellite galaxies of halo h4672.}
\end{figure*}

Figure \ref{fig_MassDist} shows the fraction of removed gas mass that is not part of any galaxy at $z=0$ normalised by the stellar mass of the galaxy of origin. 
As can be seen, our satellite sample shows two different behaviours. For some of them, such as g4341 and g4474, the removed gas is located opposite to the direction of motion and/or to the central galaxy ($\cos \theta_c$ and/or $\cos \theta_v < 0$), while others, eg. g4339 and g4471, show no strong dependence on  $\theta_v$ and $\theta_c$. The latter have experienced two or more close encounters with the central galaxy (g4339, g4343, g4471; see also  Fig. \ref{fig_orbits}). It is important to note that in observations, which in general lack precise orbital information, it is difficult to disentangle these effects \citep{rgpt20}. We can conclude that it is very difficult to associate the removed material to a galaxy by analysing only its $z=0$ distribution if the galaxy has experienced more than one passage by the pericentre of the orbit around the central galaxy (hereafter, peripassage), which can also be difficult to determine observationally. Additionally, for  galaxies with one peripassage, the distribution of material can be easily confused with that expected for galaxies on their first infall.  In the first case the material is expected to be  removed by a joint effect of the torque and ram pressure, and in the second one, it might be mainly caused by ram pressure.

\subsection{Signatures from the gas removal mechanisms}
\label{signatures}

In this section, we analyse in more detail the effects of three main processes responsible for the remotion and redistribution of the gas components: ram pressure, tidal torques and star formation activity.

In order to explore the impact of those effects, Fig.~\ref{fig_distTime_RPsSFRTT} and Fig.~\ref{fig_distTime_RPsSFRTT_append} shows 

for each galaxy, the distance to the galactic centre $r$ versus the lookback time ($T$) as a function of  the  sSFR (shades of blue), the  ram pressure (shade of green) and the  tidal torque (shades of red).
 The size of the circles indicates the amount of gas removed at each redshift.
 
In Fig.~\ref{fig_deltas} and Fig.~\ref{fig_deltas_append} the rate of change of the ram pressure, the  tidal torque, the sSFR and the amount of gas removed for each of the satellite galaxies are shown  as a function of $T$. For g4470, the time at which the last important merger occurred is also depicted (see $\mu$ in Table~\ref{table_derived}).

We also explore the densities of the intragroup gas. Figure \ref{fig_densities} shows the density profile of the hot gas ($>10^{6}$ K) for both haloes. The horizontal dotted line indicates the threshold in density at which rapid ram pressure driven quenching is expected to take place \citep[$\sim 10^{-25.3}$ kg m$^{-1}$, ][]{pallero2020}. 
We also display the nearest distance to the central galaxy reached by each  satellites in our sample.

\subsubsection{Ram Pressure}
\label{ram_pressure}

In this subsection, we explore the impact of ram pressure on the remotion of gas in satellite galaxies.

If we explore the mean values of $\langle P_{\rm ram} \rangle_{2r200}$ (Table \ref{table_derived}), we can see that, in halo h4671, galaxy g4339 is the one that endured a higher $P_{\rm ram}$, and has lost 63\% of its gas mass from the time the satellite was at $2\times r_{200}$ from the central galaxy to $z=0$, according to $f^{\rm e}_{\rm gas}$. The satellite that has lost more gas is g4343 (100\%) it has endured the second higher $P_{\rm ram}$ in the group.  The other two galaxies have experienced similar $P_{\rm ram}$ and have accordingly lost similar percentages of gas mass. In h4672, there are galaxies that have gained a large amount of gas mass along their orbits\footnote{ This extra gas mass  comes from the intragroup medium, which includes gas that was removed from other galaxies.}. Due to this there is no clear relation between the mean ram pressure and the fraction of removed gas mass. However, more information is stored in the evolution of these quantities along the orbits.

\begin{figure*}
    \includegraphics[width=0.49\textwidth]{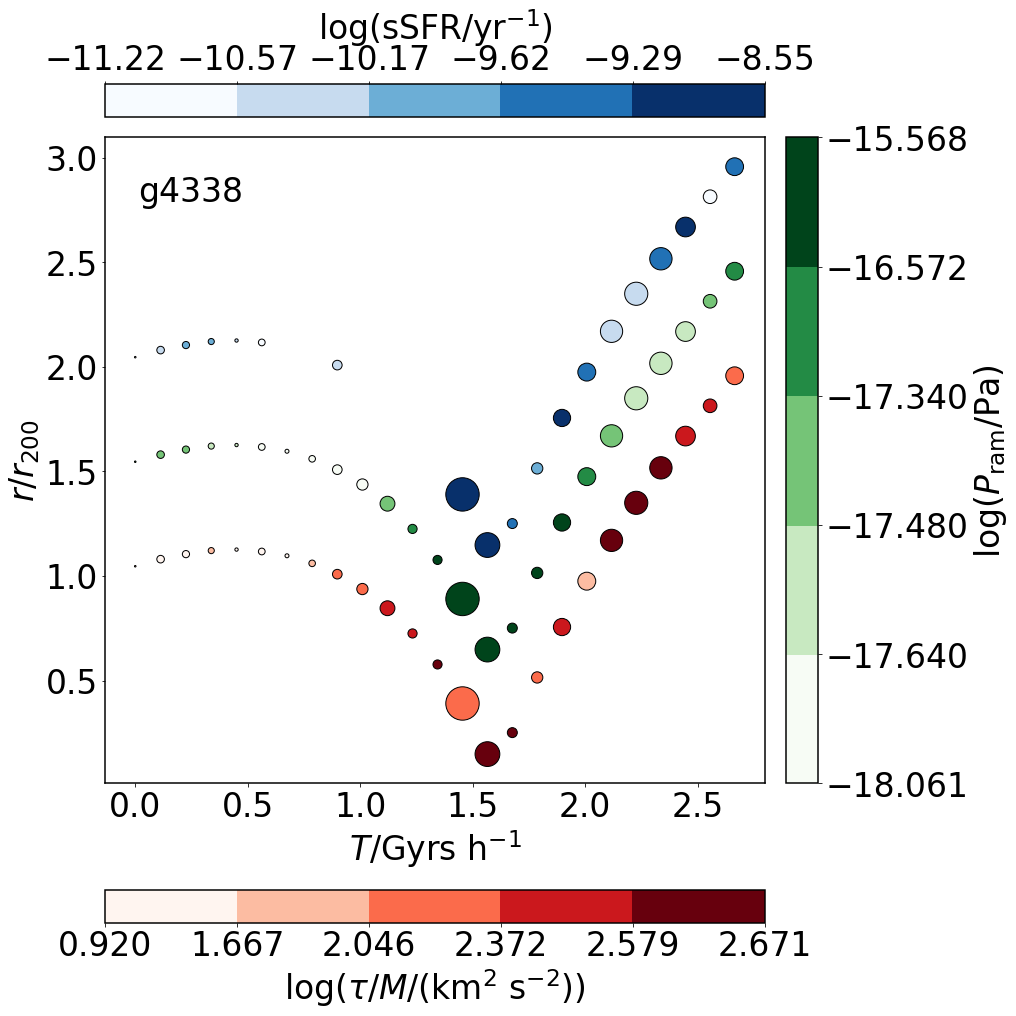}
    \includegraphics[width=0.49\textwidth]{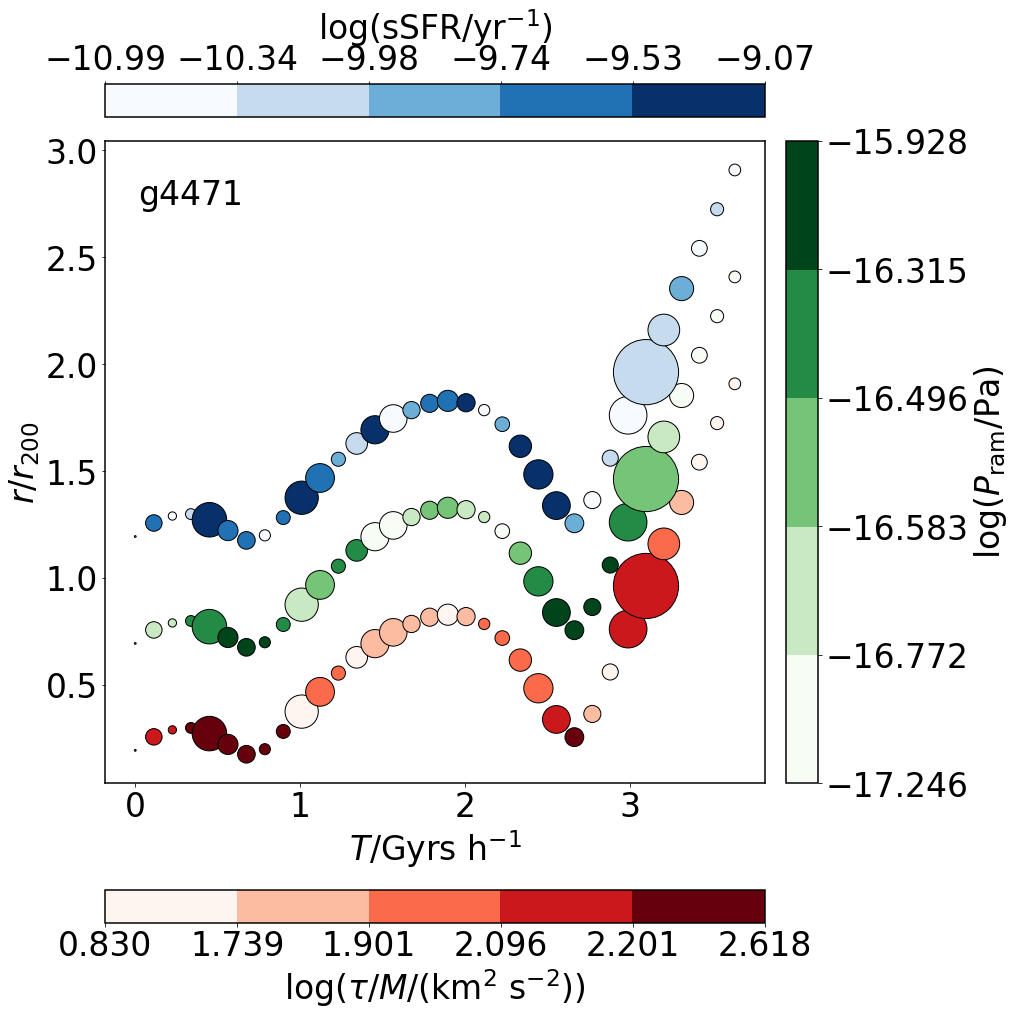}
    \caption{\label{fig_distTime_RPsSFRTT} Variation of sSFR (top colour bar and  top set of circles in blue shades), $P_{\rm ram}$ (right colour bar and  middle set of circles in green shades) and  $\tau/M$ (bottom colour bar and bottom set of circles in red shades)  along the orbits depicted by  the  $r/r_{200}$ as a function of lookback time (T).  The symbols coloured by sSFR and $P_{\rm ram}$ have been shifted in the y-axis of +1.0 $r_{200}$ and +0.5 $r_{200}$, respectively. The size of the symbols indicates $M_{\rm Rgas}$. Left: Galaxy g4338. Right: Galaxy g4471. The same plots for the other galaxies studied can be found in Fig.~\ref{fig_distTime_RPsSFRTT_append}}
\end{figure*}

In  Fig~\ref{fig_distTime_RPsSFRTT} and Fig~\ref{fig_distTime_RPsSFRTT_append} we see that $P_{\rm ram}$ increases  for decreasing  $r$ for the satellites.  For the satellites of halo h4671 the increase of ram pressure is associated with an increase of the gas mass remotion. However,  in halo h4672, this increase is not necessarily associated with a larger gas remotion. Some of the satellites experience very mild gas removal. 

\begin{figure*}
	\includegraphics[width=0.49\textwidth]{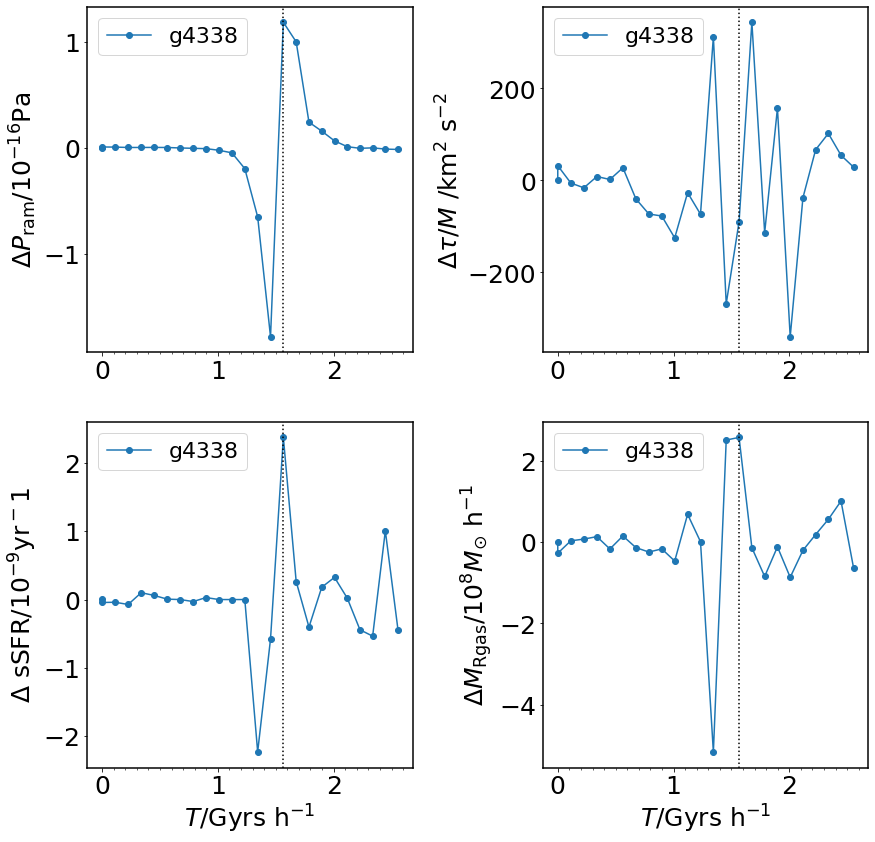}
	\includegraphics[width=0.49\textwidth]{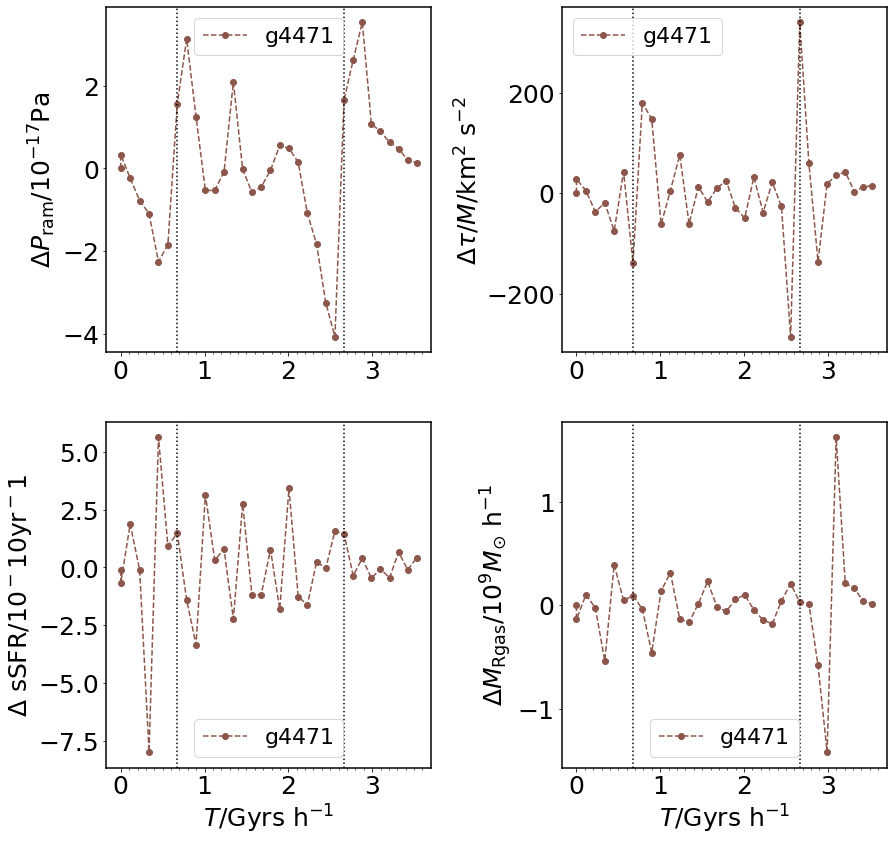}
	\caption{\label{fig_deltas}  Rate of variation of  $P_{\rm ram}$, $\tau/M$, sSFR and $M_{\rm Rgas}$ as a function of $T$ for the  galaxies analysed. The vertical lines show the times at which a local minimum in the distance to the central galaxy is reached (including $T=0$ if the distance at $z=0$ is shorter than any other local minimum). The first and second columns from the left  (blue solid lines) show the values for the galaxy g4338. The third and fourth columns  (brown dashed lines) show the values for galaxy g4471. The same plots for the rest of galaxies are presented in \ref{fig_deltas_append}.}
\end{figure*}

\begin{figure}
	\includegraphics[width=0.47\textwidth]{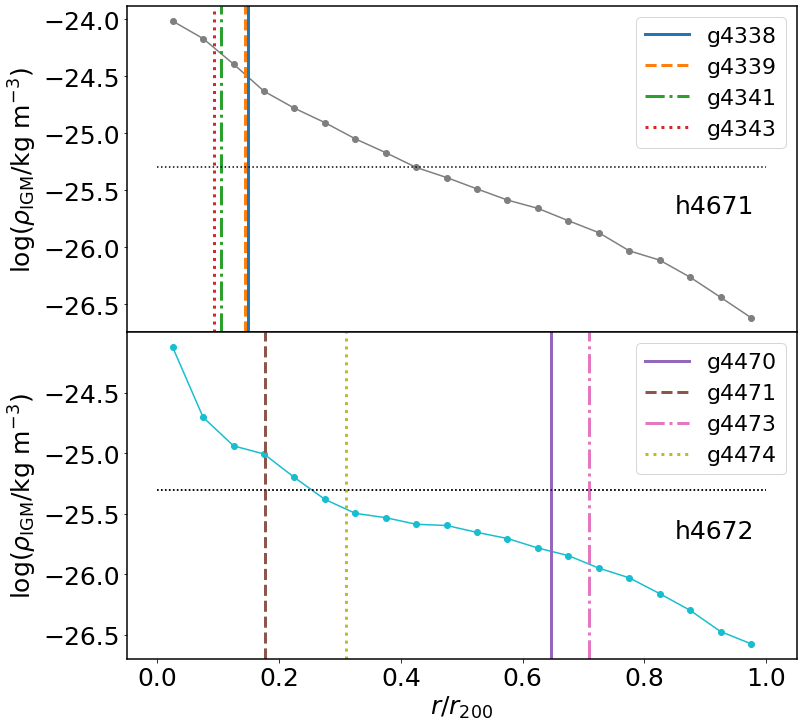}
	\caption{\label{fig_densities} Hot gas density profiles $\rho_{\rm IGM}$ as a function of $r/r_{200}$ for h4671 (top panel) and h4672 (bottom panel). The vertical lines indicate the closest distance to the central galaxy reached by each satellite galaxy analysed as indicated in the  labels. The horizontal dotted line depicts the threshold density of $10^{-25.3}$ kg m$^{-1}$.}
\end{figure}

In  Fig.~\ref{fig_deltas} and Fig.~\ref{fig_deltas_append} we can see that for galaxies in halo h4671 there is a clear relation between the time at which the rate of  variation of the ram pressure reaches a maximum  and the time at which the gas remotion is maximal. In our sample, the satellites that reach the regions within $\sim 0.2\times r_{200}$ experience $P_{\rm ram}$ values that are, in median, 2.7 times the average values within $0.2 - 1 \times r_{200}$, which are extremely low. In all galaxies in this halo, the peak values of ram pressure are also reached during the close encounter with the central galaxy.  During this period, the satellite galaxies are more actively forming stars. 
In the case of g4343, the first increase of ram pressure matches that of star formation and gas removal. However, as shown in Fig.~\ref{fig_deltas_append}, after the first peripassage, the star formation activity is very low. This indicates that there is no more gas in a condition to form stars. So even if  this galaxy has two more close encounters, the subsequent ram pressure peaks cause no impact on the star formation but it continues to retrieve the left over gas. By $z=0$, this satellite is depleted of gas reservoir.
Galaxies in halo h4672 do not show the same behaviour since they do not reach the inner regions of the halo where the density is high enough to produce significant ram pressure effects as shown in Fig.~\ref{fig_densities}. Those that get close such  as g4474 show the expected trend.

In Fig.~\ref{fig_densities} we can see that the  hot gas density profiles are  different between the two main haloes. In halo h4671 it surpasses the density threshold at a distance $\sim 0.4 \times r_{200}$, while for h4672 it surpasses the density threshold at  $\sim 0.2 \times r_{200}$. Also, for halo h4672, the central density is higher and their density profile is steeper than the those of h4671. All analysed galaxies in halo h4671 reach a shorter distance to the central galaxy than the radius at which the hot gas reaches the density threshold. These galaxies experienced  strong ram pressure such that it caused a significant effect on the galaxies, at least when they are located in the inner part of their orbits. In halo h4672, only g4471 and g4343 reach  distances closer the central region where the gas density is above the threshold.
 
The other two  satellites never get to a galactocentric position where the ram pressure is expected to have an important role in the evolution of the galaxy.

The main  differences  between  the  impact of ram pressure  on the  galaxies  in   haloes  h4671  and  h4672  can be explained  by two factors:
\begin{itemize}
   
    \item The density profile of the hot gas in h4671 is steeper than in h4672 and hence, it reaches higher densities at larger galactocentric distances.  This could be due to the fact that the central galaxy of  h4671 is more massive, even if the total virial masses of the haloes are similar. 
    \item All satellites galaxies in h6471 get very close to their central galaxies approaching  regions of  hot, high density gas. 
   
\end{itemize}


\subsubsection{Tidal Torque effects}
\label{tidal}

Figures \ref{fig_distTime_RPsSFRTT} and \ref{fig_distTime_RPsSFRTT_append} show that the episodes of strong gas remotion  generally occur close to those of high tidal torque, ram pressure and star formation rate. The times of maximum tidal torque are related to those of high ram pressure since the former are stronger when satellites are close to the central galaxies. This agrees with the stages of higher  density of the intragroup gas and larger relative velocity with respect to the central galaxies as discussed in the previous sections.
In fact, close galaxy interactions can lead to the removal of material after the first pericentre as the spiral arms react to  tidal forces, opening and reducing  binding energy \citep{rupke2010,perez2011,torre2012}. This fact may become critical during the trajectory after the first passage, when there is significant mass loss\footnote{This can be visualised in the videos that form part of the additional material.}.

These trends can be clearly seen in  the analysed galaxies of halo h4671, which show almost simultaneous occurrence of  maximal tidal torque, sSFR and gas remotion. In halo h4672,  g4673 and g4674 display similar behaviour  as can seen in Fig.~\ref{fig_distTime_RPsSFRTT_append}.
In the case of g4470, the peaks of tidal torque, sSFR and $M_{\rm Rgas}$ can be associated with a merger event that took place before $t_{\rm e}$. We note that this satellite does not get closer than $0.6 r_{200}$ of the corresponding central galaxy. Hence the gas remotion  is likely effect of pre-processing due to previous galaxy interactions.

We also quantify the variations of ram pressure, tidal torques, sSFR and gas mass remotion as a function of time by estimating the rate of change between the available snapshots as shown in Fig.~\ref{fig_deltas} and Fig.~\ref{fig_deltas_append}. 
In agreement with the trends shown in Fig.~\ref{fig_distTime_RPsSFRTT} and Fig.~\ref{fig_distTime_RPsSFRTT_append}, the impact of
ram pressure and tidal torques are clearly associated with  close encounters with the central galaxies. It can be seen that the rate of variations are stronger around these episodes. A similar behaviour is detected for the rate of variation of both sSFR and  gas mass removal, which is to be  expected  since both physical mechanisms are able to compress the gas, inducing star formation activity as well as  removing part of the gas, quenching  the galaxies. SN feedback produced by newborn stars would also contribute to eject material, reinforcing the quenching (see Section~\ref{sSFR}). By exploring the rates of variation of the sSFR and the gas mass removal, we estimate that the impact of these processes persist for around $\sim 0.7$ Gyr, on average. The relative efficiencies of these mechanisms are expected to change with the properties of the satellites galaxies, including the gas fraction and the disc orientation along the orbits, on top of the characteristics of the intragroup medium and dark matter halo onto which they fall in.

Table \ref{table_derived} shows that there is no clear relation between the time averaged values of tidal torque, $\tau/M$, and the ratio between the gas mass of the galaxy when it entered the halo and its gas mass at $z=0$, $f^{\rm e}_{\rm gas}$.  This is valid for all galaxies of both h4671 and h4672.

\subsubsection{Star formation and its effect on the intra--group medium}
\label{sSFR}

The trends in Figs.~\ref{fig_distTime_RPsSFRTT}, \ref{fig_distTime_RPsSFRTT_append}, \ref{fig_deltas} and \ref{fig_deltas_append} respond to two possible scenarios. In the first one,  ram pressure and torques show positive large rate of variations. In the second one,  the effects of these mechanisms are unimportant. 

In the first scenario, ram pressure can efficiently trigger  higher star formation activity through gas compaction into the central galaxy region, similarly to the effects of  interactions with the central galaxy \citep{cintio21}. The increase of the sSFR will be accompanied by the generation of SN feedback, which can trigger galactic outflows, contributing to gas removal. Examples of this scenario can be seen in the satellites of h4671. They reach the central regions and hence, are subject to dense environmental effects as discussed in the previous section (see also Fig. \ref{fig_densities} profile).

In the second scenario, the gas removal can be mainly associated to an enhancement of sSFR and to the action of SN feedback. Satellites galaxies g4470 and g4473 in halo h4672 are consistent with this scenario since they do not reach the central regions. Interesting, g4471, which has gained gas along its orbit,  has two peripassages by the central galaxy that induce strong star formation activity. In fact this satellite is actively forming stars at $z=0$.

We note that, in halo h4672,  g4470 exhibits a peak in sSFR and gas removal  corresponding to a merger with another galaxy before $t_{\rm e}$. Additionally, this satellite never reaches the inner regions and hence both ram pressure and tidal torque  are not strong enough to produce an impact on the galaxies (see Fig.~\ref{fig_deltas_append}).

\begin{figure}
	\begin{center}
	\includegraphics[width=0.4\textwidth]{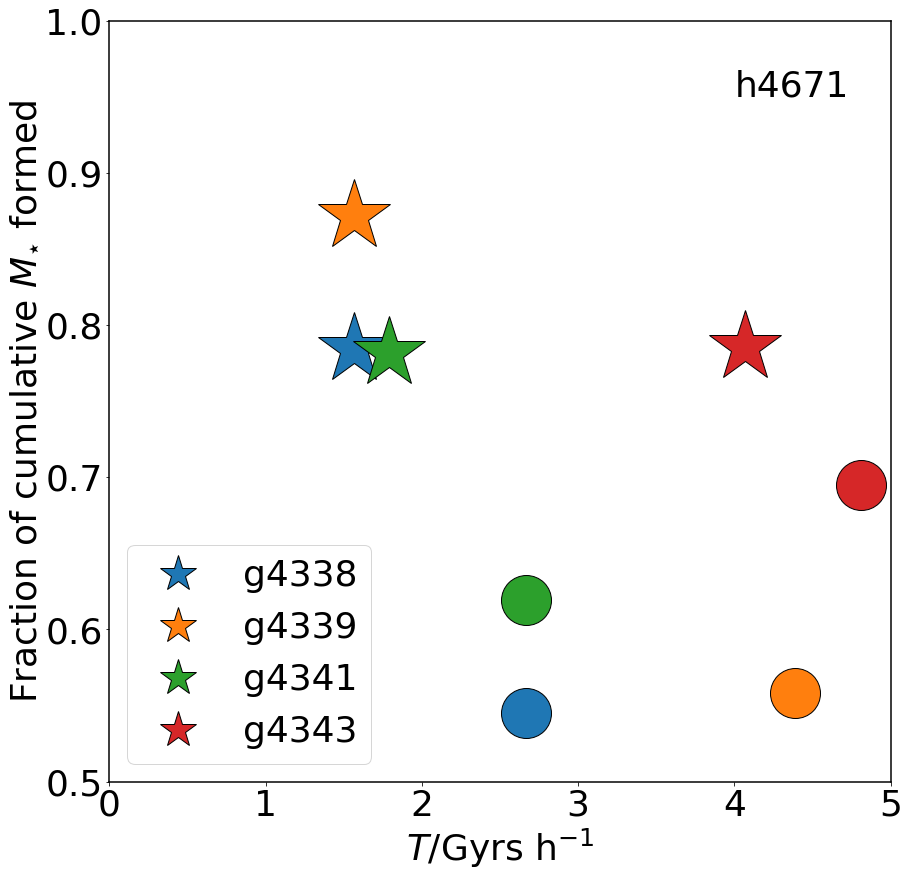}
	\includegraphics[width=0.4\textwidth]{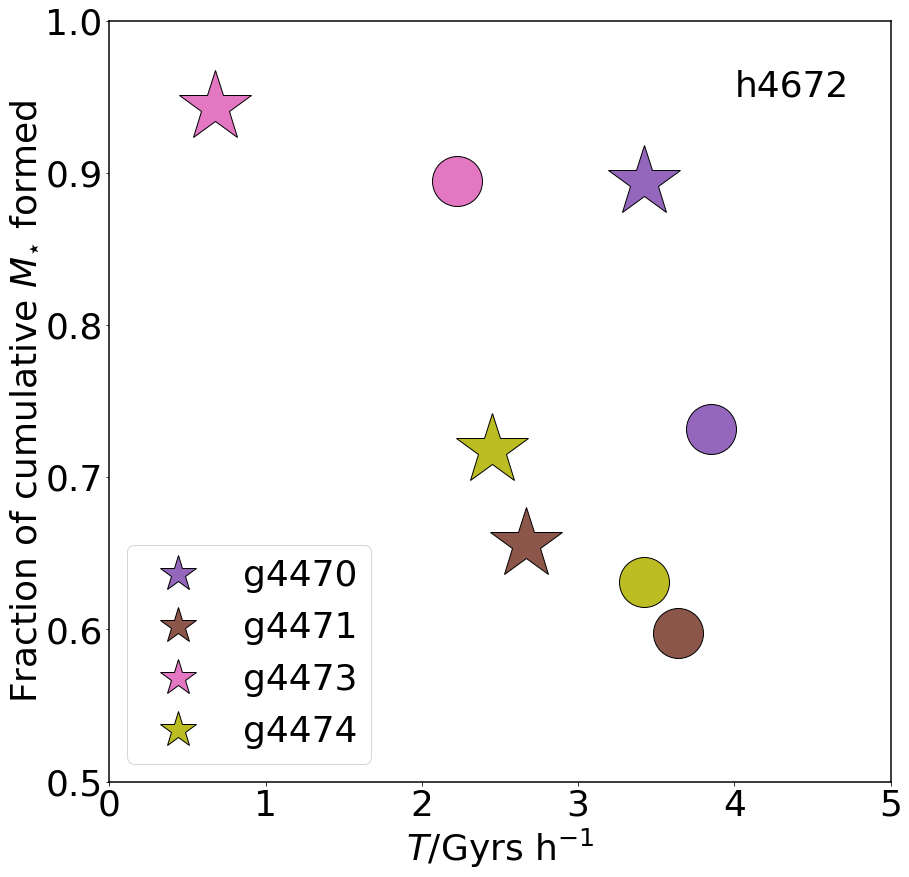}
	\end{center}
	\caption{\label{fig_StarFormation} Stellar mass fraction already formed at the time of the first pericentre (stars) and when the galaxy was at $2 \times r_{200}$ (circles) for the first time, with respect to their $M_{\star}$ at $z=0$.  Satellite galaxies in h4671 (upper panel) and  h4672 (lower panel) are shown.}
\end{figure}

In order to assess the impact of the environment mechanisms on galaxy quenching, we
analyse the stellar mass growth since the analysed galaxies first reach $2\times r_{200}$ from the central galaxy, until the time they reach the pericentre of their orbits.
To accomplish this, we  followed  \citet{Upadhyay2021}, who use N-body simulations to infer the most probable orbits of satellites galaxies in the Coma Cluster. Based on inferred star formation histories, in their figure 4, they showed the most probable percentage of stellar mass already formed at the time of entering the cluster and at the peripassage, in order to estimate  when the quenching processes took place.

We calculated the cumulative stellar mass formed between the time when the satellite galaxies were at $2\times r_{200}$ from the central galaxy, and at their first pericentre (Fig.~\ref{fig_StarFormation}). The percentage of the stellar mass already formed in our simulated satellite galaxies at this time is similar to those reported by \citeauthor{Upadhyay2021}. However, for our simulated local environment, the stellar mass formed until the first passage are considerably smaller. As it can be seen in  Fig.~\ref{fig_StarFormation}, the stellar mass fractions formed before the peripassages  vary from $\sim 60\%$ to $\sim 94\%$ of the total stellar mass at $z=0$, indicating that our satellite galaxies are not totally quenched when they get to  their first pericentre. In fact six out of eight satellites have less  than 80\% of their final stellar mass formed at this stage.

The estimation of the gas fraction at the pericentre shows that 
satellite galaxies in h6472 reach the pericentre with a larger gas fraction than those in h6471. This is expected because the latter are more affected by ram pressure due to the higher hot gas density and the fact that they get closer to the central galaxy (as discussed in previous sections). Nevertheless, all galaxies formed stars between $t_{e}$ and the pericentre  and three of them formed $\sim 90\%$ of the final stellar mass  by that time. 

These differences with \citeauthor{Upadhyay2021} could be, in part, due to the fact that our galaxies have endured  less denser environments compared with galaxies in the  Coma Cluster, which has an estimated M$_{200} = 5.1^{+4.3}_{-2.1} \times 10^{14} M_{\sun} h^{-1}$ \citep{Gavazzi2009}, around 500 times more massive than our haloes.

\section{Summary and conclusions}
\label{conc}

In this work, we have analysed the process of gas removal from simulated satellite galaxies as they enter  their $z=0$ haloes. We pay special attention to the relation between astrophysical and dynamical conditions along their orbits, focusing on ram pressure stripping, the impact of star formation feedback, tidal torque effects and galaxy interactions. 
For this purpose,  we have selected a sample of eight spiral satellite galaxies from a hydrodynamical simulation from the {\sc CIELO} Project with  initial conditions that resemble the Local Group. 

We follow the redistribution of all gas particles that were once part of our selected  satellites, but are no longer bound to them at $z=0$.
We find that most of the removed gas roughly follows the same orbital path of  the mother satellites within the potential well of the main haloes. However, we obtain that a significant fraction of the removed gas has been also acquired by other galaxies of the same halo at $z=0$ ( Fig. \ref{fig_orbits}). 
We also find that a fraction of the removed gas is striped  from the satellite galaxies during their first apocentre after the interaction with the central galaxy (see the videos linked). We argue that this is induced as the result of the opening of the arms due to the strong tidal torques exerted by the central galaxy \citep{rupke2010,perez2011,torre2012}. As a result of this mechanism, removed dust associated to the striped gas could be wrongly interpreted in observations as signatures of gas removed by ram pressure effects of infalling satellites along the halocentric direction to the satellite (see also  Fig.~\ref{fig_MassDist}).

The most important episodes of mass removal occur when the satellite galaxy discs are oriented perpendicular to their direction of motion (Fig.~\ref{fig_Orientation} and Fig~\ref{fig_Orientation_append}). Under this condition, ram pressure effects tend to be stronger. We also find larger star formation rates associated to this configuration so that, at least partially, some gas  removal could be also induced by SN feedback mechanisms. Our results indicate that  significant (more than 45$^{\circ}$) changes in the orientation of the discs are not common. Interesting,  every time that a satellite galaxy disc changes its orientation, it is located near the central galaxy of the halo (Fig.~\ref{fig_cosVSeffect} and Fig.~\ref{fig_cosVSeffect_append}).

The spatial distribution of the removed gas is very different for satellite galaxies with several peripassages by the central galaxy 
compared to those with a single peripassage.  In the later case,  a larger fraction of the removed gas is located in the opposite direction to the central galaxy or to the velocity vector, while in the case of satellites with several pericentre passages we find no relation between the location of the removed gas and the direction to neither the central galaxy nor the direction of motion (Fig.~\ref{fig_MassDist}).

From our analysis of the evolution of ram pressure (Section \ref{ram_pressure}), tidal torque  (Section \ref{tidal}) and star formation activity (Section \ref{sSFR}) in disc galaxies as they enter and move in the  potential well of a halo, we notice that these three physical mechanisms  reach peak values at similar times. This could be due to the fact that the maximum ram pressure is reached when the surrounding gas is dense and/or has a maximum relative velocity with respect to the central galaxy (Fig.~\ref{fig_distTime_RPsSFRTT} and Fig.~\ref{fig_distTime_RPsSFRTT_append}). Both conditions coincide when the satellite orbits near the central galaxy. Tidal torques reach a peak when the satellite galaxy is close to massive galaxies, either the central galaxy or  massive satellites, and the ram pressure could trigger burst of star formation close the central regions \citep{rupke2010, perez2011}.
We also observed cases for which a large remotion of material is accompanied by an increase in sSFR (eg. g4470) which could be an indication of preprocessing mechanisms due to galaxy--galaxy interactions (Fig. \ref{fig_deltas} and Fig. \ref{fig_deltas_append}).

Our results suggest that the simultaneous  action of ram pressure, tidal torques and SN feedback boost the impact they would produce individually. This could be the case for halo h4671, where the galaxies reach closer distances to the central galaxy ($< 0.2 \times r_{200}$, see figure \ref{fig_densities}). 
Interestingly, we see that when this occurs, the intragroup gas density is higher than $10^{-25.3}$  kg m$^{-1}$. This density threshold has been  associated with the quenching of the infalling galaxies onto high density environments such as clusters by previous works \citep{pallero2020}. Our findings show that this may also be valid for LG environments where  a systematic decrease of  star formation activity after  pericentre passages is detected in most of the analysed satellites. However,  star formation could be re-boosted in subsequent encounters with the central galaxies.

Our studies suggest possible pathways for future analysis of observational data  paying particular attention to disc orientations along the orbit, interaction events such as mergers which can be estimated using projected positions and line-of-sight dynamics, distribution of stripped material via analysis of background objects (as in \citealt{rgpt20}), and galaxy quenching time--scale of satellite galaxies via their spectral analysis.

In a forthcoming paper we will focus on the impact of these processes on the regulation of the star formation activity and the metallicity distribution within the satellite galaxies.

\section*{Acknowledgements}
{We thank A.Knebe for his support in the use of AMIGA. 
PBT acknowledges partial funding by Fondecyt 1200703/2020 (ANID),  ANID Basal projects ACE210002 and FB210003 and Millennium Nucleus ERIS. NP thanks the hospitality of the IATE-CONICET institute where he spent his 2021 sabbatical period, and acknowledges support from  the ANID BASAL projects ACE210002 and FB210003, and Fondecyt Regular 1191813. LB acknowledges support from CONICYT FONDECYT POSTDOCTORADO 3180359. 
Part  of this work  was  performed  in Ladgerda Cluster (Fondecyt 1200703/2020).
This project has been supported partially by the European Union Horizon 2020 Research and Innovation Programme under the Marie Sklodowska-Curie grant agreement No 734374.
This work uses simulations of the {\sc CIELO} project performed in Geryon cluster (PUC, Chile), the NLHPC (Chile) and Marenostrum4 (Barcelona Supercomputer Center, Spain). 
}

\section*{Data availability}
{The data underlying this article will be shared on reasonable request.}

\bibliographystyle{mnras}
\footnotesize
\bibliography{biblio}

\begin{thebibliography}{}
\makeatletter
\relax
\def\mn@urlcharsother{\let\do\@makeother \do\$\do\&\do\#\do\^\do\_\do\%\do\~}
\def\mn@doi{\begingroup\mn@urlcharsother \@ifnextchar [ {\mn@doi@}
  {\mn@doi@[]}}
\def\mn@doi@[#1]#2{\def\@tempa{#1}\ifx\@tempa\@empty \href
  {http://dx.doi.org/#2} {doi:#2}\else \href {http://dx.doi.org/#2} {#1}\fi
  \endgroup}
\def\mn@eprint#1#2{\mn@eprint@#1:#2::\@nil}
\def\mn@eprint@arXiv#1{\href {http://arxiv.org/abs/#1} {{\tt arXiv:#1}}}
\def\mn@eprint@dblp#1{\href {http://dblp.uni-trier.de/rec/bibtex/#1.xml}
  {dblp:#1}}
\def\mn@eprint@#1:#2:#3:#4\@nil{\def\@tempa {#1}\def\@tempb {#2}\def\@tempc
  {#3}\ifx \@tempc \@empty \let \@tempc \@tempb \let \@tempb \@tempa \fi \ifx
  \@tempb \@empty \def\@tempb {arXiv}\fi \@ifundefined
  {mn@eprint@\@tempb}{\@tempb:\@tempc}{\expandafter \expandafter \csname
  mn@eprint@\@tempb\endcsname \expandafter{\@tempc}}}

\bibitem[\protect\citeauthoryear{{Bah{\'e}} \& {McCarthy}}{{Bah{\'e}} \&
  {McCarthy}}{2015}]{Bahe2015}
{Bah{\'e}} Y.~M.,  {McCarthy} I.~G.,  2015, \mn@doi [\mnras]
  {10.1093/mnras/stu2293}, \href
  {https://ui.adsabs.harvard.edu/abs/2015MNRAS.447..969B} {447, 969}

\bibitem[\protect\citeauthoryear{{Bellhouse} et~al.,}{{Bellhouse}
  et~al.}{2017}]{Bellhouse2017}
{Bellhouse} C.,  et~al., 2017, \mn@doi [\apj] {10.3847/1538-4357/aa7875}, \href
  {https://ui.adsabs.harvard.edu/abs/2017ApJ...844...49B} {844, 49}

\bibitem[\protect\citeauthoryear{Benitez-Llambay}{Benitez-Llambay}{2015}]{sphviewer}
Benitez-Llambay A.,  2015, py-sphviewer: Py-SPHViewer v1.0.0,
  \mn@doi{10.5281/zenodo.21703}, \url {http://dx.doi.org/10.5281/zenodo.21703}

\bibitem[\protect\citeauthoryear{{Berrier}, {Stewart}, {Bullock}, {Purcell},
  {Barton}  \& {Wechsler}}{{Berrier} et~al.}{2009}]{Berrier2009}
{Berrier} J.~C.,  {Stewart} K.~R.,  {Bullock} J.~S.,  {Purcell} C.~W.,
  {Barton} E.~J.,   {Wechsler} R.~H.,  2009, \mn@doi [\apj]
  {10.1088/0004-637X/690/2/1292}, \href
  {https://ui.adsabs.harvard.edu/abs/2009ApJ...690.1292B} {690, 1292}

\bibitem[\protect\citeauthoryear{{Boselli} \& {Gavazzi}}{{Boselli} \&
  {Gavazzi}}{2006}]{Boselli2006}
{Boselli} A.,  {Gavazzi} G.,  2006, \mn@doi [\pasp] {10.1086/500691}, \href
  {https://ui.adsabs.harvard.edu/abs/2006PASP..118..517B} {118, 517}

\bibitem[\protect\citeauthoryear{{Bravo-Alfaro}, {Caretta}, {Lobo}, {Durret}
  \& {Scott}}{{Bravo-Alfaro} et~al.}{2009}]{BravoAlfaro2009}
{Bravo-Alfaro} H.,  {Caretta} C.~A.,  {Lobo} C.,  {Durret} F.,   {Scott} T.,
  2009, \mn@doi [\aap] {10.1051/0004-6361:200810731}, \href
  {https://ui.adsabs.harvard.edu/abs/2009A&A...495..379B} {495, 379}

\bibitem[\protect\citeauthoryear{{Chabrier}}{{Chabrier}}{2003}]{chabrier2003}
{Chabrier} G.,  2003, \mn@doi [\apjl] {10.1086/374879}, \href
  {https://ui.adsabs.harvard.edu/abs/2003ApJ...586L.133C} {586, L133}

\bibitem[\protect\citeauthoryear{{Condon}, {Condon}, {Gisler}  \&
  {Puschell}}{{Condon} et~al.}{1982}]{Condon1982}
{Condon} J.~J.,  {Condon} M.~A.,  {Gisler} G.,   {Puschell} J.~J.,  1982,
  \mn@doi [\apj] {10.1086/159538}, \href
  {https://ui.adsabs.harvard.edu/abs/1982ApJ...252..102C} {252, 102}

\bibitem[\protect\citeauthoryear{{Davis}, {Efstathiou}, {Frenk}  \&
  {White}}{{Davis} et~al.}{1985}]{davis1985}
{Davis} M.,  {Efstathiou} G.,  {Frenk} C.~S.,   {White} S.~D.~M.,  1985,
  \mn@doi [\apj] {10.1086/163168}, \href
  {https://ui.adsabs.harvard.edu/abs/1985ApJ...292..371D} {292, 371}

\bibitem[\protect\citeauthoryear{{De Grandi} \& {Molendi}}{{De Grandi} \&
  {Molendi}}{2001}]{DeGrandi2001}
{De Grandi} S.,  {Molendi} S.,  2001, \mn@doi [\apj] {10.1086/320098}, \href
  {https://ui.adsabs.harvard.edu/abs/2001ApJ...551..153D} {551, 153}

\bibitem[\protect\citeauthoryear{{De Young}}{{De Young}}{1978}]{deYoung1978}
{De Young} D.~S.,  1978, \mn@doi [\apj] {10.1086/156234}, \href
  {https://ui.adsabs.harvard.edu/abs/1978ApJ...223...47D} {223, 47}

\bibitem[\protect\citeauthoryear{{Di Cintio}, {Mostoghiu}, {Knebe}  \&
  {Navarro}}{{Di Cintio} et~al.}{2021}]{cintio21}
{Di Cintio} A.,  {Mostoghiu} R.,  {Knebe} A.,   {Navarro} J.,  2021, arXiv
  e-prints, \href {https://ui.adsabs.harvard.edu/abs/2021arXiv210302739D} {p.
  arXiv:2103.02739}

\bibitem[\protect\citeauthoryear{{Dolag}, {Borgani}, {Murante}  \&
  {Springel}}{{Dolag} et~al.}{2009}]{dolag2009}
{Dolag} K.,  {Borgani} S.,  {Murante} G.,   {Springel} V.,  2009, \mn@doi
  [\mnras] {10.1111/j.1365-2966.2009.15034.x}, \href
  {https://ui.adsabs.harvard.edu/abs/2009MNRAS.399..497D} {399, 497}

\bibitem[\protect\citeauthoryear{{Dutta} et~al.,}{{Dutta}
  et~al.}{2021}]{Dutta2021}
{Dutta} R.,  et~al., 2021, arXiv e-prints, \href
  {https://ui.adsabs.harvard.edu/abs/2021arXiv210910927D} {p. arXiv:2109.10927}

\bibitem[\protect\citeauthoryear{{Gavazzi}, {Adami}, {Durret}, {Cuillandre},
  {Ilbert}, {Mazure}, {Pell{\'o}}  \& {Ulmer}}{{Gavazzi}
  et~al.}{2009}]{Gavazzi2009}
{Gavazzi} R.,  {Adami} C.,  {Durret} F.,  {Cuillandre} J.~C.,  {Ilbert} O.,
  {Mazure} A.,  {Pell{\'o}} R.,   {Ulmer} M.~P.,  2009, \mn@doi [\aap]
  {10.1051/0004-6361/200911841}, \href
  {https://ui.adsabs.harvard.edu/abs/2009A&A...498L..33G} {498, L33}

\bibitem[\protect\citeauthoryear{{Gunn} \& {Gott}}{{Gunn} \&
  {Gott}}{1972}]{GunnAndGott72}
{Gunn} J.~E.,  {Gott} J.~Richard I.,  1972, \mn@doi [\apj] {10.1086/151605},
  \href {https://ui.adsabs.harvard.edu/abs/1972ApJ...176....1G} {176, 1}

\bibitem[\protect\citeauthoryear{{Hahn} \& {Abel}}{{Hahn} \&
  {Abel}}{2011}]{HahnandAbel2011}
{Hahn} O.,  {Abel} T.,  2011, \mn@doi [\mnras]
  {10.1111/j.1365-2966.2011.18820.x}, \href
  {https://ui.adsabs.harvard.edu/abs/2011MNRAS.415.2101H} {415, 2101}

\bibitem[\protect\citeauthoryear{{Haynes}, {Giovanelli}  \&
  {Chincarini}}{{Haynes} et~al.}{1984}]{Haynes1984}
{Haynes} M.~P.,  {Giovanelli} R.,   {Chincarini} G.~L.,  1984, \mn@doi [\araa]
  {10.1146/annurev.aa.22.090184.002305}, \href
  {https://ui.adsabs.harvard.edu/abs/1984ARA&A..22..445H} {22, 445}

\bibitem[\protect\citeauthoryear{{Heckman}, {Armus}  \& {Miley}}{{Heckman}
  et~al.}{1990}]{Heckman1990}
{Heckman} T.~M.,  {Armus} L.,   {Miley} G.~K.,  1990, \mn@doi [\apjs]
  {10.1086/191522}, \href
  {https://ui.adsabs.harvard.edu/abs/1990ApJS...74..833H} {74, 833}

\bibitem[\protect\citeauthoryear{{Hou}, {Parker}  \& {Harris}}{{Hou}
  et~al.}{2014}]{Hou2014}
{Hou} A.,  {Parker} L.~C.,   {Harris} W.~E.,  2014, \mn@doi [\mnras]
  {10.1093/mnras/stu829}, \href
  {https://ui.adsabs.harvard.edu/abs/2014MNRAS.442..406H} {442, 406}

\bibitem[\protect\citeauthoryear{{Hummel}, {van der Hulst}, {Kennicutt}  \&
  {Keel}}{{Hummel} et~al.}{1990}]{Hummel1990}
{Hummel} E.,  {van der Hulst} J.~M.,  {Kennicutt} R.~C.,   {Keel} W.~C.,  1990,
  \aap, \href {https://ui.adsabs.harvard.edu/abs/1990A&A...236..333H} {236,
  333}

\bibitem[\protect\citeauthoryear{{Iwamoto}, {Brachwitz}, {Nomoto}, {Kishimoto},
  {Umeda}, {Hix}  \& {Thielemann}}{{Iwamoto} et~al.}{1999}]{iwamoto1999}
{Iwamoto} K.,  {Brachwitz} F.,  {Nomoto} K.,  {Kishimoto} N.,  {Umeda} H.,
  {Hix} W.~R.,   {Thielemann} F.-K.,  1999, \mn@doi [ApJS] {10.1086/313278},
  \href {http://adsabs.harvard.edu/abs/1999ApJS..125..439I} {125, 439}

\bibitem[\protect\citeauthoryear{{Jackson} et~al.,}{{Jackson}
  et~al.}{2021}]{Jackson2021}
{Jackson} R.~A.,  et~al., 2021, \mn@doi [\mnras] {10.1093/mnras/stab077}, \href
  {https://ui.adsabs.harvard.edu/abs/2021MNRAS.502.4262J} {502, 4262}

\bibitem[\protect\citeauthoryear{{Jaff{\'e}} et~al.,}{{Jaff{\'e}}
  et~al.}{2014}]{Jaffe2014}
{Jaff{\'e}} Y.~L.,  et~al., 2014, \mn@doi [\mnras] {10.1093/mnras/stu507},
  \href {https://ui.adsabs.harvard.edu/abs/2014MNRAS.440.3491J} {440, 3491}

\bibitem[\protect\citeauthoryear{{Jaff{\'e}} et~al.,}{{Jaff{\'e}}
  et~al.}{2018}]{Jaffe2018}
{Jaff{\'e}} Y.~L.,  et~al., 2018, \mn@doi [\mnras] {10.1093/mnras/sty500},
  \href {https://ui.adsabs.harvard.edu/abs/2018MNRAS.476.4753J} {476, 4753}

\bibitem[\protect\citeauthoryear{{Jim{\'e}nez}, {Tissera}  \&
  {Matteucci}}{{Jim{\'e}nez} et~al.}{2015}]{Jimenez2015}
{Jim{\'e}nez} N.,  {Tissera} P.~B.,   {Matteucci} F.,  2015, \mn@doi [\apj]
  {10.1088/0004-637X/810/2/137}, \href
  {https://ui.adsabs.harvard.edu/abs/2015ApJ...810..137J} {810, 137}

\bibitem[\protect\citeauthoryear{{Kapferer}, {Sluka}, {Schindler}, {Ferrari}
  \& {Ziegler}}{{Kapferer} et~al.}{2009}]{Kapferer2009}
{Kapferer} W.,  {Sluka} C.,  {Schindler} S.,  {Ferrari} C.,   {Ziegler} B.,
  2009, \mn@doi [\aap] {10.1051/0004-6361/200811551}, \href
  {https://ui.adsabs.harvard.edu/abs/2009A&A...499...87K} {499, 87}

\bibitem[\protect\citeauthoryear{{Keel}, {Kennicutt}, {Hummel}  \& {van der
  Hulst}}{{Keel} et~al.}{1985}]{Keel1985}
{Keel} W.~C.,  {Kennicutt} R.~C. J.,  {Hummel} E.,   {van der Hulst} J.~M.,
  1985, \mn@doi [\aj] {10.1086/113779}, \href
  {https://ui.adsabs.harvard.edu/abs/1985AJ.....90..708K} {90, 708}

\bibitem[\protect\citeauthoryear{{Kenney} \& {Koopmann}}{{Kenney} \&
  {Koopmann}}{1999}]{Kenney1999}
{Kenney} J. D.~P.,  {Koopmann} R.~A.,  1999, \mn@doi [\aj] {10.1086/300683},
  \href {https://ui.adsabs.harvard.edu/abs/1999AJ....117..181K} {117, 181}

\bibitem[\protect\citeauthoryear{{Kenney}, {Geha}, {J{\'a}chym}, {Crowl},
  {Dague}, {Chung}, {van Gorkom}  \& {Vollmer}}{{Kenney}
  et~al.}{2014}]{Kenney2014}
{Kenney} J. D.~P.,  {Geha} M.,  {J{\'a}chym} P.,  {Crowl} H.~H.,  {Dague} W.,
  {Chung} A.,  {van Gorkom} J.,   {Vollmer} B.,  2014, \mn@doi [\apj]
  {10.1088/0004-637X/780/2/119}, \href
  {https://ui.adsabs.harvard.edu/abs/2014ApJ...780..119K} {780, 119}

\bibitem[\protect\citeauthoryear{{Kennicutt}}{{Kennicutt}}{1983}]{Kennicutt1983}
{Kennicutt} R.~C. J.,  1983, \mn@doi [\aj] {10.1086/113334}, \href
  {https://ui.adsabs.harvard.edu/abs/1983AJ.....88..483K} {88, 483}

\bibitem[\protect\citeauthoryear{{Kennicutt}}{{Kennicutt}}{1998}]{Kennicutt1998}
{Kennicutt} Robert~C. J.,  1998, \mn@doi [\araa]
  {10.1146/annurev.astro.36.1.189}, \href
  {https://ui.adsabs.harvard.edu/abs/1998ARA&A..36..189K} {36, 189}

\bibitem[\protect\citeauthoryear{{Kennicutt}, {Keel}, {van der Hulst}, {Hummel}
   \& {Roettiger}}{{Kennicutt} et~al.}{1987}]{Kennicutt1987}
{Kennicutt} Robert~C. J.,  {Keel} W.~C.,  {van der Hulst} J.~M.,  {Hummel} E.,
   {Roettiger} K.~A.,  1987, \mn@doi [\aj] {10.1086/114384}, \href
  {https://ui.adsabs.harvard.edu/abs/1987AJ.....93.1011K} {93, 1011}

\bibitem[\protect\citeauthoryear{{Knollmann} \& {Knebe}}{{Knollmann} \&
  {Knebe}}{2009}]{amiga2009}
{Knollmann} S.~R.,  {Knebe} A.,  2009, \mn@doi [\apjs]
  {10.1088/0067-0049/182/2/608}, \href
  {https://ui.adsabs.harvard.edu/abs/2009ApJS..182..608K} {182, 608}

\bibitem[\protect\citeauthoryear{{Lambas}, {Tissera}, {Alonso}  \&
  {Coldwell}}{{Lambas} et~al.}{2003}]{lambas2003}
{Lambas} D.~G.,  {Tissera} P.~B.,  {Alonso} M.~S.,   {Coldwell} G.,  2003,
  \mn@doi [\mnras] {10.1111/j.1365-2966.2003.07179.x}, \href
  {https://ui.adsabs.harvard.edu/abs/2003MNRAS.346.1189L} {346, 1189}

\bibitem[\protect\citeauthoryear{{Martin}, {Shapley}, {Coil}, {Kornei},
  {Murray}  \& {Pancoast}}{{Martin} et~al.}{2013}]{Martin2013}
{Martin} C.~L.,  {Shapley} A.~E.,  {Coil} A.~L.,  {Kornei} K.~A.,  {Murray} N.,
    {Pancoast} A.,  2013, \mn@doi [\apj] {10.1088/0004-637X/770/1/41}, \href
  {https://ui.adsabs.harvard.edu/abs/2013ApJ...770...41M} {770, 41}

\bibitem[\protect\citeauthoryear{{Martin} et~al.,}{{Martin}
  et~al.}{2019}]{Martin2019}
{Martin} G.,  et~al., 2019, \mn@doi [\mnras] {10.1093/mnras/stz356}, \href
  {https://ui.adsabs.harvard.edu/abs/2019MNRAS.485..796M} {485, 796}

\bibitem[\protect\citeauthoryear{{Mayer}, {Mastropietro}, {Wadsley}, {Stadel}
  \& {Moore}}{{Mayer} et~al.}{2006}]{Mayer2006}
{Mayer} L.,  {Mastropietro} C.,  {Wadsley} J.,  {Stadel} J.,   {Moore} B.,
  2006, \mn@doi [\mnras] {10.1111/j.1365-2966.2006.10403.x}, \href
  {https://ui.adsabs.harvard.edu/abs/2006MNRAS.369.1021M} {369, 1021}

\bibitem[\protect\citeauthoryear{{Mayer}, {Kazantzidis}, {Mastropietro}  \&
  {Wadsley}}{{Mayer} et~al.}{2007}]{Mayer2007}
{Mayer} L.,  {Kazantzidis} S.,  {Mastropietro} C.,   {Wadsley} J.,  2007,
  \mn@doi [\nat] {10.1038/nature05552}, \href
  {https://ui.adsabs.harvard.edu/abs/2007Natur.445..738M} {445, 738}

\bibitem[\protect\citeauthoryear{{Merritt}}{{Merritt}}{1983}]{Merrit1983}
{Merritt} D.,  1983, \mn@doi [\apj] {10.1086/160571}, \href
  {https://ui.adsabs.harvard.edu/abs/1983ApJ...264...24M} {264, 24}

\bibitem[\protect\citeauthoryear{{Mosconi}, {Tissera}, {Lambas}  \&
  {Cora}}{{Mosconi} et~al.}{2001}]{mosc2001}
{Mosconi} M.~B.,  {Tissera} P.~B.,  {Lambas} D.~G.,   {Cora} S.~A.,  2001,
  \mn@doi [\mnras] {10.1046/j.1365-8711.2001.04198.x}, \href
  {https://ui.adsabs.harvard.edu/abs/2001MNRAS.325...34M} {325, 34}

\bibitem[\protect\citeauthoryear{{Navarro} \& {White}}{{Navarro} \&
  {White}}{1993}]{navarro93}
{Navarro} J.~F.,  {White} S.~D.~M.,  1993, \mn@doi [\mnras]
  {10.1093/mnras/265.2.271}, \href
  {https://ui.adsabs.harvard.edu/abs/1993MNRAS.265..271N} {265, 271}

\bibitem[\protect\citeauthoryear{{Pallero}, {G{\'o}mez}, {Padilla},
  {Torres-Flores}, {Demarco}, {Cerulo}  \& {Olave-Rojas}}{{Pallero}
  et~al.}{2019}]{Pallero2019}
{Pallero} D.,  {G{\'o}mez} F.~A.,  {Padilla} N.~D.,  {Torres-Flores} S.,
  {Demarco} R.,  {Cerulo} P.,   {Olave-Rojas} D.,  2019, \mn@doi [\mnras]
  {10.1093/mnras/stz1745}, \href
  {https://ui.adsabs.harvard.edu/abs/2019MNRAS.488..847P} {488, 847}

\bibitem[\protect\citeauthoryear{{Pallero}, {G{\'o}mez}, {Padilla}, {Bah{\'e}},
  {Vega-Mart{\'\i}nez}  \& {Torres-Flores}}{{Pallero}
  et~al.}{2020}]{pallero2020}
{Pallero} D.,  {G{\'o}mez} F.~A.,  {Padilla} N.~D.,  {Bah{\'e}} Y.~M.,
  {Vega-Mart{\'\i}nez} C.~A.,   {Torres-Flores} S.,  2020, arXiv e-prints,
  \href {https://ui.adsabs.harvard.edu/abs/2020arXiv201208593P} {p.
  arXiv:2012.08593}

\bibitem[\protect\citeauthoryear{{Pedrosa} \& {Tissera}}{{Pedrosa} \&
  {Tissera}}{2015}]{Pedrosa2015}
{Pedrosa} S.~E.,  {Tissera} P.~B.,  2015, \mn@doi [\aap]
  {10.1051/0004-6361/201526440}, \href
  {https://ui.adsabs.harvard.edu/abs/2015A&A...584A..43P} {584, A43}

\bibitem[\protect\citeauthoryear{{Perez}, {Michel-Dansac}  \&
  {Tissera}}{{Perez} et~al.}{2011}]{perez2011}
{Perez} J.,  {Michel-Dansac} L.,   {Tissera} P.~B.,  2011, \mn@doi [\mnras]
  {10.1111/j.1365-2966.2011.19300.x}, \href
  {https://ui.adsabs.harvard.edu/abs/2011MNRAS.417..580P} {417, 580}

\bibitem[\protect\citeauthoryear{{Raiteri}, {Villata}  \& {Navarro}}{{Raiteri}
  et~al.}{1996}]{rait1996}
{Raiteri} C.~M.,  {Villata} M.,   {Navarro} J.~F.,  1996, \aap, \href
  {https://ui.adsabs.harvard.edu/abs/1996A&A...315..105R} {315, 105}

\bibitem[\protect\citeauthoryear{{Rasmussen}, {Ponman}  \&
  {Mulchaey}}{{Rasmussen} et~al.}{2006}]{Rasmussen2006}
{Rasmussen} J.,  {Ponman} T.~J.,   {Mulchaey} J.~S.,  2006, \mn@doi [\mnras]
  {10.1111/j.1365-2966.2006.10492.x}, \href
  {https://ui.adsabs.harvard.edu/abs/2006MNRAS.370..453R} {370, 453}

\bibitem[\protect\citeauthoryear{{Rhee}, {Smith}, {Choi}, {Contini}, {Jung},
  {Han}  \& {Yi}}{{Rhee} et~al.}{2020}]{Rhee2020}
{Rhee} J.,  {Smith} R.,  {Choi} H.,  {Contini} E.,  {Jung} S.~L.,  {Han} S.,
  {Yi} S.~K.,  2020, \mn@doi [\apjs] {10.3847/1538-4365/ab7377}, \href
  {https://ui.adsabs.harvard.edu/abs/2020ApJS..247...45R} {247, 45}

\bibitem[\protect\citeauthoryear{{Rodr{\'\i}guez}, {Garcia Lambas}, {Padilla}
  \& {Troncoso-Iribarren}}{{Rodr{\'\i}guez} et~al.}{2020}]{rgpt20}
{Rodr{\'\i}guez} S.,  {Garcia Lambas} D.,  {Padilla} N.~D.,
  {Troncoso-Iribarren} P.,  2020, \mn@doi [\mnras] {10.1093/mnras/stz3456},
  \href {https://ui.adsabs.harvard.edu/abs/2020MNRAS.492..413R} {492, 413}

\bibitem[\protect\citeauthoryear{{Rubin}, {Prochaska}, {Koo}, {Phillips},
  {Martin}  \& {Winstrom}}{{Rubin} et~al.}{2014}]{Rubin2014}
{Rubin} K. H.~R.,  {Prochaska} J.~X.,  {Koo} D.~C.,  {Phillips} A.~C.,
  {Martin} C.~L.,   {Winstrom} L.~O.,  2014, \mn@doi [\apj]
  {10.1088/0004-637X/794/2/156}, \href
  {https://ui.adsabs.harvard.edu/abs/2014ApJ...794..156R} {794, 156}

\bibitem[\protect\citeauthoryear{{Rupke}, {Kewley}  \& {Chien}}{{Rupke}
  et~al.}{2010}]{rupke2010}
{Rupke} D. S.~N.,  {Kewley} L.~J.,   {Chien} L.~H.,  2010, \mn@doi [\apj]
  {10.1088/0004-637X/723/2/1255}, \href
  {https://ui.adsabs.harvard.edu/abs/2010ApJ...723.1255R} {723, 1255}

\bibitem[\protect\citeauthoryear{{Safarzadeh} \& {Loeb}}{{Safarzadeh} \&
  {Loeb}}{2019}]{Safarzadeh2019}
{Safarzadeh} M.,  {Loeb} A.,  2019, \mn@doi [\mnras] {10.1093/mnrasl/slz053},
  \href {https://ui.adsabs.harvard.edu/abs/2019MNRAS.486L..26S} {486, L26}

\bibitem[\protect\citeauthoryear{{Scannapieco}, {Tissera}, {White}  \&
  {Springel}}{{Scannapieco} et~al.}{2005}]{scan05}
{Scannapieco} C.,  {Tissera} P.~B.,  {White} S.~D.~M.,   {Springel} V.,  2005,
  \mn@doi [\mnras] {10.1111/j.1365-2966.2005.09574.x}, \href
  {https://ui.adsabs.harvard.edu/abs/2005MNRAS.364..552S} {364, 552}

\bibitem[\protect\citeauthoryear{{Scannapieco}, {Tissera}, {White}  \&
  {Springel}}{{Scannapieco} et~al.}{2006a}]{Scannapieco2006}
{Scannapieco} C.,  {Tissera} P.~B.,  {White} S.~D.~M.,   {Springel} V.,  2006a,
  \mn@doi [\mnras] {10.1111/j.1365-2966.2006.10785.x}, \href
  {https://ui.adsabs.harvard.edu/abs/2006MNRAS.371.1125S} {371, 1125}

\bibitem[\protect\citeauthoryear{{Scannapieco}, {Tissera}, {White}  \&
  {Springel}}{{Scannapieco} et~al.}{2006b}]{scan06}
{Scannapieco} C.,  {Tissera} P.~B.,  {White} S.~D.~M.,   {Springel} V.,  2006b,
  \mn@doi [\mnras] {10.1111/j.1365-2966.2006.10785.x}, \href
  {https://ui.adsabs.harvard.edu/abs/2006MNRAS.371.1125S} {371, 1125}

\bibitem[\protect\citeauthoryear{{Sillero}, {Tissera}, {Lambas}  \&
  {Michel-Dansac}}{{Sillero} et~al.}{2017}]{sillero17}
{Sillero} E.,  {Tissera} P.~B.,  {Lambas} D.~G.,   {Michel-Dansac} L.,  2017,
  \mn@doi [\mnras] {10.1093/mnras/stx2265}, \href
  {https://ui.adsabs.harvard.edu/abs/2017MNRAS.472.4404S} {472, 4404}

\bibitem[\protect\citeauthoryear{{Skillman}, {Kennicutt}, {Shields}  \&
  {Zaritsky}}{{Skillman} et~al.}{1996}]{Skillman1996}
{Skillman} E.~D.,  {Kennicutt} Robert~C. J.,  {Shields} G.~A.,   {Zaritsky} D.,
   1996, \mn@doi [\apj] {10.1086/177138}, \href
  {https://ui.adsabs.harvard.edu/abs/1996ApJ...462..147S} {462, 147}

\bibitem[\protect\citeauthoryear{{Springel}}{{Springel}}{2005}]{springel2005}
{Springel} V.,  2005, \mn@doi [\mnras] {10.1111/j.1365-2966.2005.09655.x},
  \href {https://ui.adsabs.harvard.edu/abs/2005MNRAS.364.1105S} {364, 1105}

\bibitem[\protect\citeauthoryear{{Springel} \& {Hernquist}}{{Springel} \&
  {Hernquist}}{2003}]{springel2003}
{Springel} V.,  {Hernquist} L.,  2003, \mn@doi [\mnras]
  {10.1046/j.1365-8711.2003.06206.x}, \href
  {https://ui.adsabs.harvard.edu/abs/2003MNRAS.339..289S} {339, 289}

\bibitem[\protect\citeauthoryear{{Springel}, {White}, {Tormen}  \&
  {Kauffmann}}{{Springel} et~al.}{2001}]{springel2001a}
{Springel} V.,  {White} S. D.~M.,  {Tormen} G.,   {Kauffmann} G.,  2001,
  \mn@doi [\mnras] {10.1046/j.1365-8711.2001.04912.x}, \href
  {https://ui.adsabs.harvard.edu/abs/2001MNRAS.328..726S} {328, 726}

\bibitem[\protect\citeauthoryear{{Steinhauser}, {Haider}, {Kapferer}  \&
  {Schindler}}{{Steinhauser} et~al.}{2012}]{Steinhauser2012}
{Steinhauser} D.,  {Haider} M.,  {Kapferer} W.,   {Schindler} S.,  2012,
  \mn@doi [\aap] {10.1051/0004-6361/201118311}, \href
  {https://ui.adsabs.harvard.edu/abs/2012A&A...544A..54S} {544, A54}

\bibitem[\protect\citeauthoryear{{Steinhauser}, {Schindler}  \&
  {Springel}}{{Steinhauser} et~al.}{2016}]{Steinhauser2016}
{Steinhauser} D.,  {Schindler} S.,   {Springel} V.,  2016, \mn@doi [\aap]
  {10.1051/0004-6361/201527705}, \href
  {https://ui.adsabs.harvard.edu/abs/2016A&A...591A..51S} {591, A51}

\bibitem[\protect\citeauthoryear{{Tissera}, {White}  \&
  {Scannapieco}}{{Tissera} et~al.}{2012}]{tissera2012}
{Tissera} P.~B.,  {White} S. D.~M.,   {Scannapieco} C.,  2012, \mn@doi [\mnras]
  {10.1111/j.1365-2966.2011.20028.x}, \href
  {https://ui.adsabs.harvard.edu/abs/2012MNRAS.420..255T} {420, 255}

\bibitem[\protect\citeauthoryear{{Tissera}, {Pedrosa}, {Sillero}  \&
  {Vilchez}}{{Tissera} et~al.}{2016a}]{tissera2016a}
{Tissera} P.~B.,  {Pedrosa} S.~E.,  {Sillero} E.,   {Vilchez} J.~M.,  2016a,
  \mn@doi [\mnras] {10.1093/mnras/stv2736}, \href
  {https://ui.adsabs.harvard.edu/abs/2016MNRAS.456.2982T} {456, 2982}

\bibitem[\protect\citeauthoryear{{Tissera}, {Machado}, {Sanchez-Blazquez},
  {Pedrosa}, {S{\'a}nchez}, {Snaith}  \& {Vilchez}}{{Tissera}
  et~al.}{2016b}]{tissera2016b}
{Tissera} P.~B.,  {Machado} R. E.~G.,  {Sanchez-Blazquez} P.,  {Pedrosa} S.~E.,
   {S{\'a}nchez} S.~F.,  {Snaith} O.,   {Vilchez} J.,  2016b, \mn@doi [\aap]
  {10.1051/0004-6361/201628188}, \href
  {https://ui.adsabs.harvard.edu/abs/2016A&A...592A..93T} {592, A93}

\bibitem[\protect\citeauthoryear{{Tonnesen}, {Bryan}  \& {van
  Gorkom}}{{Tonnesen} et~al.}{2007}]{Tonnesen2007}
{Tonnesen} S.,  {Bryan} G.~L.,   {van Gorkom} J.~H.,  2007, \mn@doi [\apj]
  {10.1086/523034}, \href
  {https://ui.adsabs.harvard.edu/abs/2007ApJ...671.1434T} {671, 1434}

\bibitem[\protect\citeauthoryear{{Torrey}, {Cox}, {Kewley}  \&
  {Hernquist}}{{Torrey} et~al.}{2012}]{torre2012}
{Torrey} P.,  {Cox} T.~J.,  {Kewley} L.,   {Hernquist} L.,  2012, \mn@doi
  [\apj] {10.1088/0004-637X/746/1/108}, \href
  {https://ui.adsabs.harvard.edu/abs/2012ApJ...746..108T} {746, 108}

\bibitem[\protect\citeauthoryear{{Troncoso-Iribarren}, {Padilla}, {Santander},
  {Lagos}, {Garc{\'\i}a-Lambas}, {Rodr{\'\i}guez}  \&
  {Contreras}}{{Troncoso-Iribarren} et~al.}{2020}]{Troncoso2020}
{Troncoso-Iribarren} P.,  {Padilla} N.,  {Santander} C.,  {Lagos} C.~D.~P.,
  {Garc{\'\i}a-Lambas} D.,  {Rodr{\'\i}guez} S.,   {Contreras} S.,  2020,
  \mn@doi [\mnras] {10.1093/mnras/staa274}, \href
  {https://ui.adsabs.harvard.edu/abs/2020MNRAS.497.4145T} {497, 4145}

\bibitem[\protect\citeauthoryear{{Upadhyay}, {Oman}  \& {Trager}}{{Upadhyay}
  et~al.}{2021}]{Upadhyay2021}
{Upadhyay} A.~K.,  {Oman} K.~A.,   {Trager} S.~C.,  2021, arXiv e-prints, \href
  {https://ui.adsabs.harvard.edu/abs/2021arXiv210404388U} {p. arXiv:2104.04388}

\bibitem[\protect\citeauthoryear{{Weiner} et~al.,}{{Weiner}
  et~al.}{2009}]{Weiner2009}
{Weiner} B.~J.,  et~al., 2009, \mn@doi [\apj] {10.1088/0004-637X/692/1/187},
  \href {https://ui.adsabs.harvard.edu/abs/2009ApJ...692..187W} {692, 187}

\bibitem[\protect\citeauthoryear{{Woosley} \& {Weaver}}{{Woosley} \&
  {Weaver}}{1995}]{WW95}
{Woosley} S.~E.,  {Weaver} T.~A.,  1995, \mn@doi [\apjs] {10.1086/192237},
  \href {https://ui.adsabs.harvard.edu/abs/1995ApJS..101..181W} {101, 181}

\bibitem[\protect\citeauthoryear{{Yoon}, {Chung}, {Smith}  \&
  {Jaff{\'e}}}{{Yoon} et~al.}{2017}]{Yoon2017}
{Yoon} H.,  {Chung} A.,  {Smith} R.,   {Jaff{\'e}} Y.~L.,  2017, \mn@doi [\apj]
  {10.3847/1538-4357/aa6579}, \href
  {https://ui.adsabs.harvard.edu/abs/2017ApJ...838...81Y} {838, 81}

\bibitem[\protect\citeauthoryear{{Yoshida} et~al.,}{{Yoshida}
  et~al.}{2008}]{Yoshida2008}
{Yoshida} M.,  et~al., 2008, \mn@doi [\apj] {10.1086/592430}, \href
  {https://ui.adsabs.harvard.edu/abs/2008ApJ...688..918Y} {688, 918}

\bibitem[\protect\citeauthoryear{{Zabludoff} \& {Mulchaey}}{{Zabludoff} \&
  {Mulchaey}}{1998}]{Zabludoff1998}
{Zabludoff} A.~I.,  {Mulchaey} J.~S.,  1998, \mn@doi [\apjl] {10.1086/311312},
  \href {https://ui.adsabs.harvard.edu/abs/1998ApJ...498L...5Z} {498, L5}

\bibitem[\protect\citeauthoryear{{van der Marel}, {Fardal}, {Besla}, {Beaton},
  {Sohn}, {Anderson}, {Brown}  \& {Guhathakurta}}{{van der Marel}
  et~al.}{2012}]{vanderMarel2012}
{van der Marel} R.~P.,  {Fardal} M.,  {Besla} G.,  {Beaton} R.~L.,  {Sohn}
  S.~T.,  {Anderson} J.,  {Brown} T.,   {Guhathakurta} P.,  2012, \mn@doi
  [\apj] {10.1088/0004-637X/753/1/8}, \href
  {https://ui.adsabs.harvard.edu/abs/2012ApJ...753....8V} {753, 8}

\makeatother
\end{thebibliography}

\onecolumn

\appendix
\section{Properties of  galaxies in the sample}
\label{secapen}
\begin{figure*}
	\includegraphics[width=0.29\textwidth]{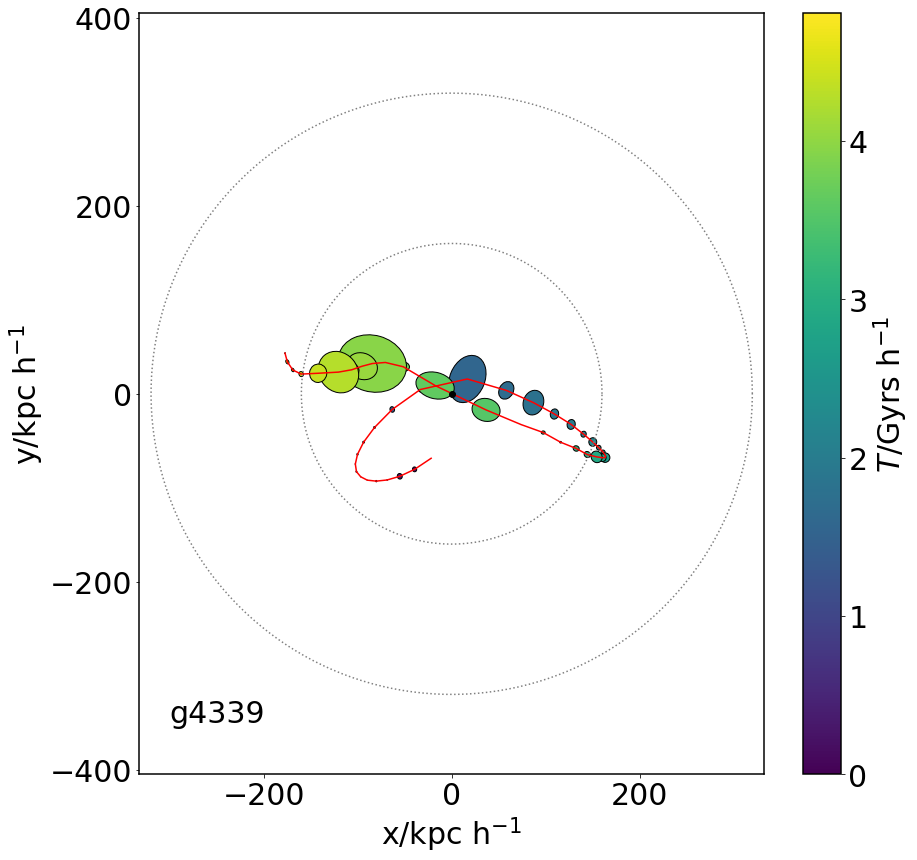}
	\includegraphics[width=0.29\textwidth]{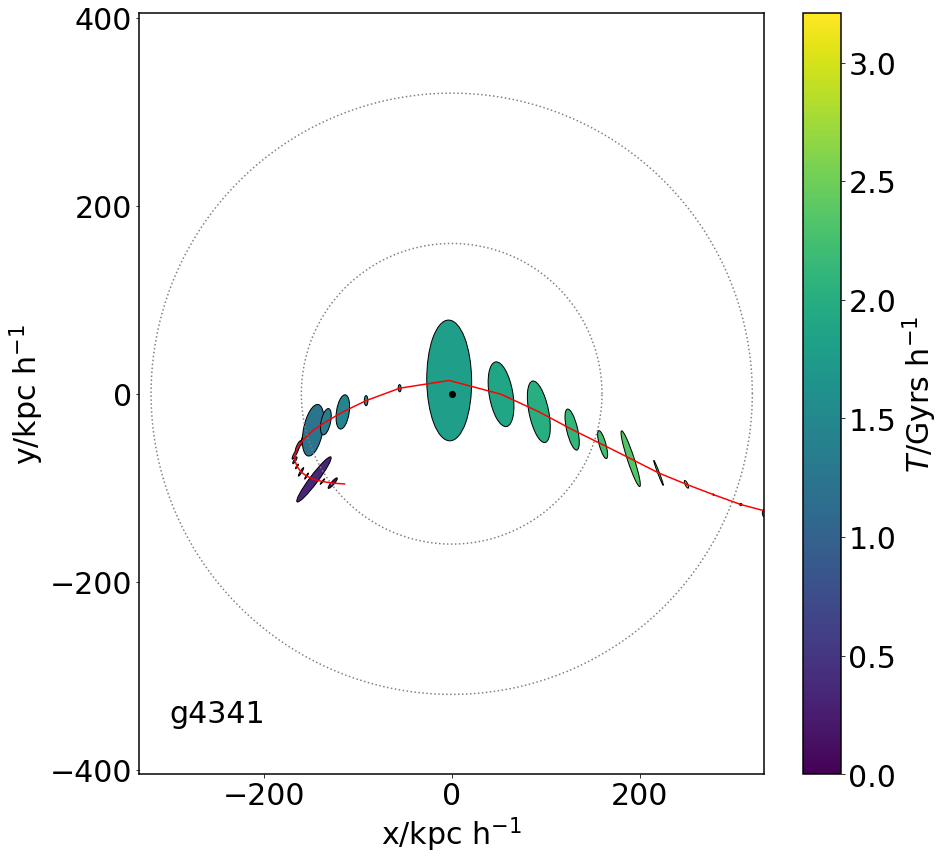}
	\includegraphics[width=0.29\textwidth]{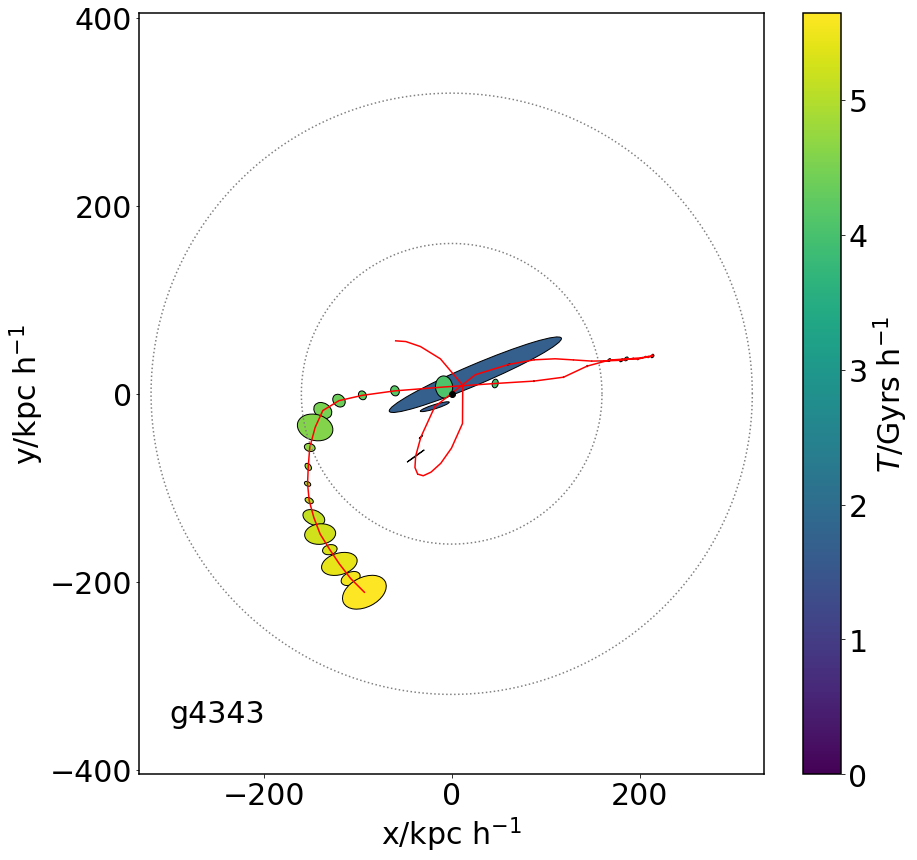}
	\includegraphics[width=0.29\textwidth]{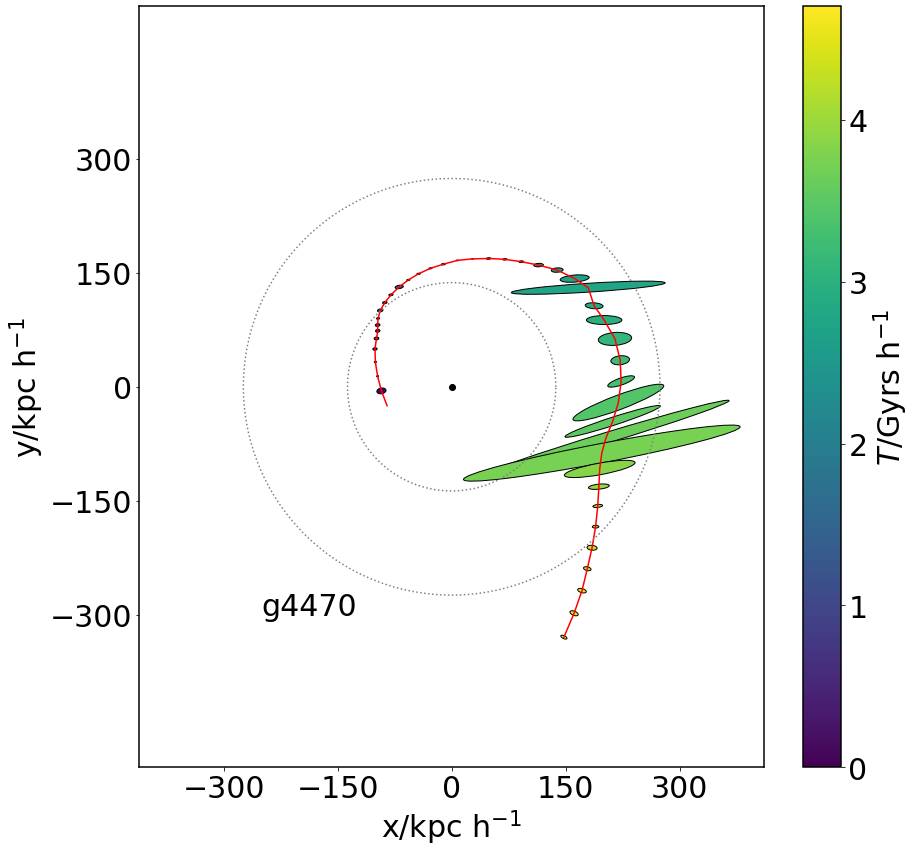}
	\includegraphics[width=0.29\textwidth]{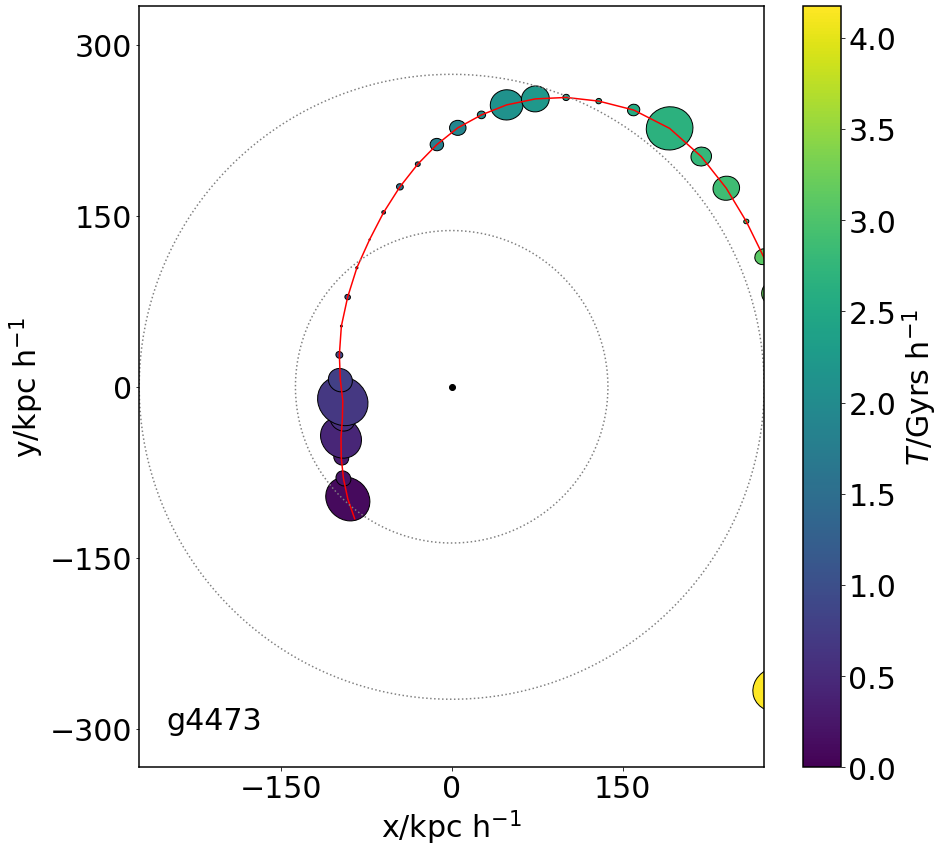}
	\includegraphics[width=0.29\textwidth]{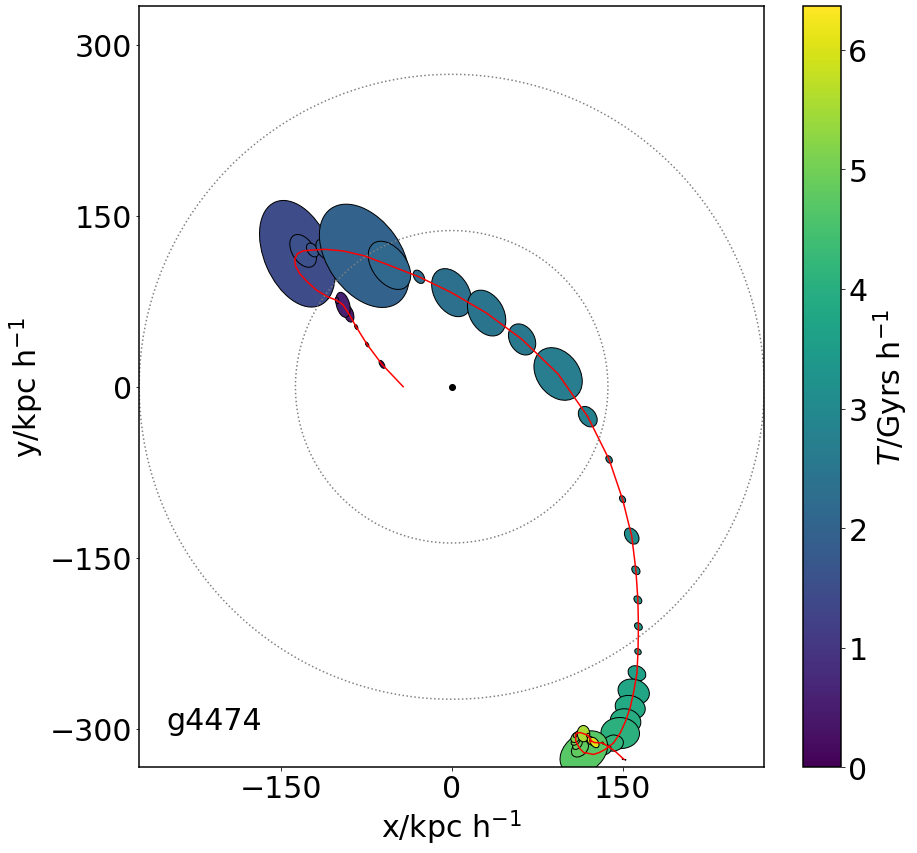}
	
	\caption{\label{fig_Orientation_append} Variation of the orientation of satellite galaxies along their orbits (as in figure \ref{fig_Orientation}). Top Left: Galaxy g4339. Top Middle: Galaxy g4341. Top Right: Galaxy g4343. Bottom Left: Galaxy g4470. Bottom Middle: Galaxy g4473. Bottom Right: Galaxy g4474.}
\end{figure*}

\begin{figure*}
	\includegraphics[width=0.29\textwidth]{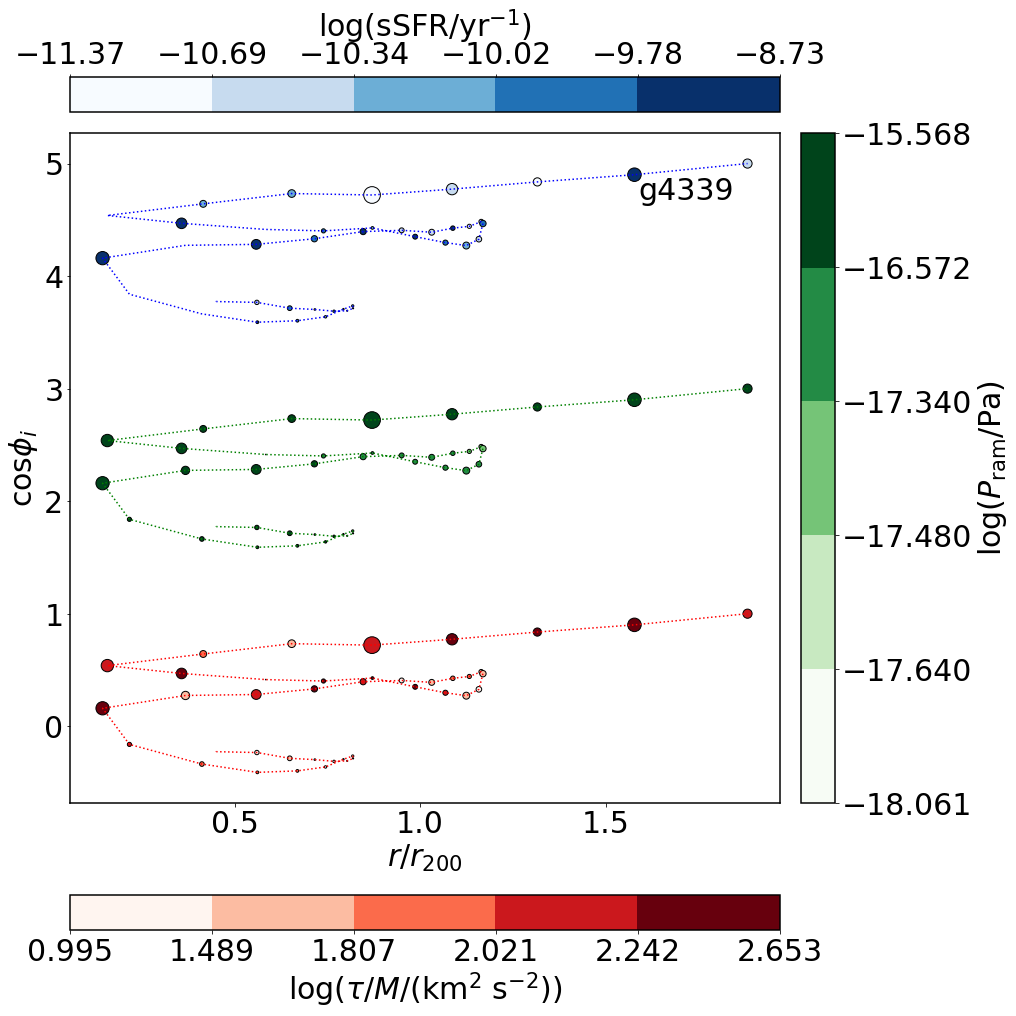}
	\includegraphics[width=0.29\textwidth]{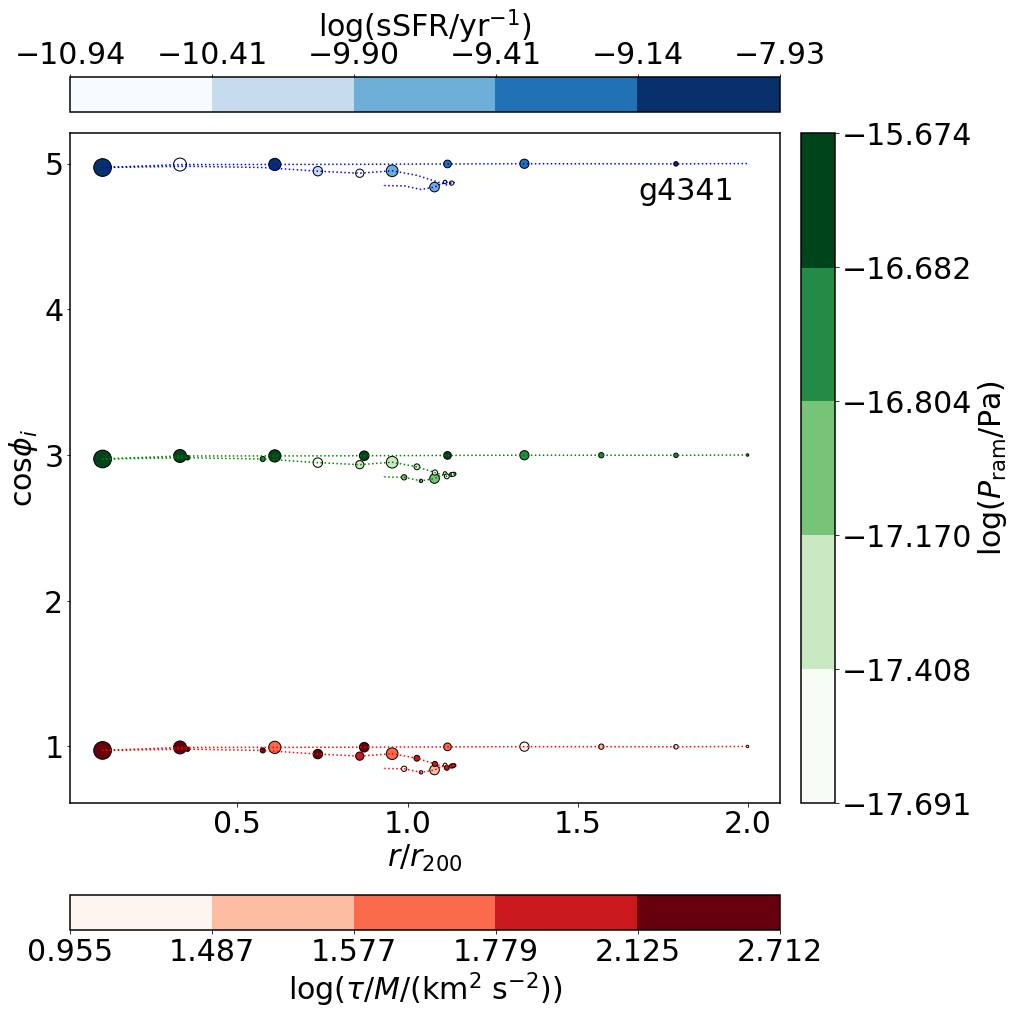}
	\includegraphics[width=0.29\textwidth]{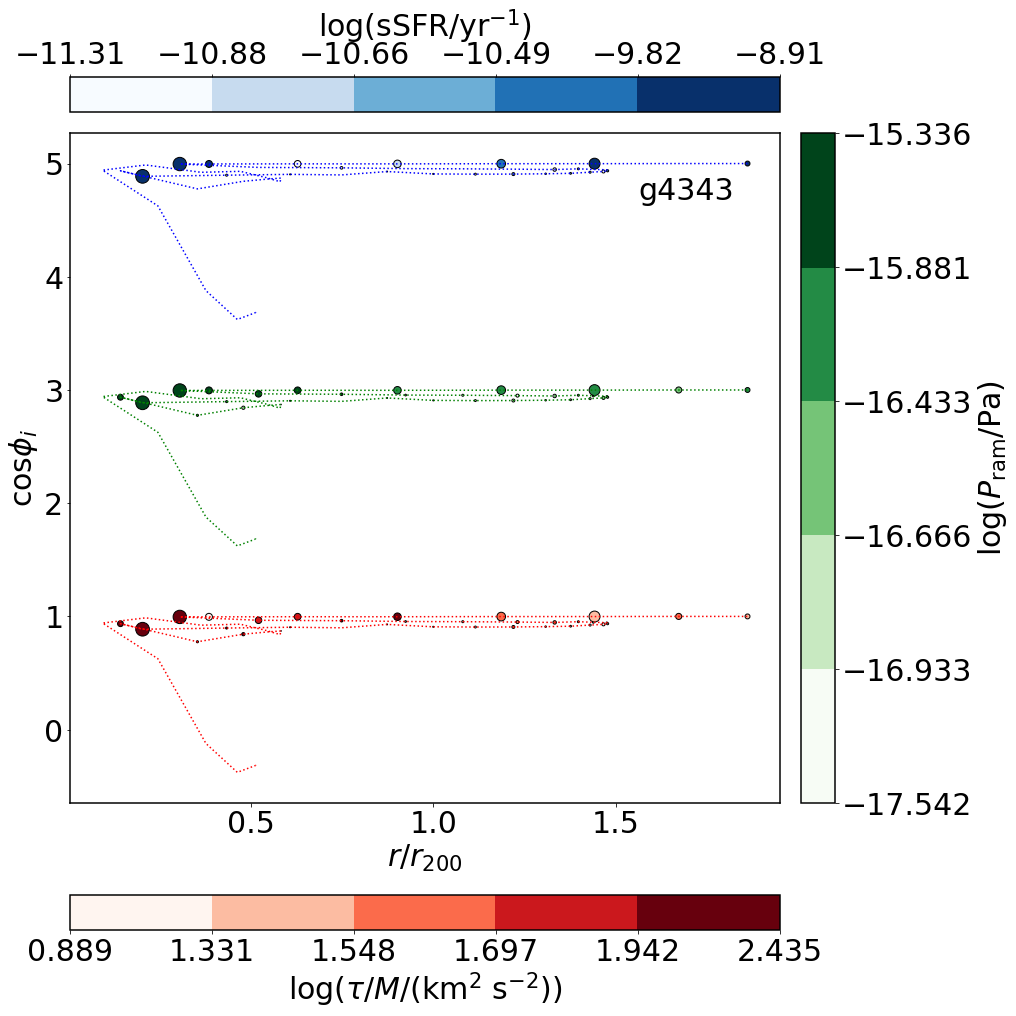}
	\includegraphics[width=0.29\textwidth]{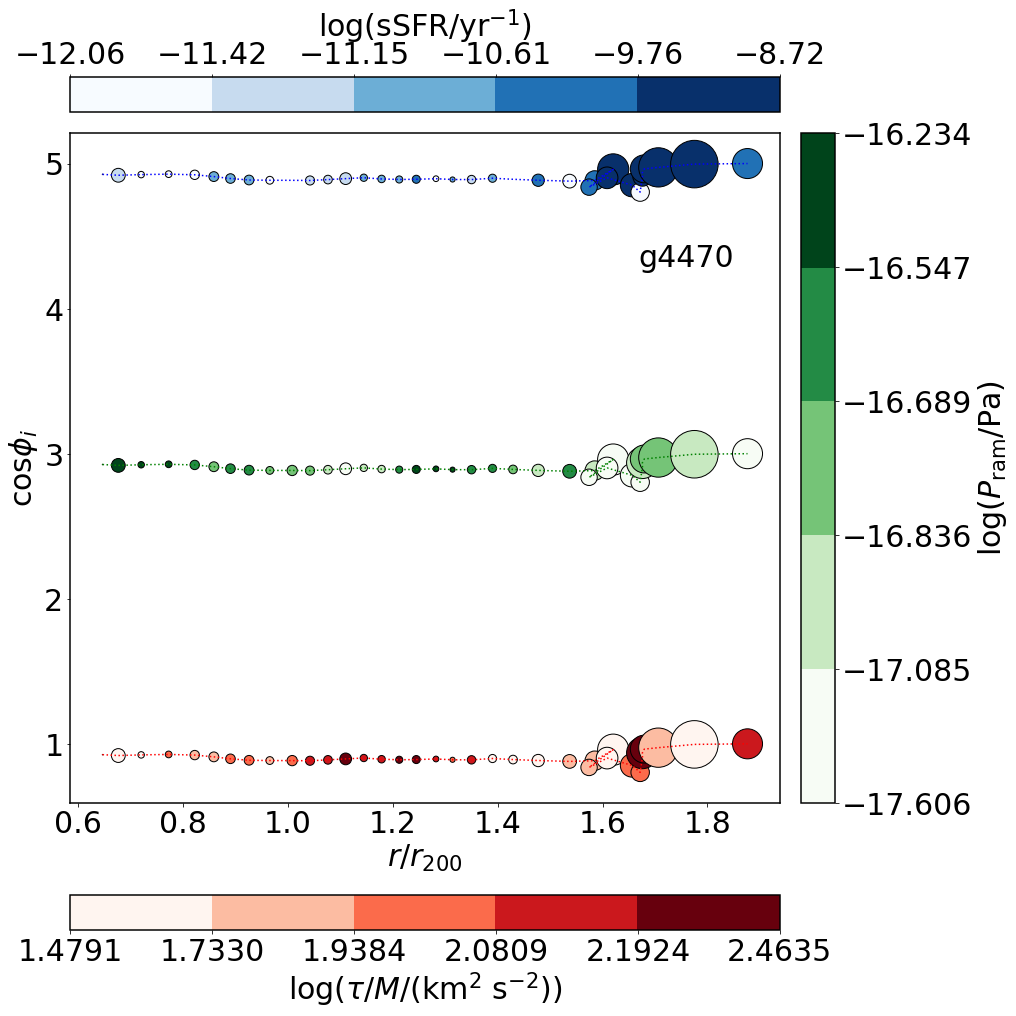}
	\includegraphics[width=0.29\textwidth]{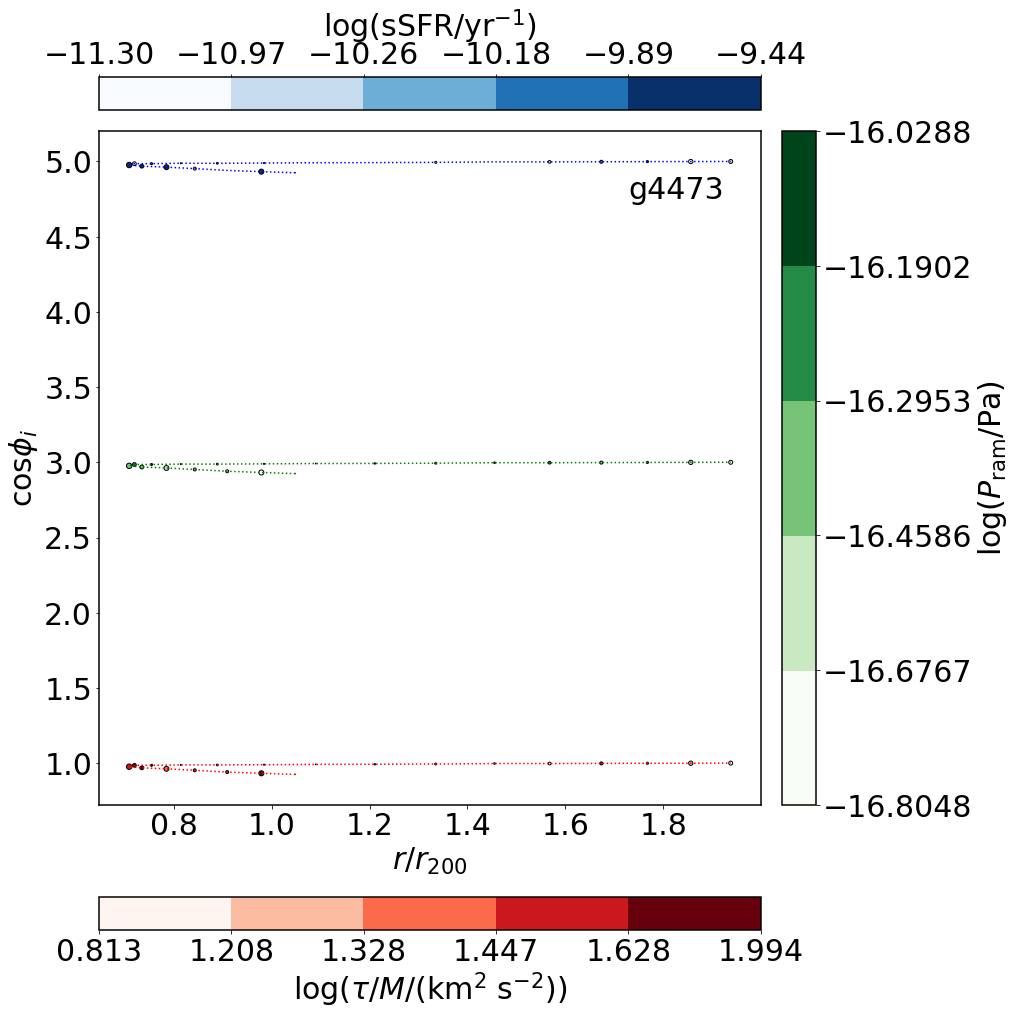}
	\includegraphics[width=0.29\textwidth]{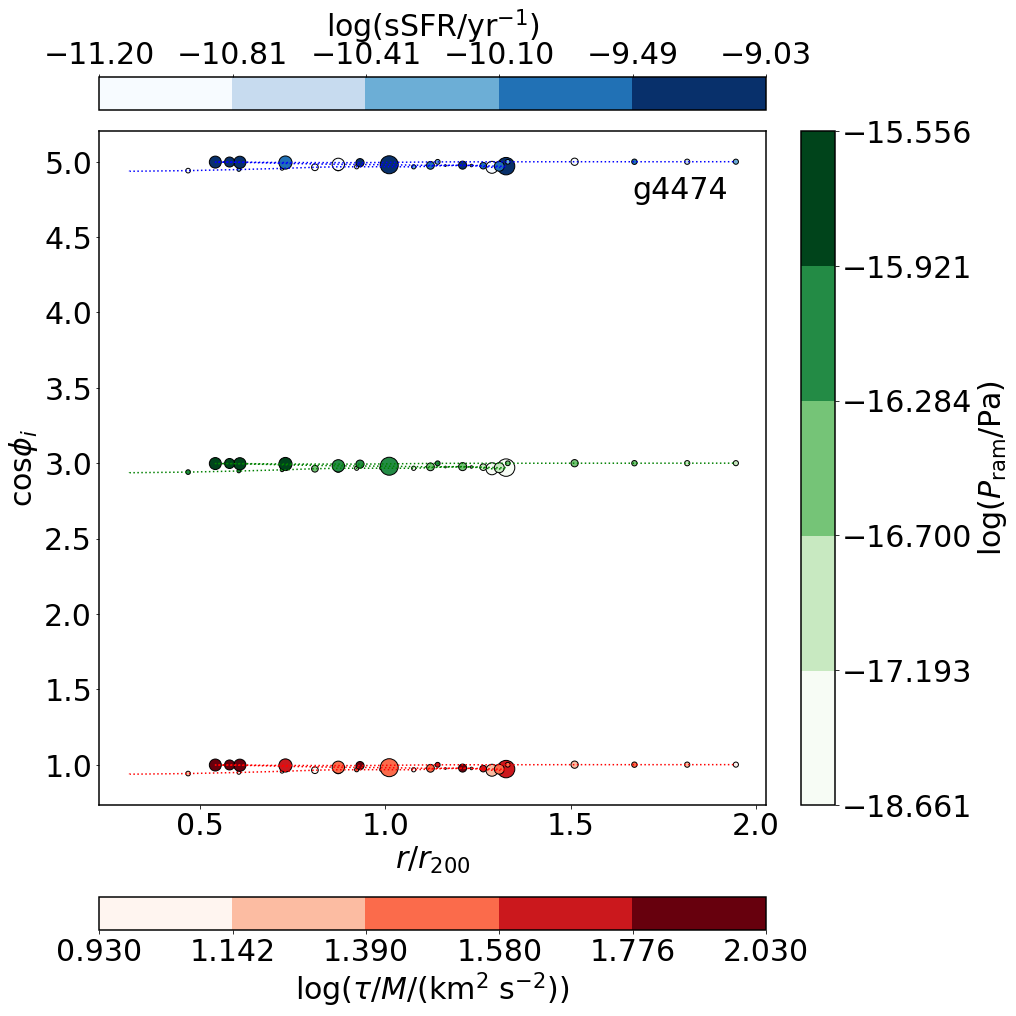}
	\caption{\label{fig_cosVSeffect_append} The angle of the orientation of satellite galaxies alog their orbits (as in figure \ref{fig_cosVSeffect}). Top Left: Galaxy g4339. Top Middle: Galaxy g4341. Top Right: Galaxy g4343. Bottom Left: Galaxy g4470. Bottom Middle: Galaxy g4473. Bottom Right: Galaxy g4474.}
\end{figure*}

\begin{figure*}
	\includegraphics[width=0.28\textwidth]{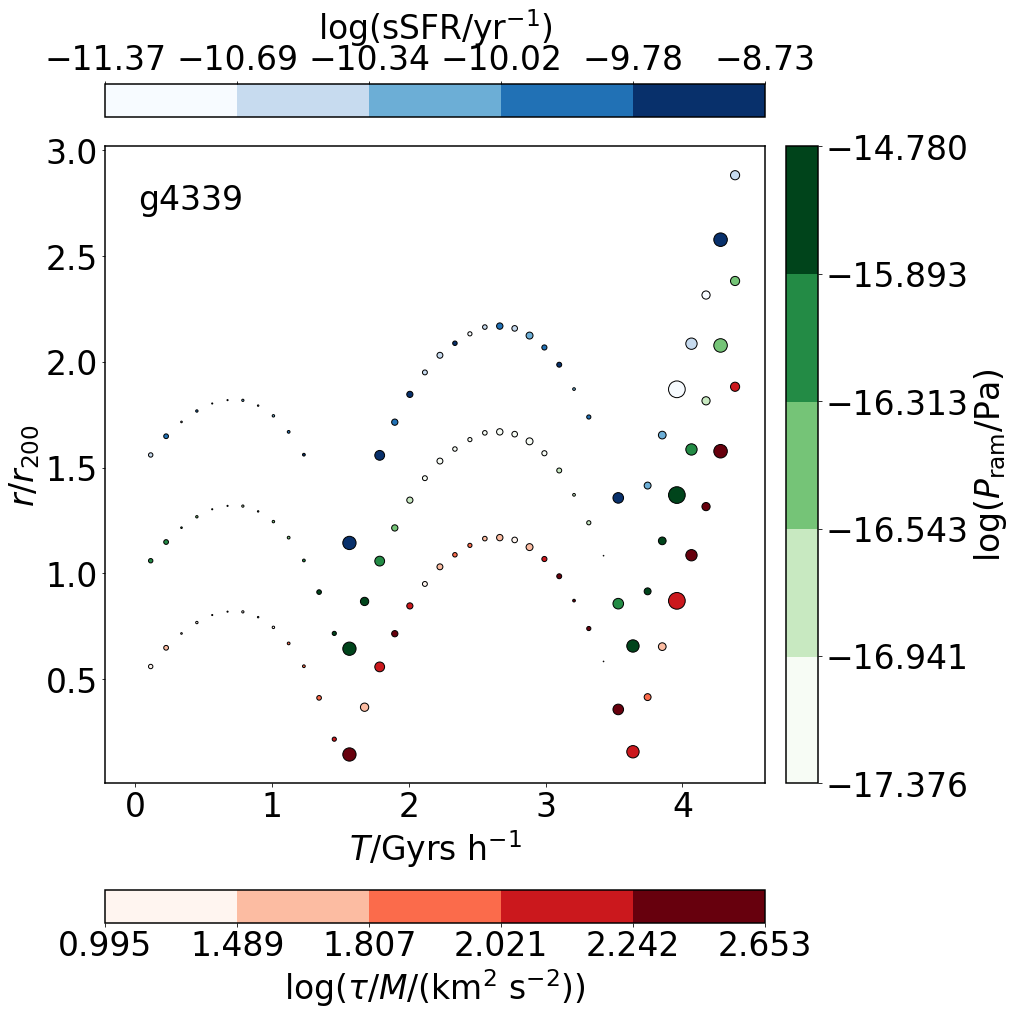}
	\includegraphics[width=0.28\textwidth]{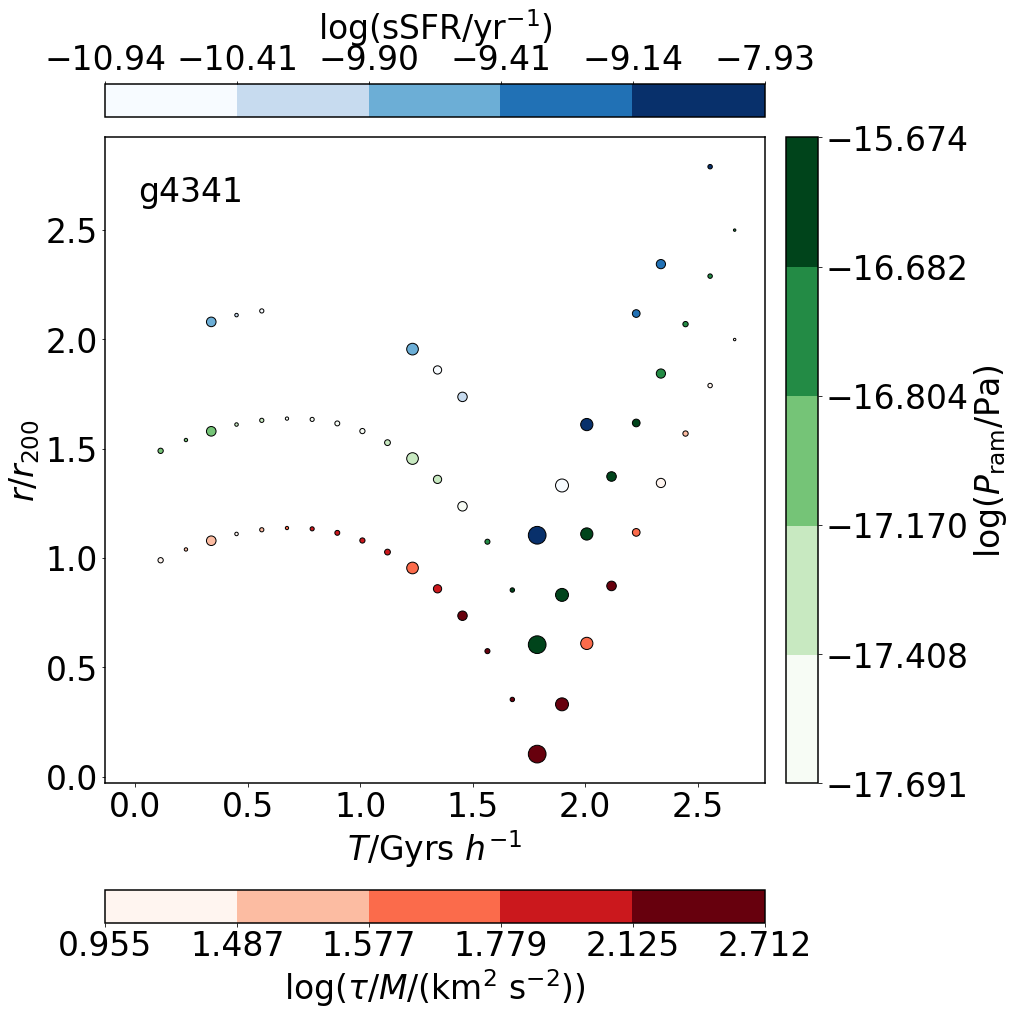}
	\includegraphics[width=0.28\textwidth]{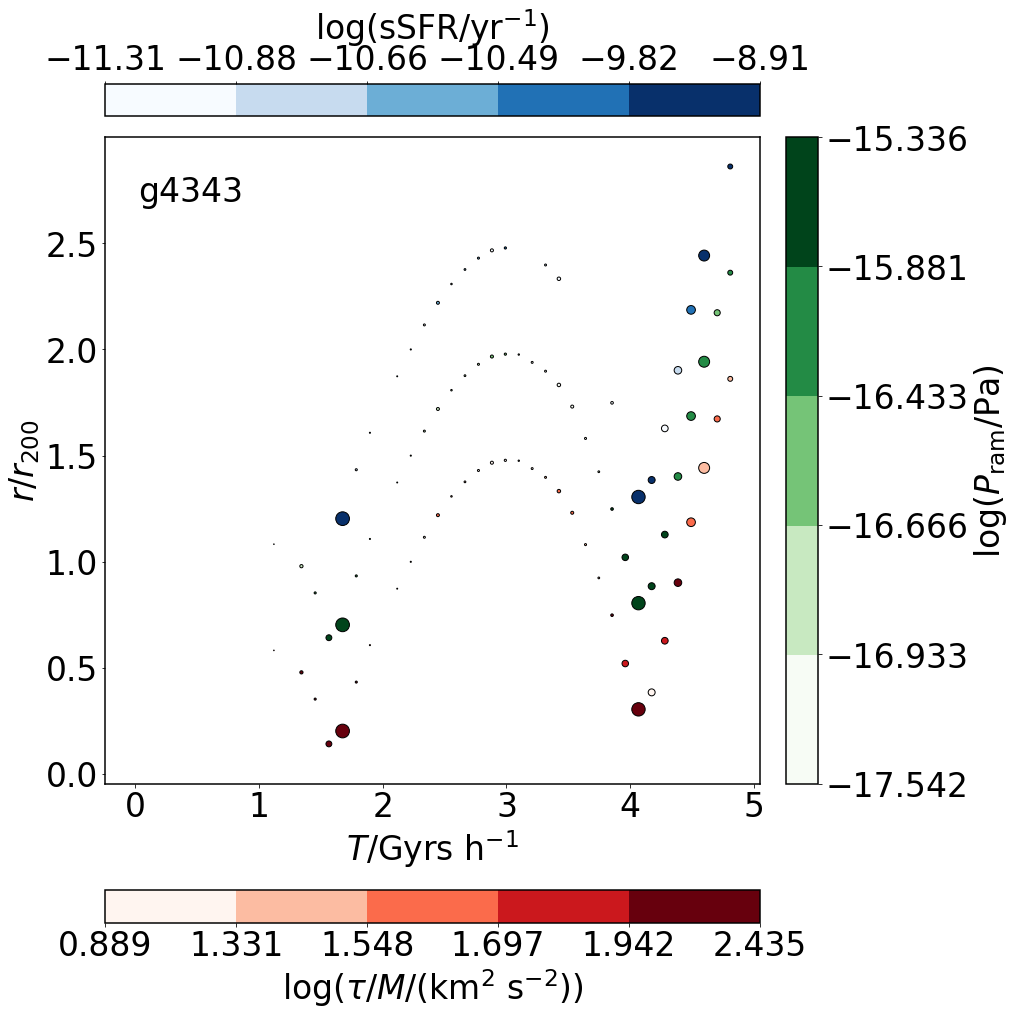}
	
	\includegraphics[width=0.28\textwidth]{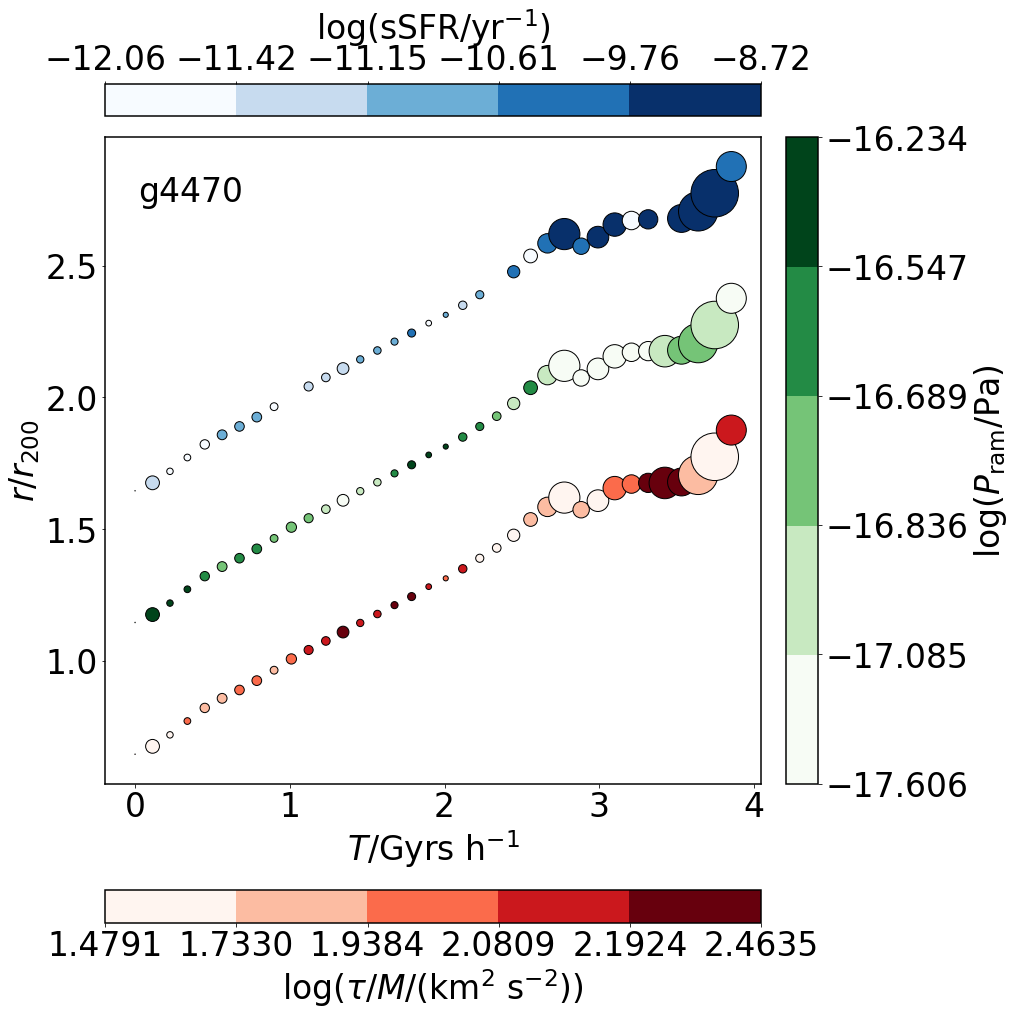}
	\includegraphics[width=0.28\textwidth]{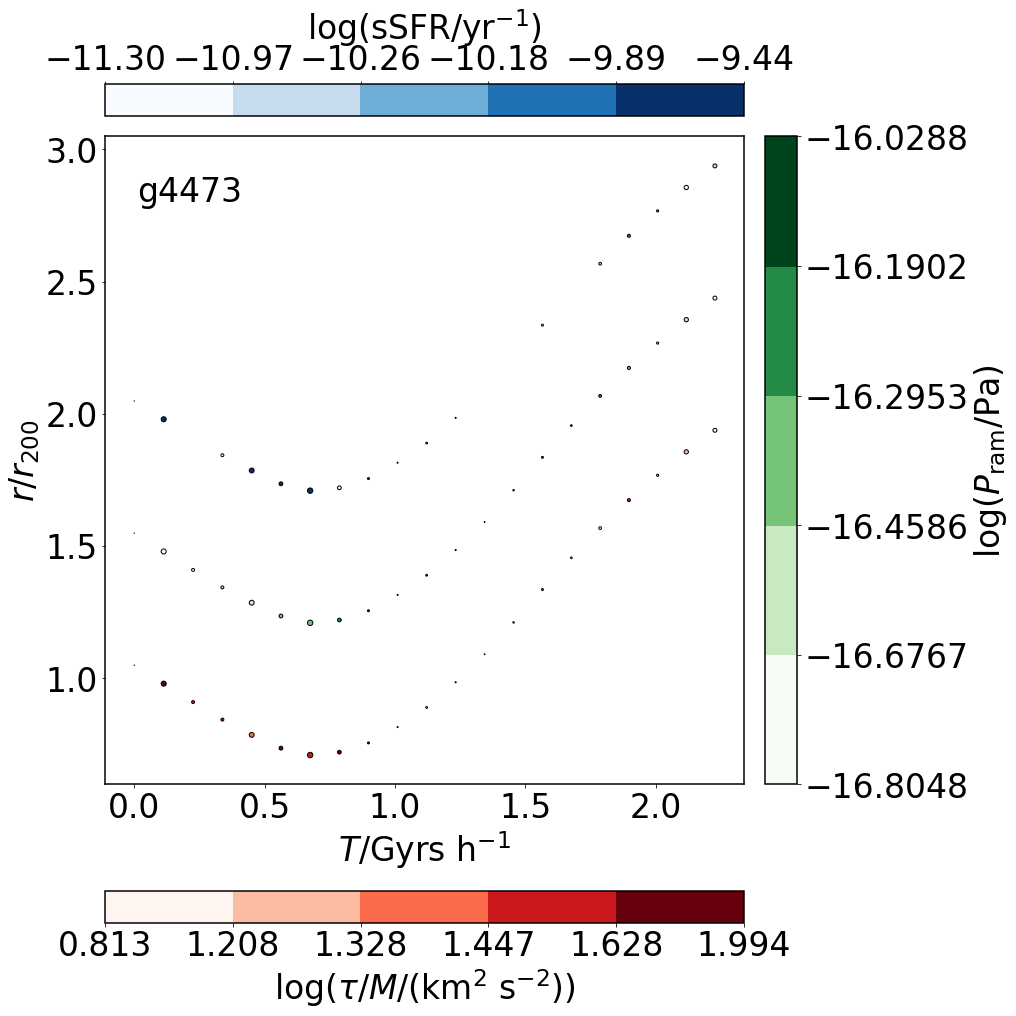}
	\includegraphics[width=0.28\textwidth]{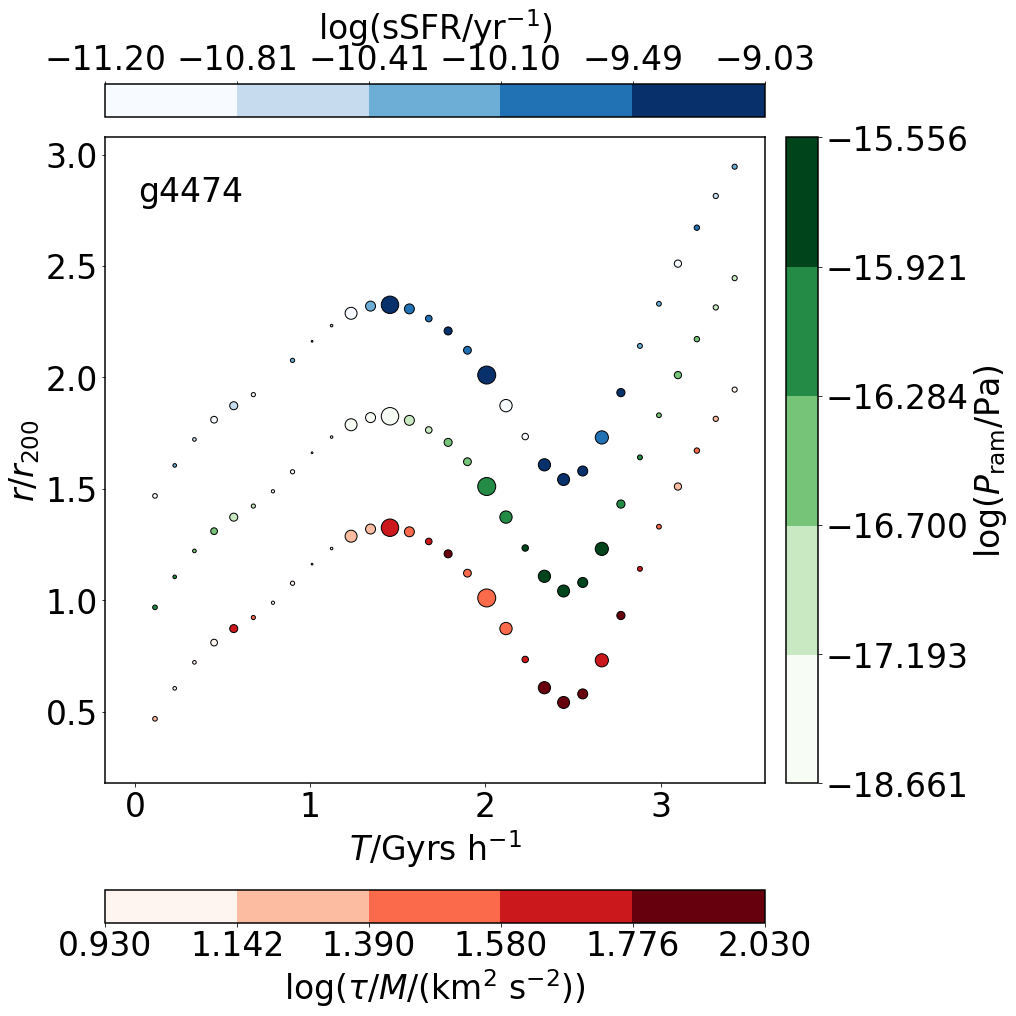}
	
	\caption{\label{fig_distTime_RPsSFRTT_append} Variation of tidal torques, ram pressure values and sSFR with time and distance for the rest of the galaxies in the sample (as in figure \ref{fig_distTime_RPsSFRTT}). Satellite galaxies in h6471 (top panels): g4339 (left), g4341 (middle) and  g4343 (right) and in h6742 (bottom panels):   g4470 (left), g4473 (middle) and g4474 (right).}
\end{figure*}

\begin{figure*}
	\includegraphics[width=0.32\textwidth]{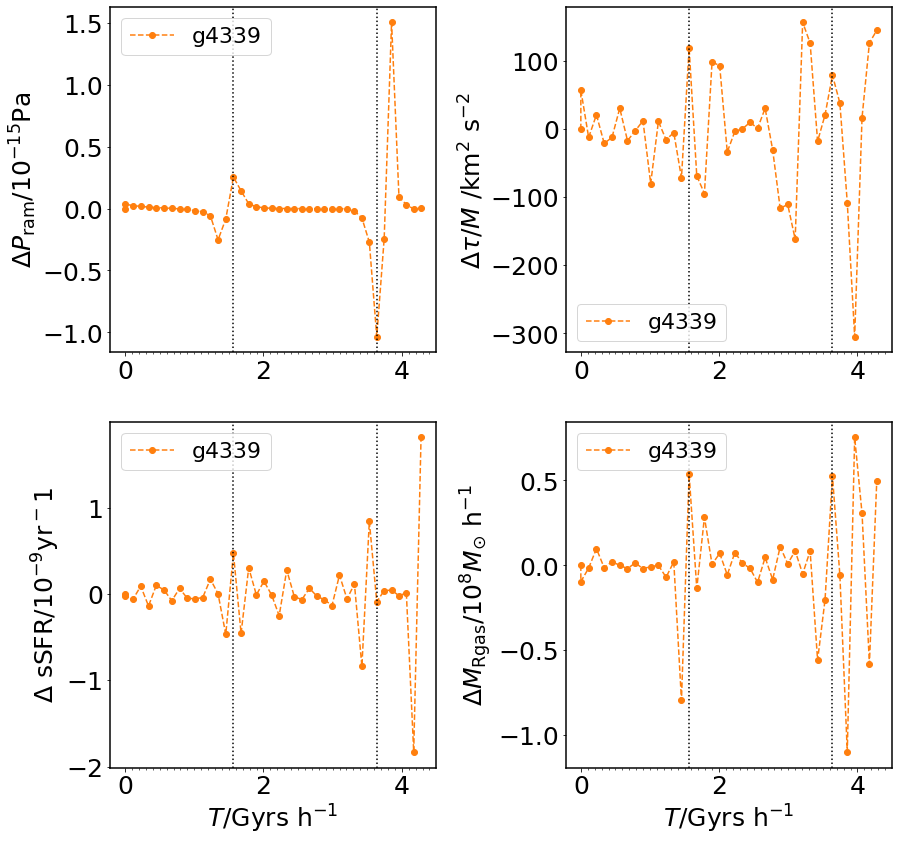}
	\includegraphics[width=0.32\textwidth]{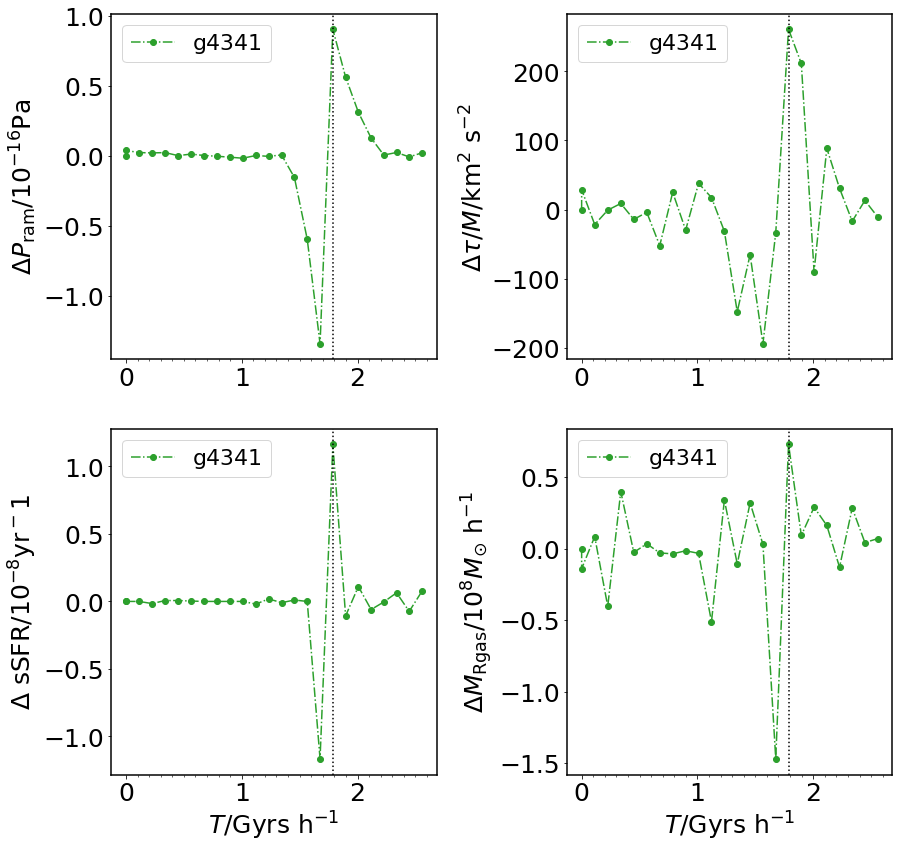}
	\includegraphics[width=0.32\textwidth]{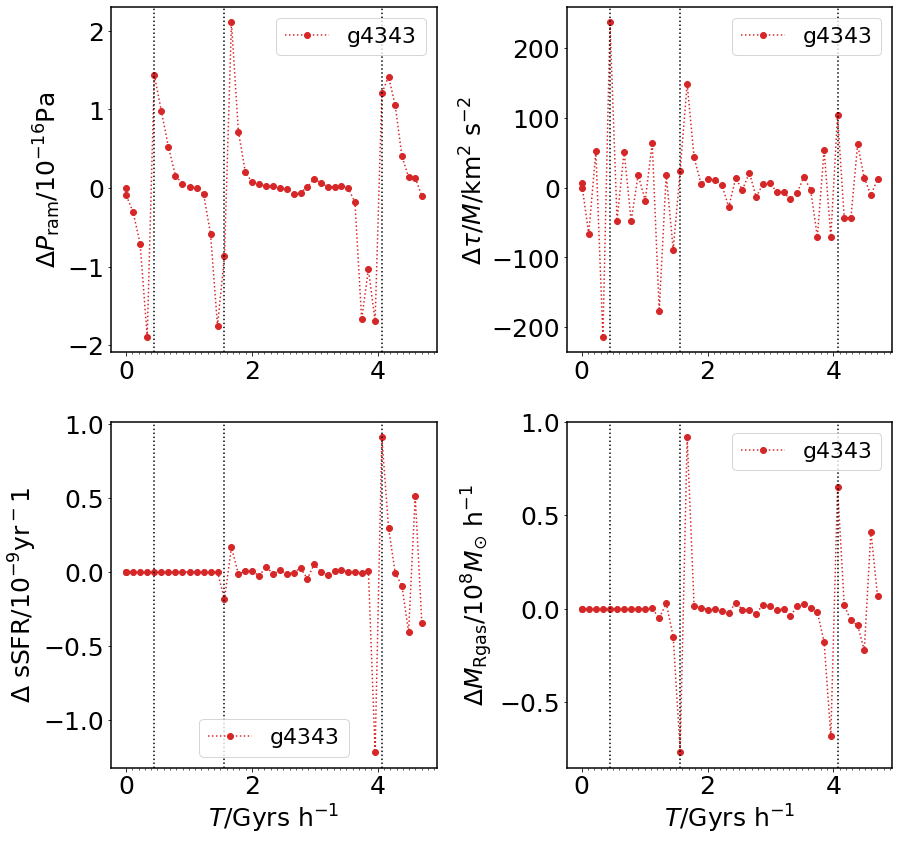}
	
	\includegraphics[width=0.32\textwidth]{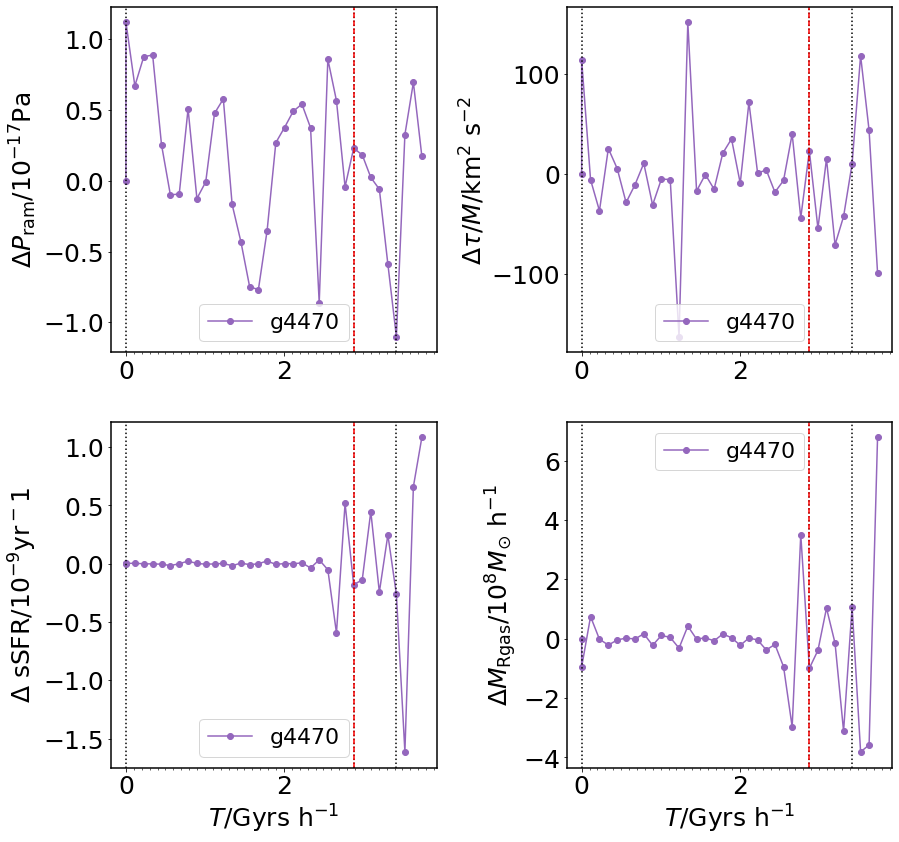}
	\includegraphics[width=0.32\textwidth]{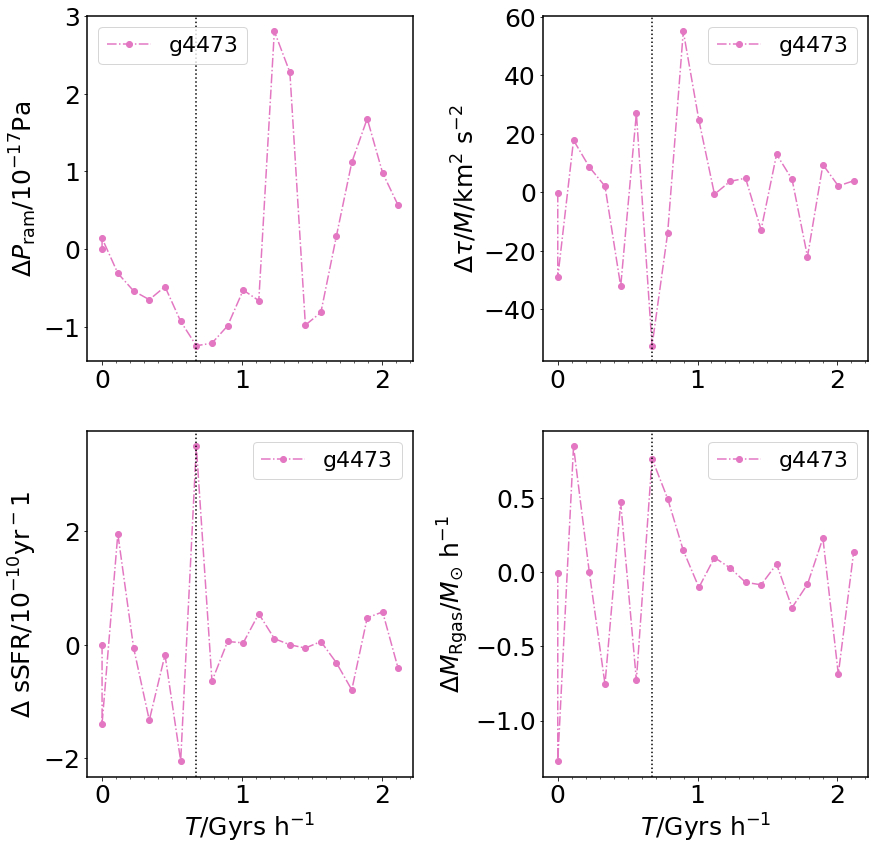}
	\includegraphics[width=0.32\textwidth]{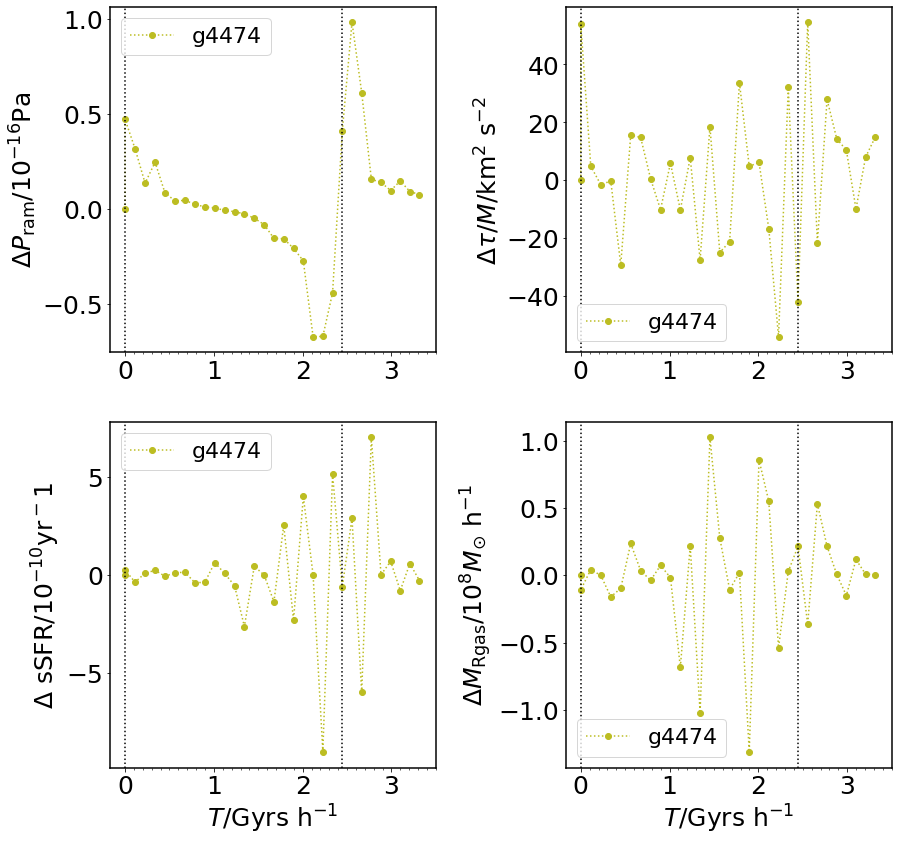}
	\caption{\label{fig_deltas_append} Rate of variation of ram pressure, tidal torque, sSFR and amount of removed gas (as in figure \ref{fig_deltas}). The top left set of plots (dashed orange lines) are for  g4339. The top middle set of plots (green dot-dashed lines) are for  g4341. The top right set of plots (red dotted lines) are for  g4343. The bottom left set of plots (purple solid lines) are for  g4470. The middle bottom set of plots (pink dot-dashed lines) are for g4473. The bottom right set of plots (olive dotted lines) are for  g4474.}
\end{figure*}

In this section, we display the properties of the satellite galaxies not shown in the main body of the paper. Figure \ref{fig_Orientation_append}, Fig.\ref{fig_cosVSeffect_append}, Fig.\ref{fig_distTime_RPsSFRTT_append} and Fig.\ref{fig_deltas_append} illustrate both the global similarities in the trends and the particularities associated to their specific properties and orbital parameters as discussed above.

\label{lastpage}
\end{document}